\documentclass[epj-spec]{svjour}
\usepackage{graphicx}
\usepackage{a4,amsmath,amssymb,srcltx}

\usepackage{amsfonts}

\begin{document}
\title{Applying GSH to a Wide Range of\\  Experiments in  Granular Media}
\author{Yimin Jiang\inst{1}
\and Mario Liu \inst{2}
\fnmsep\thanks{\email{mliu@uni-tuebingen.de}} }
\institute{Central South University, Changsha 410083, China \and
Theoretische Physik, Universit\"{a}t T\"{u}bingen,72076
T\"{u}bingen, Germany}
\abstract{Granular solid hydrodynamics ({\sc gsh}) is a continuum-me\-chanical theory for granular media, the range of which is shown in this paper. Simple, frequently analytic solutions are related to classic observations at different shear rates, including: 
(i)~static stress distribution, clogging;
(ii)~elasto-plastic motion: loading and unloading, approach to the critical state, angle of stability and repose;
(iii)~rapid dense flow: the $\mu$-rheology, Bagnold scaling and the stress minimum; (iv)~elastic waves, compaction, wide and narrow shear band. Less conventional experiments have also been considered: shear jamming, creep flow, visco-elastic behavior and nonlocal fluidization. 
With all these phenomena ordered, related, explained and accounted for,  though frequently qualitatively, we believe that {\sc gsh} may be taken as a unifying framework, providing the appropriate macroscopic vocabulary and mindset that help one coming to terms with the breadth of granular physics.  
}
\maketitle 
\tableofcontents

\section{Introduction\label{intro}} 
Being a subject of practical importance, elasto-plastic deformation of dense granular media has been under the focus of engineering research for many decades if not
centuries~\cite{schofield,nedderman,wood1990,kolymbas1,kolymbas2,gudehus2010,goddard2013}. The state of geotechnical theories, however, is confusing, at least for physicists: Innumerable
continuum mechanical models compete, employing strikingly different
expressions. In a recent
book on soil mechanics, phrases such as {\em morass of equations} and {\em
jungle of data} were employed as metaphors~\cite{gudehus2010}. 
Moreover, this competition is among theories applicable only to the slow shear rates of elasto-plastic 
deformation, while rapid dense flow (such as heap flow or
avalanches) is taken to obey yet rather different equations~\cite{hutter2007}. All this renders a  unified theory capable of accounting for granular phenomena at different rates seemingly illusory. 

This is the reason we adopted a different approach,  focusing on the physics and leaving  the rich and subtle granular phenomenology aside while constructing the theory. Our hope was to arrive at a set of equations that is firmly based in physics,  broadly applicable, and affords a well founded, correlated understanding of granular media.  

The formalism we employ is called the {\em hydrodynamic theory} (which physicists take to be the long-wave-length, continuum-mechanical  theory of condensed systems, in contrast to its more widespread usage, as a synonym for the Navier-Stokes' equations).
The hydrodynamic formalism  was pioneered by Landau~\cite{LL6} and Khalatnikov~\cite{Khal} in the context of superfluid helium, and 
introduced to complex fluids by de Gennes~\cite{deGennes}. Its two crucial points are: 
The input in physics that specifies the complete set of state variables, and the simultaneous consideration of energy and momentum conservation. As a result, there are many more constraints, and far less liberty, than the usual  approach of constitutive relations.  Moreover, being derived from physics rather than a subset of experimental data, if the theory renders some phenomena correctly, chances are that the rest is also adequately accounted for.~\footnote{We note there are also constitutive approaches which starts successfully from physics, more specifically from micromechanical properties of granular ensembles~\cite{microMech}.}

Hydrodynamic theories~\cite{hydro-1,hydro-2} have been derived for many condensed 
systems, including liquid crystals
~\cite{liqCryst-1,liqCryst-2,liqCryst-3,liqCryst-4,liqCryst-5,liqCryst-6,liqCryst-7},
superfluid $^3$He~\cite{he3-1,he3-2,he3-3,he3-4,he3-5,he3-6},
superconductors~\cite{SC-1,SC-2,SC-3}, macroscopic
electro-magnetism~\cite{hymax-1,hymax-2,hymax-3,hymax-4},
ferrofluids~\cite{FF-1,FF-2,FF-3,FF-4,FF-5,FF-6,FF-7,FF-8,FF-9,FF-10}, and
polymers~\cite{polymer-1,polymer-2,polymer-3,polymer-4}.
We contend that a hydrodynamic theory is also useful and possible for granular media: Useful, because it should help to illuminate and order their
complex behavior; possible, because a significant portion is already accomplished. We call it ``granular solid hydrodynamics,'' abbreviated as  {\sc gsh}. 

The structure of  {\sc gsh} is, as far as we can see, adequate and complete. Starting from two basic notions, {\it two-stage irreversibility} and {\it variable transient elasticity}, we have set up the theory in~\cite{granR2,granR3,granRgudehus,granR4}. In this paper, we focus on 
applying these equations to varying circumstances, a large collection of 
experiments. In fact,  no other continuum mechanical theory comes even close. ({\sc gsh} is summarized in Sec.\ref{GSH}. It is not a  derivation, only meant to keep this paper self-contained.)   

There are two aspects of {\sc gsh} that we need to communicate: the ideology of its approach  and the number of experiments it accounts for. Some of our starting points, such as energy conservation or the validity of thermodynamics, are not generally accepted in the granular community. 
We have detailed our reasons why we believe our postulates are appropriate in~\cite{granR2,granR3,granRgudehus,granR4}, and shall not repeat them here.  One of our hopes for the present paper is that the second aspect of  {\sc gsh},  impressive and easily accessible, is also quietly convincing -- or at least thought-provoking, for those who still have doubts about the basic approach of  {\sc gsh}.

\section{A Brief Presentation of GSH \label{GSH}    } 
As any hydrodynamic theory,  {\sc gsh} has two parts, structure and parameters. The first is derived from general principles, but the second -- values and  functional dependence of the energy and transport coefficients -- are inputs, obtained either from a microscopic theory (a tall order in any dense systems), or in a trial-and-error iteration, in which the ramifications of postulated dependences are compared to experiments and simulations. 
Many details of granular phenomena depend on these parameters, and we are still in the midst of the iteration evaluating them. More specifically, we have an energy expression that is both simple and realistic, but the transport coefficients are in a less  satisfactory state: Their dependence on $T_g$, obtained from more general considerations,  seems quite universal, but the density dependence is not. Varying with details possibly including rigidity, shape and friction of the grains, they are material-specific and hard to arrive at in the absence of more systematic data. These need to be given by a complete range of experiments in uniform geometries employing only  {\it one kind of grains}. Nevertheless, in spite of the tentative character of the density dependence assumed below, our results do show at least qualitative agreement with experiments and realistic constitutive models.

\subsection{The State Variables}
A {\it complete set of state variables} is one that
uniquely determines a macroscopic state of the system. If it is given, there is no room
for ambiguity or ``history-dependence." Conversely, any such
dependences indicate that the set is incomplete. In the hydrodynamic
theory, a physical quantity is a state variable if the
energy density $w$ depends on it. In {\sc gsh},  the state variables are, in 
addition to the usual ones (the density $\rho$, the momentum density $\rho 
v_i$, the true entropy $s$): the granular entropy $s_g$ and the elastic strain 
$u_{ij}$. Entropy $s_g$, along with $T_g\equiv\partial w/\partial s_g$,  
quantifies granular jiggling and is closely associated with the averaged velocity 
fluctuation $\delta\bar v\equiv\sqrt{\langle v_i^2\rangle -\langle 
v_i\rangle^2}$. (It would be wrong to take $T_g\sim\delta\bar v^2$, because 
any kinetic theory fails for $T_g\to0$, when enduring contacts dominate, 
see~\cite{GGas}, also~\cite{granR2,granR3,granR4}.) 

The elastic 
strain $u_{ij}$ is associated with the deformation of the grains (or in DEM-jargon: their {\it overlap}). We  do not consider the true entropy  $s$ below, although it is undoubtedly a state variable, because effects such as thermal expansion are not at present under our focus. 
Fabric anisotropy $f_{ij}$, the number of average contacts in different 
directions, is a useful microscopic characterization of granular states. But there is insufficient evidence that it is macroscopically {\it independent}. To keep {\sc 
gsh} as simple as possible, our working hypothesis is that it is not. In~\cite{luding2011}, Magnanimo and Luding employ $f_{ij}$ to account for 
the anisotropic velocity of elastic waves, because their theory uses linear elasticity and does not have stressed-induced anisotropy. {\sc gsh} does and 
yields  velocities very close to the measured ones, without $f_{ij}$, see~\cite{ge4}. We note that anisotropy of elastic waves that persists for 
isotropic stress and  $u_{ij}$ would be a sign that $f_{ij}$ is an independent variable.

Denoting the (rest-frame or internal) energy density as $w=w(\rho, s_g, u_{ij})$,
we define the conjugate variables as: 
\begin{equation}\label{2-2} \mu\equiv\frac{\partial w}{\partial\rho},\quad T_g\equiv\frac{\partial
w}{\partial s_g},\quad \pi_{ij}\equiv-\frac{\partial w}{\partial u_{ij}},
\end{equation} 
calling $\mu$ is the  chemical potential, $T_g$ the 
granular temperature, and $\pi_{ij}$ the elastic stress. These are given once the energy $w$ is. (See~\cite{granR2,granR3,granRgudehus,granR4} for a a treatment including the true entropy $s$ and temperature $T\equiv{\partial
w}/{\partial s}$.)
 %


There are  three spatial scales in any granular media: (a)~the macroscopic, (b)~the mesoscopic or inter-granular, and (c)~the microscopic or inner granular. Dividing all degrees of freedom (DoF) into these three categories, we treat those of (a) differently from (b,c). Macroscopic
DoF, such as the slowly varying stress, flow and density fields, are
employed as state variables, but inter- and inner granular DoF are treated summarily: 
Only their contributions to the energy are considered and taken,
respectively, as granular and true heat. So we do not account for the motion of a jiggling grain, only include its fluctuating kinetic and elastic energy as contributions to the granular 
heat, $\int T_g{\rm d}s_g$. 
Similarly, phonons are part of true heat, $\int T{\rm d}s$. There are a handful of macroscopic DoF~(a), many inter-granular ones (b), and innumerable inner granular ones (c). So the statistical tendency to equally distribute the energy among all DoF implies an energy decay: (a) $\to$ (b,c) and (b) $\to$ (c). 
(In kinetic theories, assuming $T_g\gg T$ holds, the (b) $\to$ (c) decay is replaced by a constant restitution coefficient~\cite{granR4}.)
This is what we call {\em two-stage irreversibility}.


With $v_{ij}\equiv\frac12(\nabla_iv_j+\nabla_jv_i)$,
$v^*_{ij}$ its traceless part, $v_s^2\equiv v^*_{ij}v^*_{ij}$, the balance equation for $s_g$ (closely related to the energy balance in the kinetic theory~\cite{luding2009}) is 
\begin{equation}\label{2c-4} 
\partial_ts_g+\nabla_i(s_gv_i-\kappa\nabla_iT_g)=
(\eta_g v_s^2+\zeta_gv^2_{\ell\ell}-\gamma T_g^2)/T_g. 
\end{equation}
Here, $s_gv_i$ is the convective, and $-\kappa\nabla_iT_g$ the diffusive, flux. $\eta_g v_s^2$ accounts for viscous heating, for the increase of $T_g$ because a
macroscopic shear rate jiggles the grains. A compressional rate
$\zeta_gv^2_{\ell\ell}$ does the same, though not as effectively~\cite{granL3}. The term $-\gamma T_g^2$ accounts for the relaxation of $T_g$, the  (b) $\to$ (c) decay of energy.  

Our second notion, {\em variable transient elasticity}, addresses the interplay between elaticity and plasticity.
The free surface of a granular system at rest is frequently tilted. When
perturbed, when the grains jiggle and $T_g\not=0$, the tilted surface will
decay and become horizontal. The stronger the grains jiggle
and slide, the faster the decay is. We take this as indicative of a system
that is elastic for $T_g=0$, transiently elastic for $T_g\not=0$,
with a stress relaxation rate $\propto T_g$. 
A relaxing stress is typical of any viscous-elastic system such as polymers~\cite{polymer-1}. 
The unique circumstance here is that the relaxation rate is not a material
constant, but a function of the state variable $T_g$. As we shall see, it
is this {\em variable transient elasticity} -- a simple fact at heart -- that underlies the complex behavior of granular plasticity. This is an insight that yields a most economic way to capture granular rheology.

Employing the strain rather than stress as a state variable yields a simpler description, because the former is a geometric quantity, the latter a physical one (that includes material constants such as the stiffness). Yet one cannot use the standard strain $\varepsilon_{ij}$, because the relation between stress and $\varepsilon_{ij}$ lacks uniqueness when the system is plastic. 
Engineering theories frequently divide the strain into two fields, elastic $u_{ij}$ and plastic
$\varepsilon^{p}_{ij}$, with the first accounting for
the reversible and second for the irreversible part. They then employ $\varepsilon_{ij}$ and $\varepsilon^{p}_{ij}$ as two independent variables to account for the elasto-plastic motion~\cite{Houlsby,Houlsby2}.
We believe that, on the contrary, the elastic strain $u_{ij}$
is the sole state variable. As convincingly argued by Rubin~\cite{rubin}, there is a unique relation between $u_{ij}$ and the elastic stress $\pi_{ij}$. We take $u_{ij}$ as the portion of the strain that deforms the grains, changes the elastic energy $w=w(u_{ij})$, and builds up an elastic stress $\pi_{ij}$. Employing $u_{ij}$ preserves useful features of elasticity, especially the  relation, $\pi_{ij}=-\partial w(u_{ij})/\partial u_{ij}$, cf.\cite{granR2}. 

This is easy to understand via
an simple analogy. The wheels of a car driving up a snowy hill will grip the ground  part of the time, slipping otherwise. When the wheels grip, the car moves and its gravitational energy $w$ is increased (same as only $u_{ij}$ increases the elastic energy). Dividing the wheel's rotation $\theta$ into a gripping $\theta^{(e)}$ and a slipping $\theta^{(p)}$ portion, we may compute the torque on the wheel as $\partial w/\partial\theta^{(e)}$ [same as $\pi_{ij}=-\partial w(u_{ij})/\partial u_{ij}$]. How much the wheel turns or slips, how large $\theta$ or $\theta^{(p)}$ are, is irrelevant for the torque.
The equation for $u_{ij}$ is
\begin{equation}\label{2c-6a}
\partial_tu_{ij}-v_{ij}+\alpha_{ijk\ell}v_{k\ell}=-(\lambda_{ijk\ell}T_g)\,u_{k\ell},
\end{equation} 
cf.\cite{granR2} for the general expression including the objective derivative. (In contrast to the total strain, the change in the elastic one $u_{ij}$ remains small, rendering the additional terms irrelevant -- unless one wants to describe, say, a rotating sand pile.) If 
$T_g$ is finite, grains jiggle and briefly
lose or loosen contact with one another, during which their deformation is
partially lost. Macroscopically, this shows up as a relaxation of
$u_{ij}$, with a rate that grows with $T_g$, and vanishes for
$T_g=0$, with the lowest order term in a $T_g$-expansion being
$\lambda_{ijk\ell}T_g$. Within its range of stability, the energy $w$ is convex, and $-\pi_{ij}\equiv\partial w/\partial u_{ij}$ is a monotonic function of $u_{ij}$. So  $-\pi_{ij}, u_{ij}$
decrease and relax at the same time, in accordance to Eq~(\ref{2c-6a}). 

Conservation of momentum, $\partial_t(\rho v_i)+\nabla_j(\sigma_{ij}+\rho v_iv_j)=g_i\rho$ and mass, $\partial_t\rho=-\nabla_i(\rho v_i)$, close the set of equations. The Cauchy stress $\sigma_{ij}$ is (see~\cite{granR2,granR3,granRgudehus,granR4}): 
\begin{align}\label{ST}
\sigma_{ij}=\pi_{ij}-\alpha_{k\ell ij}\pi_{k\ell}+ (P_T-\zeta_g v_{\ell\ell})\delta_{ij}-\eta_gv^*_{ij},\\
P_T\equiv-\partial(w/\rho)/\partial (1/\rho)=Ts+T_gs_g+\mu\rho-w,
\end{align}
where $P_T$ (that  will turn out to be the kinetic pressure) and $\pi_{ij}$ are given by Eqs.(\ref{2-2}). The total stress $\sigma_{ij}$, though generally valid, is explicit only if $w$ is given. The terms $\propto\zeta_g,\eta_g$ are the viscous stress; the tensor  $\alpha_{ijk\ell}$ is an off-diagonal Onsager coefficient that couples the stress components and softens them. The above expressions yield the structure of {\sc gsh}. 
Next, we specify the energy and transport coefficients.

\subsection{The Energy\label{granEn}} 
Due to a lack of interaction among the grains, the energy density $w$ vanishes when the grains are neither deformed nor jiggling.  Assuming $w=w_T(\rho,s_g)+w_\Delta(\rho,u_{ij})$, we have $w_T\to0$ for $s_g\to0$, and  $w_\Delta\to0$ for $u_{ij}\to0$.  So, considering  slightly excited, stiff grains (such that the lowest order terms in $u_{ij},s_g$ suffice), we take
\begin{align}
\label{2b-2}  w_T={s_g^2}/{(2\rho b)},\quad 
w_\Delta=\sqrt{\Delta }(2 {\mathcal B} \Delta^2/5+ {\mathcal A}u_s^2),
\\\label{2b-2a} 
\pi_{ij}=\sqrt\Delta({\cal B}\Delta+{\cal A}
{u_s^2}/{2\Delta})\delta _{ij}-2{\cal A}\sqrt\Delta\, u_{ij}^*, 
\\\label{2b-2b} 
P_\Delta=\sqrt\Delta({\cal B}\Delta+{\cal A}
{u_s^2}/{2\Delta}),\quad \pi_s=-2{\cal A}\sqrt\Delta\, u_s,
\\\label{2b-1}
{4P_\Delta}/{|\pi_s|}=2({\cal B}/{\cal
A})(\Delta/u_s)+{u_s}/{\Delta},\qquad\qquad \end{align} 
where $\Delta\equiv -u_{\ell\ell}$, $P_\Delta\equiv\pi_{\ell\ell}/3$, $u_s^2\equiv
u^*_{ij}u^*_{ij}$, $\pi_s^2\equiv \pi^*_{ij}\pi^*_{ij}$, with
$u^*_{ij},\pi^*_{ij}$ the respective traceless tensors.
$w_T$ is an expansion in $s_g$. The quadratic term is the lowest order one
because $s_g\sim T_g=0$ is an energy  minimum. (As we shall soon see,  the $s_g^2$-term is in fact sufficient to account for fast dense flow and the gaseous state.)

Calling something a temperature, we also give it the dimension kelvin or energy. Taking $[s_g]=$ 1/vol, $[T_g]=$ energy, implies $[1/\rho b]=$ energy $\times$ vol. But we note the following point: Equilibration, or equality  of temperatures, is usually a ubiquitous process, and what requires all temperatures to possess the same dimension.  However, granular media in ``thermal contacts" do not usually equilibrate -- in the sense that the energy distribution is independent of details, and the energy flux vanishes. Given two different granular systems, 1 and 2, with only 1 being driven, there are,  in the steady state, four temperatures: $T^1,T_g^1,T^2,T_g^2$, with an ongoing energy transfer:  $T_g^2\to T^2$ and $T^1_g\to T^1,T_g^2$, such that none of the temperatures is equal to another. The differences depend on details such as the  contact area and the respective restitution coefficients. Only when the driving stops, will they eventually become equal, but this is well approximated by $T_g^1=T_g^2=0$. Therefore, there is no harm in giving $s_g$ or $T_g$ any dimension -- as long as $T_gs_g$ is an energy density. 

Given Eq.(\ref{2b-2}) with $b=b(\rho)$, there is quite generally a pressure contribution $P_T$,  
\begin{equation}
-P_T\equiv\left.\frac{\partial (w_T/\rho)}{\partial 1/\rho}\right|_{s_g}=\left.\frac{\partial [(w_T-T_gs_g)/\rho]}{\partial 1/\rho}\right|_{T_g}=\frac{T_g^2\rho^2}{2}\frac{\partial b}{\partial\rho}.
\end{equation} 
We choose  $b=b(\rho)$ such that it yields the kinetic pressure $\propto w_T$ for the rarefied limit $\rho\to0$, and the usual form $\propto
w_T/(\rho_{cp}-\rho)$ in the dense limit $\rho\to\rho_{cp}$, see~\cite{Bocquet,luding2009},
\begin{equation}\label{2b-5}
b=b_1\rho^{a_1}+b_0\left[1-\frac{\rho}{\rho_{cp}}\right]^a,\quad
P_T=\frac{w_T}{b}\left[\frac{ab\cdot\rho/\rho_{cp}}{1-\rho/\rho_{cp}}-a_1b_1\rho^{a_1}\right]\equiv g_p(\rho)T_g^2, \end{equation}
with $a\approx0.1$ a small positive number, and $-a_1=2/3, 1$ for two and three dimensions, respectively. For $\rho\to0$, we have $b\approx b_1$, $P_T\approx-a_1w_T$, with $w_T=\frac12\rho\delta\bar v^2=\frac32T_k\rho/m$ in three dimensions (where $\delta\bar v^2\equiv{\langle v_i^2\rangle -\langle v_i\rangle^2}$, $T_k$ denotes the temperature of the kinetic theory, and $P_T=T_k\rho/m$ the usual kinetic pressure). In the dense limit, the first term in $P_T$ dominates, and the pressure is as desired $\propto w_T/(\rho_{cp}-\rho)$. (The term $\propto b_1$ is new, and not in \cite{granR2,granR3}.)

Without equilibration, there is no thermometers that measures $T_g$. It is 
therefore useful to relate $T_g$ to $\delta\bar v$, a quantity that is directly 
measurable, at least in simulations. This is easily done for two limits, because 
$w_T=\frac12\rho\delta\bar v^2$ or $\delta\bar v=T_g\sqrt{b}$ in the rarefied 
one; and $w=\rho\delta\bar v^2$ or $\delta\bar v=T_g\sqrt{b/2}$ in the dense 
one. (For $\rho\to\rho_{cp}$, granular jiggling occurs in a network of linear oscillators, which oscillate weakly around the static stress. So there is on 
average as much potential energy as kinetic one.)  We note that, for given $\delta\bar v$, the energy $w_T$ remains finite in both limits, although $b$ 
diverges and $T_g$ vanishes for $\rho\to0$. 

The expression for $w_\Delta$, with ${\cal A,B}>0$, is the elastic contribution. Given by the energy of linear elasticity multiplied by $\sqrt\Delta$, the form is clearly inspired by the Hertzian contact, though its connection to granular elasticity goes beyond that, and includes both {\it stress-induced anisotropy} and the {\it convexity transition} (see below). The elastic stress $\pi_{ij}$ has been validated for the following circumstances, achieving at least semi-quantitative agreement: 
\begin{itemize}
\item  Static stress distribution in three classic geometries: silo, sand pile,
point load on a granular sheet, calculated employing $\nabla_i\pi_{ij}=\rho g_i$, see~\cite{ge1,ge2,granR1}. 
\item  Incremental stress-strain relation, starting from varying static stresses~\cite{kuwano2002,ge3}.
\item  Propagation of anisotropic elastic waves at varying static stresses~\cite{jia2009,ge4}. 
\end{itemize}

 {\it Stress-induced anisotropy}: In  linear elasticity, $w\propto u_s^2$, the velocity of an elastic wave $\propto\sqrt{\partial^2w/\partial u_s^2}\,$ does not depend on $u_s$, or equivalently, the stress. For any exponent other than 2, the velocity depends on the stress, and is anisotropic if the stress is.  
We note that $u_{ij}$ and $\pi_{ij}$ from the expression of Eq.(\ref{2b-2a}) are colinear, in the sense that $u_{ij}^*/u_s=\pi_{ij}^*/\pi_s$ holds (but not $\varepsilon_{ij}$). They also have the same principal axes. More recently, we have employed a slightly more complicated $w_\Delta$ that includes the third strain invariant~\cite{3inv}. Here, colinearity is lost, but strain and stress still share the same principle axis. 

{\it Convexity Transition}: 
In a space spanned by stress components and the density, there is a surface  that 
divides two regions in any granular media, one in which the grains are necessarily agitated, 
another in which they may be in a static, non-dissipating state. The most obvious such surface exists with respect to the density -- when it is too small, 
grains loose contact with one another and cannot stay static. Same holds if the shear stress is too larger for given pressure, say when the slope of a sand pile is too steep. 
Note the collapse occurs in a completely static system. This is qualitatively different from the critical state, because the latter, and the approach to it, takes place in a dissipating system, at given rate and  $T_g$. These two require different descriptions, static versus dynamic. We consider the static description here, and shall return to the critical state in  Sec~\ref{critical state}. 

In Eq.(\ref{2b-4}), we introduce two material parameters, $\rho_{\ell p}$ and $\rho_{cp}$. Calling the first {\it the random-loose density}, we take it to be the lowest density at which any elastic state may be maintained, where elastic solutions are stable. The second, termed {\it random-close density}, is taken as the highest one at which grains may remain uncompressed. For lack of space, grains cannot rearrange at $\rho_{cp}$, and do not execute any plastic motion.

The divide between two regions, one in which elastic solutions are stable, and another in which they are not, in which infinitesimal perturbations suffice to destroy the solution, is the surface where the second derivative of the elastic energy changes its sign, where it turns from convex to concave. 
The elastic energy of Eq~(\ref{2b-2}) is convex only for 
\begin{equation}\label{2b-3} u_s/\Delta\le\sqrt{2{\cal B}/{\cal A}} \quad
\text{or}\quad \pi_s/P_\Delta\le\sqrt{2{\cal A}/{\cal B}},
\end{equation} 
turning concave if the condition is violated. (The second condition may be derived by considering Eq~(\ref{2b-1}),  
showing $P_\Delta/\pi_s=\sqrt{{\cal B}/2{\cal A}}$ is minimal for
$u_s/\Delta=\sqrt{2{\cal B}/{\cal A}}$.) 
Assuming ${\cal B}/{\cal A}$ is density-independent (typically 5/3),  denoting $\bar\rho\equiv(20\rho_{\ell p}-11\rho_{cp})/9$, 
we take   
\begin{equation}\label{2b-4} {\cal B}={\cal B}_0
[(\rho-\bar\rho)/(\rho_{cp}-\rho)]^{0.15},
\end{equation}
with ${\cal B}_0>0$ a constant. This expression accounts for three  granular characteristics:
\begin{itemize}
\item The energy is concave for any density smaller than $\rho_{\ell p}$. 
\item The energy is convex between $\rho_{\ell p}$ and $\rho_{cp}$, ensuring the stability of any elastic solutions in this region. In addition, the density dependence of sound
velocities (as measured by Harding and Richart~\cite{hardin}) is well
rendered by  $\sqrt{{\cal B}(\rho)}$. 
\item  The elastic energy diverges, slowly, at
$\rho_{cp}$, approximating the observation that the system becomes an order of magnitude stiffer there.
\end{itemize}
One may be bothered by the small exponent of 0.15, questioning whether we imply an accuracy over a few orders of magnitude. We do not: Since $\cal B$ loses its convexity at $\rho_{\ell p}$, the density is never close to $\bar\rho$ (note $\bar\rho<\rho_{\ell p}<\rho_{cp}$, with  $\rho_{cp}-\rho_{\ell p}\approx\rho_{\ell p} -\bar\rho$). And although $\rho$ may in principle be close to $\rho_{cp}$, it is very difficult to reach, and the slow divergence is not really relevant. Given $ {\cal B}(\rho)$, there is also a contribution $\propto\Delta^{2.5}$ to $P_T$ from  $w_\Delta$. It is neglected because it is (for small $\Delta$) much smaller than the elastic one, $P_\Delta\propto\Delta^{1.5}$.

\subsection{The Dynamics \label{dynamics}} 
Dividing $u_{ij}$ into its trace $\Delta\equiv-u_{\ell\ell}$ and traceless part $u_{ij}^*$, and specifying the matrices $\alpha_{ijk\ell},\lambda_{ijk\ell}$ with two elements each, $\alpha,\alpha_1,\lambda,\lambda_1$, the equation of motion~(\ref{2c-6a}) is written as
\begin{align}
\label{2c-7}
\partial_t\Delta+(1-\alpha )v_{\ell\ell} -\alpha_1u^*_{ij}v^*_{ij}
=-\lambda_1T_g\Delta, 
\\\label{2c-8} 
\partial_tu^*_{ij}-(1-\alpha )v^*_{ij}
= -\lambda T_gu^*_{ij},
\\\label{2c-9}
\partial_tu_s-(1-\alpha )v_s= -\lambda T_gu_s.
\end{align} 
The third equation is valid only if strain and rate are colinear, $u^*_{ij}/|u_s|=v^*_{ij}/|v_s|$. This is frequently the case  for a steady rate, because any 
 non co-linear component of $u_{ij}$ relaxes to zero quickly. 
The coefficient  $\alpha$ describes softening  (if $0<\alpha <1$), or more precisely a reduced gear ratio: The same shear rate yields a smaller deformation, $\partial_tu_{ij}=(1-\alpha)v_{ij}+\cdots$, but 
acts also at a smaller stress, $\sigma_{ij}=(1-\alpha)\pi_{ij}\cdots$, see Eqs.(\ref{2c-2},\ref{2c-2a}).  $\alpha_1$ accounts for the fact that shearing granular media will change the compression $\Delta$, implying {\em dilatancy} and {\em contractancy}. (More Onsager coefficients are permitted by symmetry, but  excluded here to keep the equations simple.) 
The Cauchy or total stress is now
%
\begin{align} \label{2c-2}
P\equiv\sigma_{\ell\ell}/3=(1-\alpha )P_\Delta+P_T-\zeta_gv_{\ell\ell},
\\\label{2c-2a}
\sigma^*_{ij}=(1-\alpha)\pi_{ij}^*-\alpha_1u^*_{ij}P_\Delta -\eta_gv^*_{ij},
\\ \label{2c-2b}
\sigma_{s}=(1-\alpha )\pi_s+\alpha_1u_sP_\Delta
+\eta_gv_s. 
\end{align} 
Again, the third equation (with $\sigma_s^2\equiv\sigma_{ij}^*\sigma_{ij}^*$) is
valid only if $\pi_{ij}^*$, $u_{ij}^*$ and $v^*_{ij}$ are colinear,
$\pi_{ij}^*/|\pi_s|=-u_{ij}^*/|u_s|=-v^*_{ij}/|v_s|$, often the case in steady state. 
The pressure $P$ and shear stress $\sigma_s$ contain elastic contributions
$\propto\pi_s,P_\Delta$ from Eq~(\ref{2b-2b}), and seismic (ie. $T_g$-dependent) ones: $P_T\propto T_g^2$ from Eq~(\ref{2b-5}), and the viscous stress $\propto\eta_g,\zeta_g$. The  coefficients $\alpha ,\alpha_1$ soften and mix the stress components. The term preceded by $\alpha_1$ is smaller by an order in the elastic strain, and may be neglected, as we shall do in this paper, if $\alpha_1$ is not too large.

The transport coefficients $\alpha,\alpha_1,\eta_g,\zeta_g$ are functions of the state variables, $u_{ij}$, $T_g$ and $\rho$. As explained above, 
they are to be obtained from experiments, in a trial-and-error iteration. And the specification below is what we at present believe to be the appropriate ones. Generally speaking, we find strain dependence weak -- plausibly so because the elastic strain is a small 
quantity. One expand in it, keeping only the constant terms. We also expand in $T_g$, but mostly eliminate the constant terms, as we take granular media to be fully elastic for $T_g\to0$, so the force balance $\nabla_j\sigma_{ij}=\rho{\rm g}_i$ reduces to its elastic form,  $\nabla_j\pi_{ij}=\rho{\rm g}_i$. This implies 
$\alpha,\alpha_1,\eta_g,\zeta_g,\kappa_g\to0$ for $T_g\to0$. In addition, we
take $\alpha,\alpha_1$ to saturate at an elevated $T_g$, such that rate-independence is established. Hence
\begin{align}\label{2c-3}
\eta_g=\eta_1T_g,\,\, \zeta_g=\zeta_1T_g,\,\,
\kappa=\kappa_1T_g,
\\\nonumber
\alpha/\bar\alpha =
\alpha_1/\bar\alpha_1={T_g}/({T_\alpha+T_g}),
\end{align} 
with $\bar\alpha,\bar\alpha_1,\eta_1,\zeta_1,\kappa_1,T_\alpha$ functions of $\rho$ only. Expanding $\gamma$ in $T_g$ yields $\gamma=\gamma_0+\gamma_1 T_g$. 
We keep $\gamma_0$, because the reason  leading to Eqs~(\ref{2c-3}) does not apply, and because $\gamma_0\not=0$ ensures a smooth transition from the hypoplastic to the quasi-elastic regime, see Eq~(\ref{TgVs2}) below. For lack of better information, we take $T_\alpha$ and $\gamma_0/\gamma_1$ to be of the same magnitude. 

Since granular media are elastic at $\rho_{cp}$, we have $\bar\alpha, \bar\alpha_1, \lambda, \lambda_1 \to 0$ for $\rho\to\rho_{cp}$, 
such that Eqs.(\ref{2c-7},\ref{2c-8}) assume the elastic form, while $\gamma_1$, the relaxation rate for $T_g$, and $\eta_1$, the viscosity,  diverge. Accordingly, we take (with $a_1,a_2,a_3,a_4,a_5>0$):
\begin{align}\label{DD}
r\equiv1-\rho/\rho_{cp},\quad \bar\alpha=\bar\alpha_0 r^{a_1},\quad \bar\alpha_1=\bar\alpha_{10}r^{a_2},\qquad \\ \lambda/\lambda_0=\lambda_1/\lambda_{10}=r^{a_3},\quad 
\eta_1=\eta_{10} r^{-a_4},\quad  \gamma_1=\gamma_{10} r^{-a_5}.  \nonumber
\end{align}
(Close to $\rho_{cp}$,  the dependence on $\rho_{cp}-\rho$ is the sensitive one, and we ignore any weaker ones on $\rho$ directly.) 
We stand behind the temperature dependence with much more confidence than that of the density, for two reasons: First, $\rho$ is not a small quantity that one may expand in, and we lack the general arguments employed to extract the $T_g-$dependence. Second, not coincidentally, the $\rho$ dependence does not appear universal: $a_4=a_5=1$ seems to fit glass beads  data, while $a_4=0.5$, $a_5=1.5$ appear more suitable for polystyrene beads~\cite{denseFlow}.   For the rest of the paper, when discussing the density dependence qualitatively, we shall use what we call {\it the exemplary values}:
$a_1=a_2=a_3=a_4=a_5=1$.

At given shear rates $v_s$, the stationary state of Eq~(\ref{2c-4}) -- with viscous heating  balancing $T_g$-relaxation and $\partial_ts_g=0$ -- is quickly
arrived at ($\lesssim10^{-3}$ s), implying 
\begin{align}\label{2c-6}
{\gamma_1} \,h^2\, T_g^2=v_s^2\,{\eta_1}+v^2_{\ell\ell}\,{\zeta_1},
\\
\nonumber\text{where}\,\,\,
h^2\equiv1+\gamma_0/(\gamma_1T_g).
\end{align} 
If the density is either constant or changing slowly, implying  $v^2_{\ell\ell}\approx0$, we have a quadratic regime for small $T_g$ and low $v_s$, and a linear one at elevated $T_g,v_s$:
\begin{align}
\label{TgVs} 
T_g=|v_s|\sqrt{\eta_1/\gamma_1}\quad\,\,\text{for}\quad \gamma_1T_g\gg\gamma_0,
\\\label{TgVs2} 
T_g=v_s^2\,\,({\eta_1/\gamma_0})\quad\text{for}\quad \gamma_1T_g\ll\gamma_0.
\end{align}
As discussed in detail in the next section, the linear regime is the {\it hypoplastic} one, in which the system displays elasto-plastic behavior and the hypoplastic model holds. In the quadratic regime, because $T_g\propto v_s^2\approx0$ is quadratically small and negligible,  the behavior is {\it quasi-elastic}, with slow, consecutive visit of static stress distributions. Note $h=1$ in the hypoplastic regime, $h\to\infty$ in the quasi-elastic one.   

We revisit Eq.(\ref{2c-4}), implementing the following simplifications: (1)~$\nabla_iT_g$ is assumed to be small and linearized in; so terms such as $(\nabla_iT_g)^2$ are eliminated. (2)~$T_g$'s convective term is taken to be negligible, as is $v_{\ell\ell}\approx0$, because density change is typically both small and slow. 
(3)~An extra source term $\gamma_1 h^2T_a^2$ is added to account for an ambient temperature $T_a$, which are external perturbations such as given by tapping or  a sound field.  Eq.(\ref{2c-4}) then reads
\begin{equation}
\label{Tg1} 
b\rho\partial_tT_g-\kappa_1T_g\nabla^2T_g=
\eta_1 v_s^2-\gamma_1 h^2(T_g^2-T_a^2). 
\end{equation}
Generally speaking, any source contributing to $T_g$ is already included. For instance, given a sound field and its compressional rate  $v_{\ell\ell}^{s}$, there is the term on the right hand side 
of Eq~(\ref{2c-4}),  $\zeta_1 (v_{\ell\ell}^{s})^2$. Coarse-graining it, we may set 
$\langle\zeta_1 (v_{\ell\ell}^{s})^2\rangle \equiv\gamma_1h^2T_a^2$. So adding such a term is simply a convenient way to account for a non-specific source. 
Finally, we rewrite Eqs.(\ref{2c-7},\ref{2c-8},\ref{2c-9},\ref{Tg1}) as coupled relaxation equations,  dimensionally streamlined with 3 time and 1 length scales,
\begin{align}\label{Tg2}
\partial_tT_g=-R_T[T_g(1-\xi_T^2\nabla^2)T_g-T_c^2-T_a^2],
\\\label{Tcf}T_c\equiv f|v_s|,\,  f^2\equiv\frac1{h^2}{\frac{\eta_1}{\gamma_1}},\, 
R_T\equiv\frac{\gamma_1h^2}{b\rho},\,
\xi_T^2\equiv\frac{\kappa_1}{\gamma_1h^2};
\\\label{eqD}
\partial_t\Delta+(1-\alpha)v_{\ell\ell}=-\lambda_1T_g[\Delta-(T_c|u_s|/T_gu_c)\Delta_c],
\\\label{eqU}
\partial_t u_{ij}^*=-\lambda T_g[u_{ij}^*-(T_c/T_g)u_{ij}^*|_c\,], 
\\\label{eqUs}
\partial_t u_s=-\lambda T_g[u_s-(T_c/T_g)u_c], 
\\\label{critV}
u_c\equiv\frac{1-\alpha}{\lambda f},\,\,
\frac{u_{ij}^*|_c} {u_c}\equiv\frac{v_{ij}^*}{|v_s|},\,\,
\frac{\Delta_c}{u_c}\equiv \frac{\alpha_1}{\lambda_1f}\frac{u_{ij}^*}{|u_s|}\frac{v_{ij}^*}{|v_s|}.
\end{align} 
For constant shear rate and $T_a,v_{\ell\ell}=0$, 
we have $T_g=T_c$, $\Delta=\Delta_c$, $u_s=u_c$, $u_{ij}^*=u_{ij}^*|_c$, with $\Delta_c, u_s,u_{ij}^*|_c$ rate-independent. 
It is customary in soil mechanics to refer to this steady state as  {\it critical}, though it is unrelated to critical phenomena in physics.
The relaxation rate $R_T$ in dense media has an inverse time scale of order ms or less. In comparison, the rates $\lambda T_g,\lambda_1 T_g\propto v_s$ are small for the shear rates typical of soil-mechanical experiments, $\lambda T_g=1$/s for $v_s=10^{-2}$/s.  The length scale $\xi_T$ is a few granular diameters. Rate-independence derives from $T_g\propto T_c\equiv f|v_s|$, and is destroyed by any $T_a\ne0$. 
[We note that $u_c,T_c>0$, but $u_s,v_s$ may be negative. Eq.(\ref{eqUs}) is obtained by multiplying Eq.(\ref{eqU}) with $u_{ij}^*/|u_s|$ and assuming $u_s>0$, $u_{ij}^*/|u_s|=v_{ij}^*/|v_s|=$ const, which is eg. not right in the load/unload experiment, as  $u_{ij}^*/|u_s|=-v_{ij}^*/|v_s|$ right  after a  rate reversal, see Sec.~\ref{Load and Unload}.]

With the differential equations derived, the energy density and transport coefficients in large part specified, {\sc gsh} is a well-defined theory. It contains clear ramifications and provides little 
leeway for retrospective adaptation to observations.  
As we shall see  in the following sections, a wide range of granular phenomena is encoded in these equations.

\subsection{Three Rate Regimes\label{sum}}

Depending on the interaction between particles, granular experiments are divided into three regimes: In the first, the particles are static and elastically deformed; in  the second, they move slowly, rearranging by overcoming frictional forces;  in the third, they interact by collisions. Although this interaction, of mesoscopic nature, is not manifest in a macroscopic theory, {\sc gsh} does have three regimes echoing its variation, and the control parameter is how strongly the grains jiggle -- quantified as  the granular temperature $T_g$:

\begin{itemize}
\item At vanishing shear rates, grains do not jiggle, $T_g\to0$. The stress stems from deformed grains and is elastic in origin. Static stress distribution and the incremental stress-strain relation are phenomena of this regime. 
Deviations from  full elasticity, $u_{ij}=\varepsilon_{ij}$ and $\sigma_{ij}=\pi_{ij}$, being quadratically small,  $\alpha,\alpha_1,\eta_g,\zeta_g,\kappa_g\propto T_g\propto v_s^2$, are frequently negligible. This is what we call the {\em quasi-elastic regime}.  

\item At slow rates, $T_g\gg\gamma_0/\gamma_1$ is somewhat elevated, see Eq.(\ref{TgVs}). The elastic
stress may now relax, implying plasticity: When the grains jiggle and briefly loosen contact with one another, the grains' deformation and the associated stress will get partially lost, irreversibly. We call this  regime {\em hypoplastic}, because this is where the hypoplastic model~\cite{kolymbas1} and other rate-independent constitutive relations are valid. Typical phenomena are the {critical state}~\cite{schofield}, and the different loading/unloading curves. Friction is a result in {\sc gsh},  not an input, and it derives from the  combined effect of elastic deformation and stress relaxation. (In spite of our borrowed usage of {\it hypoplasticity}, the reversible part of the stress is derived from an energy potential.)

In the hypoplastic regime, we have $T_g=T_c\equiv f|v_s|$, $\alpha=\bar\alpha$, $\alpha_1=\bar\alpha_1$. The equations~(\ref{eqD},\ref{eqU})  for the elastic strain are explicitly rate-independent, and the stress, generally given by Eqs.(\ref{2c-2},\ref{2c-2a},\ref{2c-2b}), is simplified, because the kinetic pressure $P_T\propto T_g^2$ and the viscous stress $\eta_1T_g v_s$, both quadratic in the rate, are negligibly small. The stress is $\sigma_{ij}=(1-\alpha)\pi_{ij}$, where the factoris typically between 0.2 and 0.3. The complex
elasto-plastic motions, observed mainly in triaxial apparatus, take place in this regime. 

\item At high shear rates, large $T_g$ and low densities, we are in the regime of {\it rapid dense flow}. The jiggling is so strong that it  gives rise to a kinetic pressure and viscous shear stress. They compete with the elastic one as rendered by the $\mu$-rheology~\cite{midi}.  
We still have $T_g\propto v_s$ at higher rates, but it is no longer small. Therefore, the kinetic pressure $P_T$ and the viscous stress become significant and compete with the elastic contribution. Both the total pressure and the shear stress may now be written as $e_1+e_2v_s^2$, with $e_1,e_2$ functions of the density. 
The {\it Bagnold regime} is given for  $e_2v_s^2\gg e_1$, 
where all stress components depend quadratically on the rate. Typically, since $e_1\gg e_2v_s^2$ for any realistic $v_s$, it is not easy to go continuously from the rate-indepedent to the Bagnold  regime at given density. However, a discontinuous transition is possible at given pressure, because $\rho$ decreases with $v_s$, eventually going below $\rho_{\ell p}$. There is then no elastic solution, $\pi_{ij}\equiv0$, or $e_1=0$. And the system is in a pure Bagnold regime.
\end{itemize}
For reasons discussed in detail in~\cite{granR4}, it is difficult to observe the transition from the hypoplastic regime to the quasi-elastic regime. And it has in fact not yet been done systematically. This is probably why soil mechanics textbooks take the hypoplastic regime to be the lowest rate one, referring to it as quasi-static. This is, we believe,  conceptually  inappropriate, because motions in the hypoplastic regime are irreversible and strongly dissipative, not consecutive visits of neighboring static states with vanishing dissipation.  Therefore, experiments at the very low end of shear rates are highly desirable. (When pressed, we need to guess. And we expect 
the {\em quasi-elastic regime} to start somewhere below $10^{-5}$/s,  with the rate-independent {\em hypoplastic regime}  above $10^{-3}$/s.) 

Next, we employ the equations presented above to account for granular phenomena, first in the hypoplastic regime, in which the complexity of granular behavior is most developed and best documented. Then we consider dense flow, including the $\mu$-rheology and the Bagnold scaling. This is followed by the nonuniform phenomena of  elastic waves, shear band and compaction. Finally, the quasi-elastic regime of vanishing rates is considered, exploring why it is hard to observe, and how best to overcome the difficulties.

\section{The Hypoplastic Regime\label{hypoplastic motion}}

Granular behavior in the hypoplastic regime are taken to generally possess rate-indepen\-dence -- meaning for given strain rates, the increase in the stress $\Delta\sigma_{ij}$ depends only on the increase in the strain, $\Delta\varepsilon_{ij}=\int v_{ij}{\rm d}t$, not the rate. As a result, engineering theories typically have rate-independence built in from the beginning. We note that it is not at all a robust feature of granular behavior. For instance, it is lost when the system is subject to an ambient temperature $T_a$ [such as given by a sound field, see the discussion around Eq.(\ref{Tg1})]: The critical stress then becomes strongly rate-dependent, vanishing for large $T_a$. And it does not extend into the higher rates of dense flow. Therefore,  rate-independence is a phenomenon that cries out for an explanation, an understanding.

Moreover, it is crucial to distinguish between rate- and stress-controlled experiments.  When the rate is given, $T_g$ quickly settles into its steady state value $T_c$, see Eq.(\ref{Tg2}), then the relaxation of $u_s$, accounting for the approach to the critical state, is independent of the rate, see Eq.(\ref{eqUs}). A rather different experiment is to hold the shear stress $\sigma_s$ fixed, starting with an elevated $T_g$. This $T_g$ will relax until it is zero, and the system static. There is also a rate in this case, referred to as {\it creep} sometimes -- the one that compensates the stress relaxation at a finite $T_g$. being proportional to $T_g$, this rate relaxes toward zero at the same time. Rate-independence is therefore a misplaced concept here. 

Stress-controlled experiments cannot be performed in triaxial apparatus with stiff steel walls, because the correcting rates employed by the feedback loop to keep the stress constant are of
hypoplastic magnitudes. As a result, much $T_g$ is excited that distorts its relaxation, and the situation is one of consecutive constant rates, not of constant stress. Instead, one may employ a soft spring to couple the granular system with its driving device, to enable small-amplitude stress corrections without exciting much $T_g$. We consider rate-controlled experiments in Sec.\ref{critical state}, \ref{Load and Unload},  and \ref{Constitutive Relations}, stress-controlled ones in Sec.\ref{aging}, and experiments subject to an ambient temperatures  $T_a$ in Sec.\ref{visElaBeh}.

\subsection{The Critical State\label{critical state}} 
Grains with enduring
contacts are deformed, which gives rise to
an  elastic stress. The deformation is
slowly lost when grains rattle and jiggle, because
they lose or loosen contact with one another. As a
consequence, a constant shear rate not only increases
the deformation, as in any elastic medium, but also
decreases it, because grains jiggle when being sheared
past one another.  A steady state exists in which both
processes balance, such that the deformation remains
constant over time -- as does the stress. This is the critical state. 
Moreover, the increase in deformation is $\propto v_s$, the  relaxation is $\propto T_g$. As  $T_g\propto v_s$ for elevated granular temperature, the steady-state, especially the  critical stress, are rate-independent. In this section, we show how {\sc gsh} mathematically codify this physics.

\subsubsection{Stationary Elastic Solutions\label{Stationary Elastic Solution}}
The critical state is given by the stationary solution $T_g=T_c, \Delta=\Delta_c, u_s=u_c$, with
\begin{equation}\label{3b-3a} u_c=\frac{1-\alpha}{\lambda}\frac{v_s}{T_g}=\frac{1-\alpha}{\lambda f},
\quad \frac{\Delta_c}{u_s}=\frac{\alpha_1}{\lambda_1}\frac{v_s}{T_g}=\frac{\alpha_1}{\lambda_1f},
\end{equation} 
see Eqs.(\ref{Tg2},\ref{eqD},\ref{eqU}). Because further shearing does not lead to any stress increase, this state is frequently referred to as {\it ideally plastic}~\cite{critState}. Note  $u_c,\Delta_c$ are rate-independent (for $\alpha=\bar\alpha$, $\alpha=\bar\alpha$, $T_a=0$) and functions of the density. Same holds for the critical stress, cf Eqs.(\ref{2b-2a},\ref{2b-2b},\ref{2b-1}),
\begin{align}
\label{3b-4a}
P_c=(1-\bar\alpha)P^c_\Delta,\quad
\sigma_c=(1-\bar\alpha)\pi_c, 
\\\label{3b-3c}
P^c_\Delta\equiv P_\Delta(\Delta_c,
u_c),\,\, 
\pi_c\equiv\pi_s(\Delta_c, u_c), \\\label{3b-3cA}
{P^c_\Delta}/{\pi_c}=({{\cal B}}/{2{\cal
A}}){\Delta_c}/{u_c}+{u_c}/{4\Delta_c}. 
\end{align} 
The loci of the critical states thus calculated~\cite{GSH&Barodesy}  (though employing the slightly more general energy of~\cite{3inv}) greatly resembles those calculated using either hypoplasticity or barodesy~\cite{barodesy}
The critical ratio $\sigma_c/P_c$ -- same as the Coulomb yield of Eq~(\ref{2b-3}) -- is also frequently associated with a friction angle. Since one is relevant for vanishing $T_g$, while the other requires an elevated $T_g\propto |v_s|$, it is appropriate to
identify one as the static friction angle, and the other as the
dynamic one. The latter is smaller than
the former,  because the critical state is elastic,
and must stay below Coulomb yield, 
$\lambda_1f/\bar\alpha_1<\sqrt{2{\cal B/A}}$, if it is viable.
Textbooks on soil mechanics state that the friction angle is
 independent of the density -- although they do not, as a rule,
distinguish between the dynamic and the static one. We assume, for lack of better  information, that both are, or $2(a_3-a_2)=a_5-a_4$, see Eq~(\ref{DD}).
Separately, both $\Delta_c$ and $u_c$ should increase with $\rho\to\rho_{cp}$, same holds for $P_c$ and $\sigma_c$.

\subsubsection{Approach to the Critical State at Constant Density\label{approach critical}} 
Solving Eqs~(\ref{3b-2},\ref{3b-3}) for $u_s,\Delta$, at constant
$\rho, v_s$, with $h=\alpha/\bar\alpha=\alpha_1/\bar\alpha_1=1$, and the initial conditions: $\Delta=\Delta_0, u_s=0$, the relaxation into the critical state is given as
\begin{align}\label{3b-6}
u_s(t)=u_c(1-e^{-\lambda f\varepsilon_s}),\quad \varepsilon_s\equiv v_st,
\\\nonumber \Delta(t)=\Delta_c(1+f_1\,e^{-\lambda f
\varepsilon_s}+f_2e^{-\lambda_1f\varepsilon_s}), \\\nonumber
f_1\equiv\frac{\lambda_1}{\lambda-\lambda_1},\quad
f_2\equiv\frac{\Delta_0}{\Delta_c}-\frac{\lambda}{\lambda-\lambda_1}.\end{align}
Clearly, this is an exponential decay for $u_s$, and a sum of two decays for $\Delta$. It is useful, and quite demystifying, that a simple, analytical solution in terms of the elastic strain exists. Because $\lambda\approx3.3\lambda_1$, the decay of $u_s$ and $f_1$ are 
faster than that of $f_2$. Note $f_2$ may be
negative, and $\Delta(t)$ is then not monotonic. The associated pressure
and shear stress are those of Eqs~(\ref{3b-4a},\ref{3b-3c},\ref{3b-3cA}). For a negative $f_2$, neither the pressure nor the shear stress is monotonic. 
For the system to complete the approach to the critical state, the yield surface [such as given by Eq.(\ref{2b-3})] must not be breached during the non-monotonic course. If it happens, there is an instability, and the most probable result are shear bands, see Sec~\ref{nsb}, \ref{sb} below. Then the uniform critical state will not be reached.

\subsubsection{Approach to the Critical State at Constant Pressure\label{pressure approach}}
\begin{figure}[t]
\includegraphics[scale=.8]{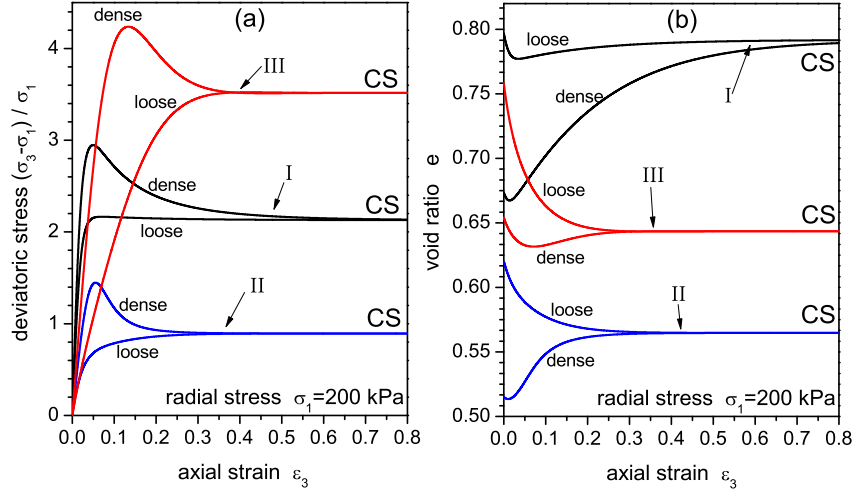}
\caption{Three
approaches to the critical state: These are the results of 
{\sc gsh}  calculations employing the parameter sets I,II,III as specified in the text. Shear stress $q\equiv(\sigma_3-\sigma_1)/\sigma_1$ and void ratio 
$e\equiv\rho_g/\rho-1$ (with $\rho_g$ the grain's density)
versus the strain $\varepsilon_3$ in triaxial tests
(cylinder axis along 3), at given $\sigma_1$ and
strain rate $\varepsilon_3/t$, for an initially dense
and loose sample. }\label{fig3}
\end{figure}
\begin{figure}[]
\includegraphics[scale=.8]{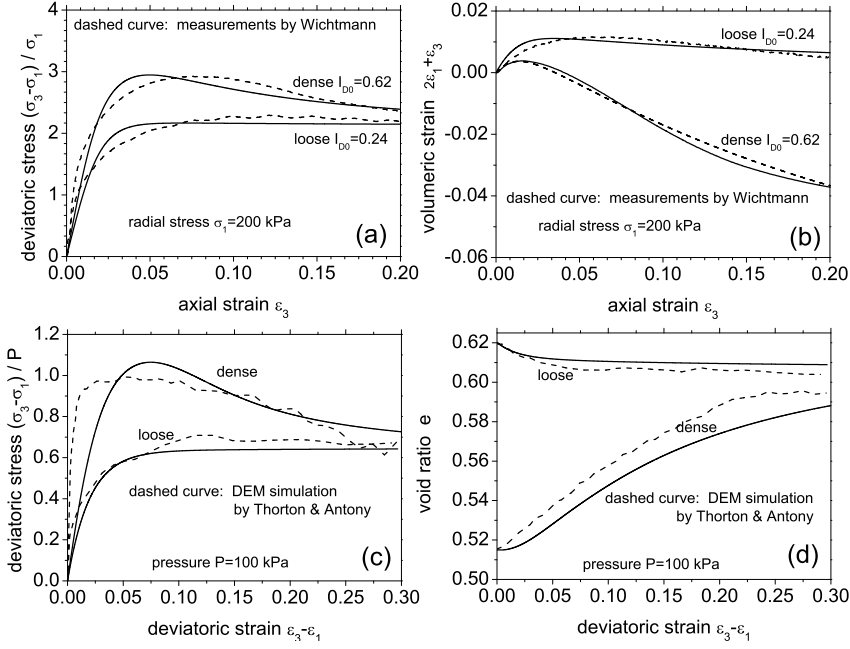}
\caption{A {\sc gsh} calculation employing I for comparing to the Wichtmann's experiment, and II to the simulation by Thornton and Antony, in the plots as originally given in~\cite{wichtmann,thornton}.} \label{fig3a}
\end{figure}

Frequently, the critical state is not approached at constant density, but at constant pressure $P$ (or a stress eigenvalue $\sigma_i$). The circumstances are then more complicated. As $\Delta,u_s$ approach
$\Delta_c,u_c$, the density compensates to keep
$P(\rho,\Delta,u_s)=$ const. Along with $\rho$, the coefficients
$\alpha,\alpha_1,\lambda,\lambda_1,f$ (all functions of $\rho$), also change
with time. In addition, with $\rho$ changing, the compressional flow
$v_{\ell\ell}=-\partial_t\rho/\rho$ no longer vanishes (though it is still small). Analytic solutions
do not seem feasible now, but numerical ones are, see Fig~\ref{fig3}, which compares three sets of parameters by plotting the deviatory stress versus axial strain at given $\sigma_1$. Clearly, any could serve as a textbook illustration of the approach to the critical state.  The parameters, see Eqs.(\ref{DD}), labeled as I, II, III, are:
\begin{itemize}
\item ${\cal B}_0=2, 0.22, 0.05$ GPa,\quad  ${\cal B/A}=5/3, 8, 5/3$,
\quad $\bar\rho/\rho_{cp}=0.615, 0.650, 0.667$, 
\item $\bar\alpha_0=1.04, 0.85, 16.25$,\quad $\bar\alpha_{10}=400, 30, 719$,
\item $\lambda_0\sqrt{\eta_{10}/\gamma_{10}}=272, 250, 2375$,\quad $\lambda/\lambda_1=3.8, 3.8, 3$, 
\item $a_1=0.15, 0.15, 1.6$,\,\, $a_2=1,0. 15, 1.6$,\,\, $a_3=0.6, 0.53, 1.6$,
\,\,  $a_4=a_5=0,0,-1$. 
\end{itemize} 
Fig \ref{fig3a} compares I to the  (drained monotonic triaxial) experiment by Wichtmann~\cite{wichtmann}, II to the simulation by Thornton and Antony~\cite{thornton}, both in the plots as originally given. The comparison of III to the barodesy model~\cite{barodesy} may be found in~\cite{GSH&Barodesy}. 

Generally speaking, we have three scalar state variables: $\rho,u_s,\Delta$, each with an equation of motion that depends on the rates $v_s,v_{\ell\ell}$ and  the variables themselves. In addition, 
$P,\sigma_s$ are functions of $\rho,u_s,\Delta$. In the last section, both rates were given, 
$v_{\ell\ell}=0$, $v_s=$ const. As a result, we have $\rho=$ const, while $\Delta(t)$ and $u_s(t)$ were calculated taking the coefficients $\alpha(\rho),\alpha_1(\rho),\lambda(\rho),
\lambda_1(\rho),f(\rho)$ as constant. The stress components were then obtained as dependent 
functions. A pressure-controlled experiment means that only the shear rate $v_s$ is given. Holding $P(\rho,u_s,\Delta)=$ const (or analogously  $\sigma_1$) implies
the density $\rho$ (and with it also $v_{\ell\ell}=-\partial_t\rho/\rho$) is a dependent function, $\rho=\rho(P,u_s,\Delta)$. Now, in the equations of motion for $u_s$ and $\Delta$, one first eliminates
$v_{\ell\ell}$ employing  $v_{\ell\ell}=-\partial_t\rho/\rho$, then eliminates both $\partial_t\rho/\rho$ and the $\rho$-dependence
of $\alpha(\rho),\alpha_1(\rho),\lambda(\rho),\lambda_1(\rho),f(\rho)$ employing
$\rho=\rho(P,u_s,\Delta)$. This changes the differential equations -- which are then solved numerically.

Many well-known features of Fig~\ref{fig3} can be understood assuming the solutions of Eq~(\ref{3b-6}) remain valid, say because the initial density is close to 
the critical one, hence it does not change much during the approach to the critical state. As a 
result, we may approximate $\alpha(\rho),\alpha_1(\rho),\lambda(\rho),\lambda_1(\rho),f(\rho)$ as constant, and take $v_{\ell\ell}\approx0$.  In addition, we assume, for simplicity, $\lambda\gg\lambda_1$, or ${\lambda}/({\lambda-\lambda_1})\approx1$ (instead of  $\,\approx1.5$). Then $f_2$ has
the same sign as $\Delta_0-\Delta_c$. The initial values are $\rho_0,\Delta_0$ and $u_s=0$, implying $P\propto{\cal B}(\rho_0)\Delta_0^{1.5}, \sigma_s=0$. For $P$ given and ${\cal B}(\rho)$ monotonically increasing with $\rho$, the pair
$\Delta_0-\Delta_c$ and $\rho_0-\rho_c$ have reversed signs. Therefore, we have a monotonic change of density for $f_2>0$, $\Delta_0>\Delta_c$, $\rho_0<\rho_c$,  and non-monotonic change otherwise. At
the beginning, the faster relaxation of $f_1$ dominates, so $\Delta$ always
decreases, and $\rho$ always increases, irrespective of $\rho_0$. After
$f_1$ has run its course, $\rho$ goes on increasing for  $\rho<\rho_0$ ({\em contractancy}) but switches to decreasing for  $\rho>\rho_0$ ({\em dilatancy}), until the critical state is reached. 
The shear stress $\sigma_s\propto \sigma_1-\sigma_2$ always increases first
with $u_s$, until $u_s$ is close to $u_c$. The subsequent behavior depends on
what $\Delta$ does. With $P\propto{\cal B}(\rho_0)\Delta_0^{1.5}$ given, $\sigma_s\propto{\cal B}\Delta^{0.5}\propto P/\Delta$ keeps growing if $\Delta$ decreases [loose
case, $f_2>0$], but becomes smaller again, displaying a peak, if $\Delta$ grows [dense
case, $f_2<0$]. 

\subsubsection{Shear Jamming}
A {\it jammed state} is one that can stably sustain a finite stress, especially  an anisotropic one. It is therefore characterized by values for $\Delta, u_s$ that satisfy the stability conditions $u_s/\Delta\le\sqrt{2{\cal B}/{\cal A}}$, or Eq.(\ref{2b-3}).
An unjammed state violates either this or another stability conditions, such as $\phi_{\ell p}<\phi<\phi_{cp}$ (where $\phi\equiv\rho/\rho_g$, with $\rho_g$ the bulk density, is the packing fraction). Typically, the critical state is approached starting from an isotropic stress, $\Delta=\Delta_0, u_s=0$. But  the approach solution Eq.(\ref{3b-6}) is also valid if the initial elastic shear strain is finite, $u_s\ne0$. Writing the solution to first order in the shear strain $\varepsilon_s\equiv v_st$,
\begin{equation}
u_s(t)=u_c\lambda f\varepsilon_s,\quad \Delta(t)=\Delta_0(1-\lambda_1f\varepsilon_s),
\end{equation}
we see a growing $u_s$ and a decreasing $\Delta$ for the initial stage. This is the reason that, if $\Delta_0$ is sufficiently small, the system will become unstable first, before it re-enters the stable region, converging eventually onto the critical state. 
Shear-jamming at constant density, as observed in~\cite{SJ1} and simulated in~\cite{SJ2}, is exactly this process, starting from the initial value  $\Delta, u_s=0$, or equivalently, from vanishing elastic pressure and shear stress, $P_\Delta,\pi_s=0$. So the system is unstable at the beginning, until $\Delta$ is sufficiently large to satisfy  Eq.(\ref{2b-3}), and the system is securely jammed. Further steady shearing then pushes the system into the critical state. 

\subsubsection{The Critical State with External Perturbations \label{external perturbation}}
If one perturbs the system, say by exposing it to weak vibrations,  or by tapping it periodically, such as in a recent experiment~\cite{vHecke2011}, the critical state is modified, and a rate-dependence of the critical shear stress is observed. The stress decreases with the shaking amplitude, and increases with the shear rate, such that the decrease is compensated at 
higher rates. Clearly, engineering theories with built-in rate-independence cannot account for this observation. {\sc gsh}, on the other  hand, if it indeed 
provides a wide-range description of granular behavior,  should be able to. 

\begin{figure}[t]
\includegraphics[scale=.8]{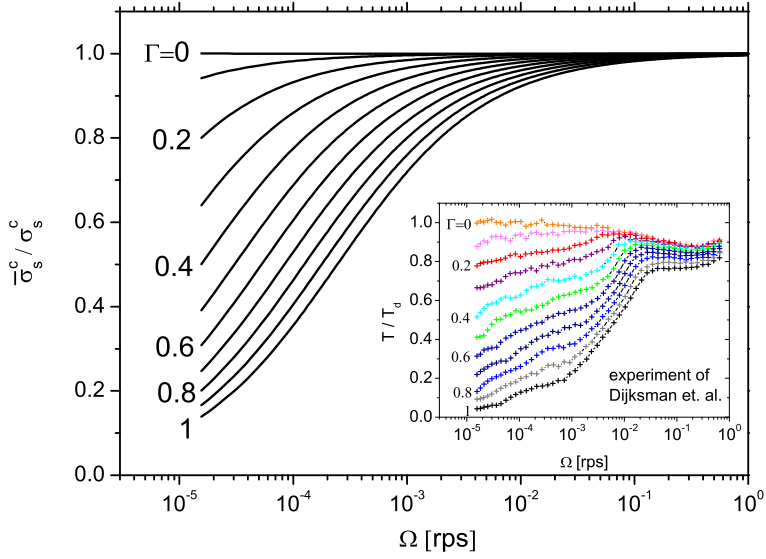}
\caption{Suppression of the critical shear stress $\sigma^c_s$ by vibration as given by Eq.(\ref{3b-8}), assuming $\Gamma=\alpha T_a,\,\, \Omega=\beta v_s^3$ (see text for details). Inset is the experimental curve of~\cite{vHecke2011}, with the torque $\tau$ denoted as $T$, as in~\cite{vHecke2011}.  (The stress dip at large $\Omega$, neglected here, is explained in~\cite{StressDip}.) }\label{fig3b}
\end{figure}

The consideration of the critical state in the previous three sections takes any granular temperature $T_g$ to be a 
result of the given shear rate, hence  $T_g=T_c\equiv|v_s|f$. This is no longer the case here, as sound field or tapping also contributes to $T_g$. And we have~Eq.(\ref{Tg2}), 
\begin{equation}\label{3b-7}
T_g^2=T_c^2+T_a^2,
\end{equation}
This is the reason the steady state values are reduced to $\bar u_c\equiv(T_c/T_g)u_c$,  $\bar \Delta_c\equiv(T_c/T_g)^2\Delta_c$, see Eqs.(\ref{eqD},\ref{eqU},\ref{eqUs}), with
\begin{equation}\label{3b-8}\frac{\bar u_c^2}{u_c^2}=
{\frac{\bar \Delta_c}{\Delta_c}}={\frac{\bar \sigma_c}{\sigma_c}}=\frac1{1+{ T_a^2}/{T_c^2}}.
\end{equation}
If there is no tapping, $T_a=0$, we retrieve the unperturbed values, $\bar u_c=u_c$, $\bar\Delta_c=\Delta_c$, $\bar\sigma_c=\sigma_c$. With tapping,  $\bar u_c, \bar\Delta_c, \bar\sigma_c$ decrease for increasing $T_a$, and increase with increasing shear  rate $T_c\equiv f|v_s|$. see Fig~\ref{fig3b}. 
(Note we have only considered the critical state at given shear rate, not the approach to it. So the result holds both at given density and pressure.)

The above consideration is the basic physics of the observation reported in~\cite{vHecke2011}. It helps to put rate-independence, frequently deemed a fundamental property of granular media, into the proper context. 
A more detailed comparison is unfortunately made difficult by the highly nonuniform experimental geometry. 
Nevertheless, some comparison, even if unabashedly qualitative, may still be useful. 
In~\cite{vHecke2011}, the torque $\tau$ on the disk on top of a split-bottom shear cell is related to its rotation velocity $\Omega$ and the shaking acceleration $\Gamma$. Now, $\tau$ and $\sigma_c$,  $\Omega$  and $v_s$, $\Gamma$ and $T_a$, are clearly related pairs, see also Sec~\ref{compaction}.
Assuming the lowest order terms suffice in an expansion, we take $\sigma_c\propto\tau$ and $\Gamma=c_1T_a$ with $c_1\sqrt{\eta_1/\gamma_1}=$20~s (noting $T_g$ is dimensionless with an appropriate $b$).   If $v_s$ were uniform, $\Omega\propto v_s$ would also hold. Since it is not, $\Omega\propto v_s^n$ with $n>1$ seems plausible, because with additional degrees of freedom such as position and width of the shear band, the system has for given $\Omega$ more possibilities to decrease its strain rate $v_s$. We take $\Omega=c_2v_s^3$ with $c_2=1{\rm rs}^2$ [implying a replacement of $T_a/v_s$ with $\Gamma/\sqrt[3]\Omega$ in Eq.(\ref{3b-8})]   for the fit of Fig \ref{fig3}, but emphasize that qualitative agreement exists irrespective of $n$'s value.  
In~\cite{vHecke2011}, a stress dip was  in addition observed at higher rates, see Fig \ref{fig3}. This is also accounted for by {\sc gsh}, see Sec~\ref{udf} and~\cite{StressDip}.

\subsection{Load and Unload\label{Load and Unload}} 
\begin{figure}[t] \begin{center}
\includegraphics[scale=0.33]{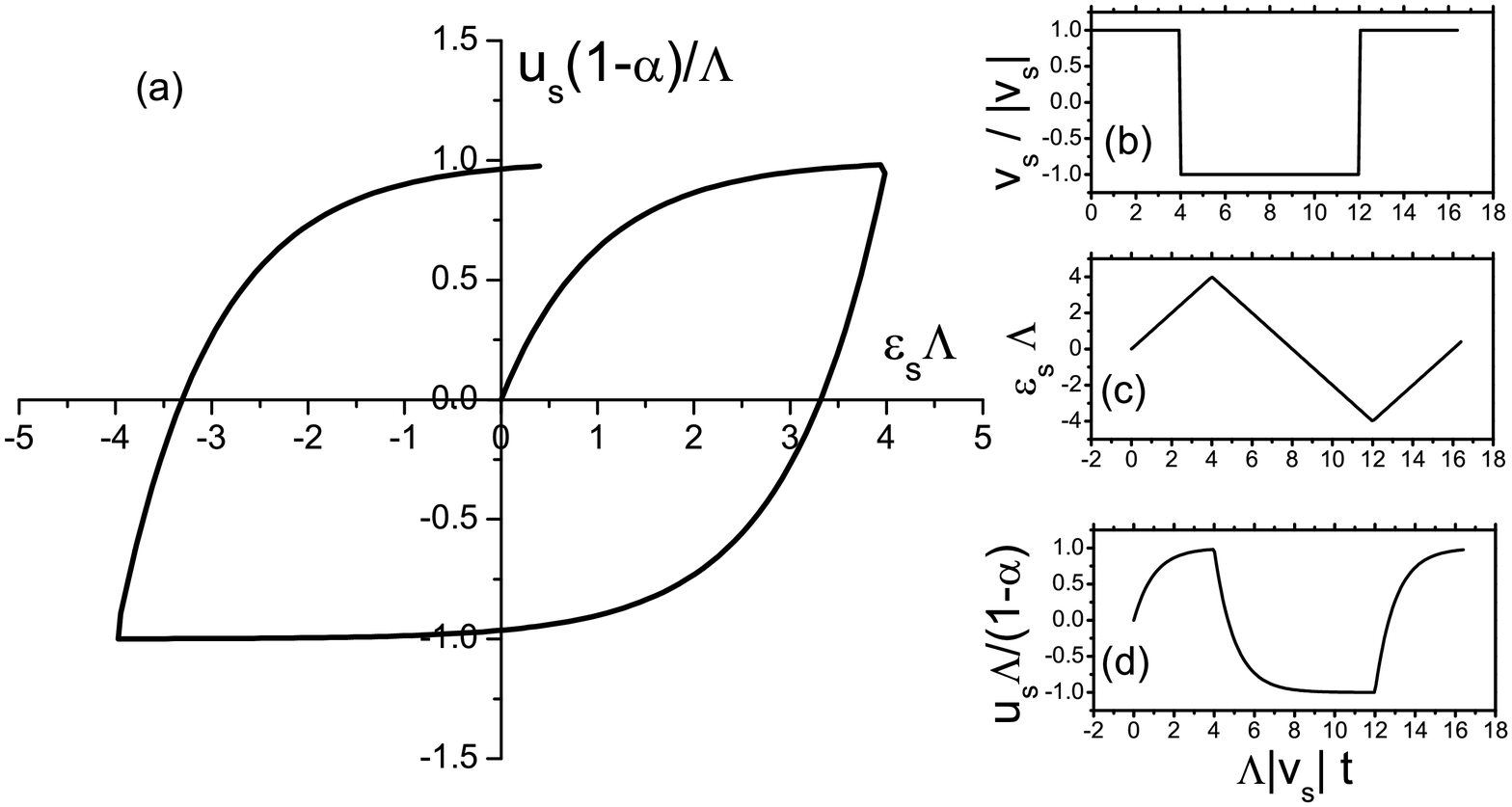}
\end{center}\vspace{-2cm} 
\caption{\label{fig2}The hysteretic change of the shear stress $\propto u_{s}$ with the strain, as given by Eq~(\ref{3b-3}). The sign of $v_{s}(t)$,  $\varepsilon _{s}\equiv\int_{0}^{t}v_{s}(t^{\prime })dt^{\prime }$, and $u_{s}(t)$ are given respectively in (b), (c) and (d).} 
\end{figure}

The simple reason for the difference between load and unload is that the stationary values $\Delta_c,u_{ij^*}|_c$ of Eqs.(\ref{eqD},\ref{eqU}) are altered when the shear rate $v_{ij}^*$ is reversed. The relaxation then proceed towards these new values, see the final paragraph of Sec.\ref{dynamics}. It is simple and deterministic and not in anyway {\it history-dependent}. We insert $T_g= f|v_s|$ into Eqs~(\ref{2c-7}, \ref{2c-9}), 
\begin{align}\label{3b-2}
\partial_t\Delta=v_s\,\alpha_1u_s-|v_s|\,\lambda_1f\Delta, 
\\\label{3b-3}
\partial_tu_s=v_s\,(1-\alpha )-|v_s|\,\lambda fu_s,
\end{align} 
%
to see that loading ($v_s=|v_s|>0$) and unloading ($v_s=- |v_s|<0$) have different slopes: $\partial_tu_s/v_s=(1-\alpha)\mp(\lambda fu_s/h)$.
Referred to as {\em
incremental nonlinearity} in soil mechanics, this phenomenon is the reason why no backtracing takes place under reversal of shear rate: Starting from isotropic stress, $u_s=0$, see Fig~\ref{fig2}, the gradient is at first $(1-\alpha)$, becoming smaller as $u_s$ grows, until it
is zero, in the stationary case $\partial_tu_s/v_s=0$. Unloading now, the
slope is $(1-\alpha)+(\lambda fu_s/h)$, steeper than it has ever been. It is again 
$(1-\alpha)$ for $u_s=0$, and vanishes for $u_s$ sufficiently negative. Same scenario holds for $\partial_t\Delta/v_s$. 
The stress components $P,\sigma_s$ are calculated employing Eqs~(\ref{2b-2b},\ref{2c-2},\ref{2c-2b}) for given $\Delta, u_s$. This consideration holds only for given density, it is more complicated if the pressure is given instead, same as in  Sec~\ref{critical state}), but the basic physics remains the same.

In systematic studies employing discrete numerical simulation, Roux and coworkers have accumulated great knowledge about granular physics, see eg.~\cite{roux}. They distinguish between two types of strain, I and II, identifying two regimes in which either dominates. This result agrees well with the above consideration, as the relaxation term in Eq.(\ref{3b-3}), being $\propto u_s$ is small if $u_s\propto\sigma_s$ is. Slow relaxation means the system is less plastic, more elastic and the difference between load and unload is small.

%


\subsection{Constitutive Relations\label{Constitutive Relations}} 
Granular dynamics is frequently modeled employing  the strategy of
{\em rational mechanics}, by postulating a function $\mathfrak{C}_{ij}$ -- of the  stress $\sigma _{ij}$, strain rate $v_{k\ell}$, and density $\rho $ --  such that the constitutive relation, ${\partial}_{t}\sigma _{ij}=\mathfrak{C}_{ij}(\sigma_{ij}, v_{k\ell}, \rho)$ holds. (More generally, ${\partial}_{t}$ is to be replaced by an appropriate objective derivative.) It forms, together with the continuity  equation $\partial _{t}\rho +\nabla _{i}\rho v_{i}=0$, momentum conservation, $\partial _{t}(\rho v_{i})+\nabla_{j}(\sigma _{ij}+\rho v_iv_j)=0$, a closed set of equations for $\sigma _{ij}$, the velocity $v_{i}$, and the density $\rho $ (or void ratio $e$). Both hypoplasticity and barodesy considered below belong to this category. 
(We do not consider elasto-plastic theories, but do note that, as shown by Einav~\cite{h^2}, they all form a special limit of the hypoplastic ones)
These models  yield, in their range of validity, a realistic account of the complex elasto-plastic motion, providing us with highly condensed and intelligently organized empirical data. This enables us to validate {\sc gsh} and reduce the latitude in specifying  the energy and transport coefficients.  

The drawbacks are, first of all, the apparent freedom in fixing ${\mathfrak C}_{ij}$ -- constrained only by the data one considers, not by energy conservation or entropy production (that were crucial in deriving {\sc gsh}).  This is probably the reason why there are many competing engineering models. And this liberty explodes when one includes gradient terms, hence most models refrain from the attempt to account for nonuniform situations, say elastic waves. 

Second, dispensing with the the variables $T_g$ and $u_{ij}$, one reduces the model's range of validity. For instance, they hold only for  $T_g=T_c\equiv f|v_s|$ and not for a $T_g$ that is either too small or oscillates too fast.  Also, as the analytical solution of the approach to the critical state shows, considering $u_{ij}$ is a highly simplifying intermediate step. The case for $u_{ij}$ is even stronger when considering proportional paths and the barodesy model, see below.

\subsubsection{The Hypoplastic Model\label{Hypoplasticity}} 

The {\em hypoplastic model} starts from the rate-independent
constitutive relation, 
\begin{equation}\label{3b-1}
\partial_t\sigma_{ij}=H_{ijk\ell}v_{k\ell}+
\Lambda_{ij}\sqrt{v_s^2+\epsilon v_{\ell\ell}^2}, \end{equation} postulated by Kolymbas~\cite{kolymbas1}, where
$H_{ijk\ell},\Lambda_{ij},\epsilon$ are functions of the stress and void ratio. The simulated granular response is realistic for deformations at constant or slowly changing rates.
Taking $h=1$, $\alpha =\bar \alpha$, $\alpha_1=\bar\alpha_1$,  $P_T, \eta_1T_g
v^0_{ij}\to0$, {\sc gsh} easily reduces to the hypoplastic model. This is because $\sigma_{ij}$ of Eqs~(\ref{2c-2},\ref{2c-2a}) is then, same as $\pi_{ij}$, a function of $u_{ij}, \rho$,  and we may write $\partial_t\sigma_{mn}=({\partial\sigma_{mn}}/{\partial u_{ij}})\partial_tu_{ij}+
({\partial\sigma_{mn}}/{\partial\rho})\partial_t\rho$.
Replacing $\partial_t\rho$ with $-\rho v_{\ell\ell}$,
$\partial_tu_{ij}$ with Eq~(\ref{2c-8}), using Eq~(\ref{TgVs}) to eliminate $T_g$, we arrive at an equation with the same structure as Eq~(\ref{3b-1}). Our derived expressions for $H_{ijk\ell},\Lambda_{ij}$ is different from the postulated ones, and somewhat simpler, but they yield very 
similar results, especially {\em response ellipses}~\cite{granL3}. (Response ellipses are
the strain increments as the response of the system, given unit stress
increments in all directions starting from an arbitrary point in the stress
space, or vice versa, stress increments as the response for unit strain increments.)

\subsubsection{Proportional Paths and Barodesy}

Barodesy is a recent model, again proposed by Kolymbas~\cite{barodesy}. It is more modular and better organized than hypoplasticity, with different parts 
in $\mathfrak C_{ij}$ taking care of specific aspects of granular deformation, especially that of {\em proportional paths}. We take {\sc p}$\varepsilon${\sc p} and {\sc 
p}$\sigma${\sc p} to denote, respectively, proportional strain and stress paths.  Their behavior is summed up by the Goldscheider rule ({\sc gr}): 
\begin{itemize}
\item A {\sc p}$\varepsilon${\sc p} starting from the stress $\sigma_{ij}=0$ is associated with a  {\sc p}$\sigma${\sc p}. (The initial value $\sigma_{ij}=0$ is a mathematical idealization, neither easily realized nor part of the empirical data. We take it  {\em cum grano salis}.) 
\item A {\sc p}$\varepsilon${\sc p} starting from $\sigma_{ij}\not=0$
leads asymptotically to the same {\sc p}$\sigma${\sc p} obtained when starting at $\sigma_{ij}=0$. 
\end{itemize}
Any constant strain rate $v_{ij}$ is a {\sc p}$\varepsilon${\sc p}. In the principal strain axes  $(\varepsilon_1,\varepsilon_2,\varepsilon_3)$, a constant $v_{ij}$ 
means the system moves with a constant rate along its direction, with $\varepsilon_1/\varepsilon_2=v_1/v_2,\, \varepsilon_2/\varepsilon_3=v_2/v_3$ 
independent of time. {\sc gr} states there is an associated stress path that is also a straight line in the principal stress space, that there are 
pairs of strain and stress path. And if the initial stress value is not on the right line, it will converge onto it.  

Again, if {\sc gsh} is indeed a broad-ranged theory on granular behavior, we should be able to  understand {\sc gr}  with it, which is indeed the case. But we need to generalize the stationary solution as given by Eq.(\ref{critV}) to include $v_{\ell\ell}\not=0$ (using $^{ni}$ to imply non-isochoric), 
\begin{equation}
\label{eq74}
u_c=\frac{1-\alpha}{\lambda f},\quad \frac{\Delta_c^{ni}}{u_c}=\frac{\alpha _1}{\lambda _{1}f}+\frac{1-\alpha}{u_c\lambda_1f}\frac{v_{\ell\ell}}{v_s},
\end{equation}
with ${\sigma_{ij}^{\ast }}/{\sigma_s}={u_{ij}^{\ast }|_c}/{u_c}={v_{ij}^{\ast }}/{v_s}$.  
If the strain path is isochoric, $v_{\ell\ell}=0, \rho=$ const, both the deviatoric strain and stress are dots that remain stationary-- these are the  critical state  considered in 
Sec~\ref{critical state}. If however $v_{\ell\ell}\not=0$, with the density $\rho[t]$ changing accordingly, ${u_{ij}^{\ast }|_c}={u_c}(\rho)\,{v_{ij}^{\ast }}/{v_s}$ and ${\sigma_{ij}^{\ast }}= {\sigma_s}(\rho)\,{v_{ij}^{\ast }}/{v_s}$ will walk down a straight line along ${v_{ij}^{\ast }}/{v_s}$, with a velocity determined, respectively, by $u_c(\rho[t])$ and $\sigma_s(\rho[t])$. 
Given an initial strain deviating from that prescribed by Eq~(\ref{eq74}),  $u_0\not=u_c,\Delta_0\not=\Delta_c^{ni}$, Eqs~(\ref{eqD},\ref{eqU}) clearly state that the deviation will relax, implying the strain and the associated stress will converge onto the prescribed line.  This is all very well, but {\sc gr} states that it is the total stress that possesses a {\sc p}$\sigma${\sc p}. With
$\pi_{ij}=P_\Delta(\rho)[\delta_{ij}+(\pi_s/P_\Delta)v_{ij}^*/v_s]$, 
this fact clearly hinges on $(\pi_s/P_\Delta)$ -- a function of $\Delta/u_s$, see  Eq~(\ref{2b-1}) --  not depending on the density. As long as $v_{\ell\ell}\ll v_s$, we have  $\Delta_c^{ni}/u_c\approx{\alpha _1}/{\lambda _{1}f}$, which we did assume in  Eq~(\ref{DD}) is density-independent,  to render the dynamic friction angle (that of the critical state) independent of $\rho$.

When looking at $\mathfrak C_{ij}$, it is easy to grasp that  the construction of a constitutive relation requires vast experience in handling granular media. That we could substitute this deep knowledge with the equations of {\sc gsh} that are  just as capable of accounting for elasto-plastic motion, is eye-opening. It suggests that sand, in its qualitative behavior, may be, after all, neither overly complicated, nor such a rebel against general principles.  

\begin{figure}[t]\begin{center}
\includegraphics[scale=2]{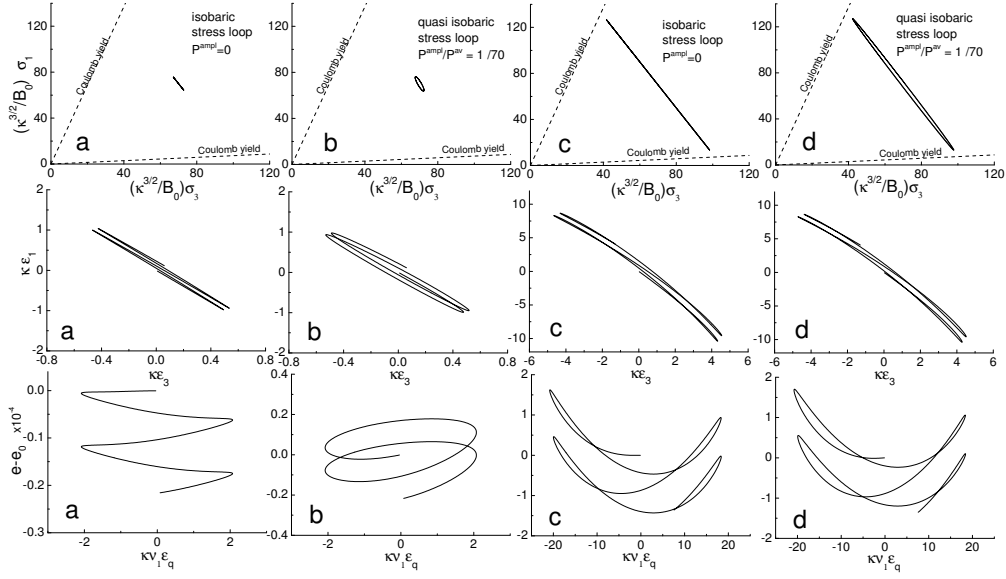}
\end{center}
\caption{Upper row: radial stress $\sigma
_{1}$ versus axial stress $\sigma _{3}$, rescaled by
$B_{0}\kappa ^{-3/2}$ (with $\kappa\equiv\sqrt{\zeta
_{1}\gamma_{1}}/\rho b$). Middle row: radial strain
$\varepsilon _{1}=\int v_{xx}dt$ versus axial strain
$\varepsilon _{3}=\int v_{zz}dt$. Lower row: $e-e_0$
(with $e_0$ the initial void ratio) versus shear
strain $\varepsilon_{q}=\int(v_{zz}-v_{xx})dt$,
rescaled by $\nu _{1}\kappa $. The stress loads are
isobaric for (a,c), and nearly (or quasi-) isobaric
for (b,d); the cyclic amplitude is small for (a,b) and
large for (c,d). The associated strain loci and void
ratio are: sawtooth-like for (a),  coil-like for (b),
butterfly-like (or double-looped) for (c,d).}
\label{butterfly1}\end{figure}

\begin{figure}[tbh]
\hspace{-1cm}
\includegraphics[scale=0.7]{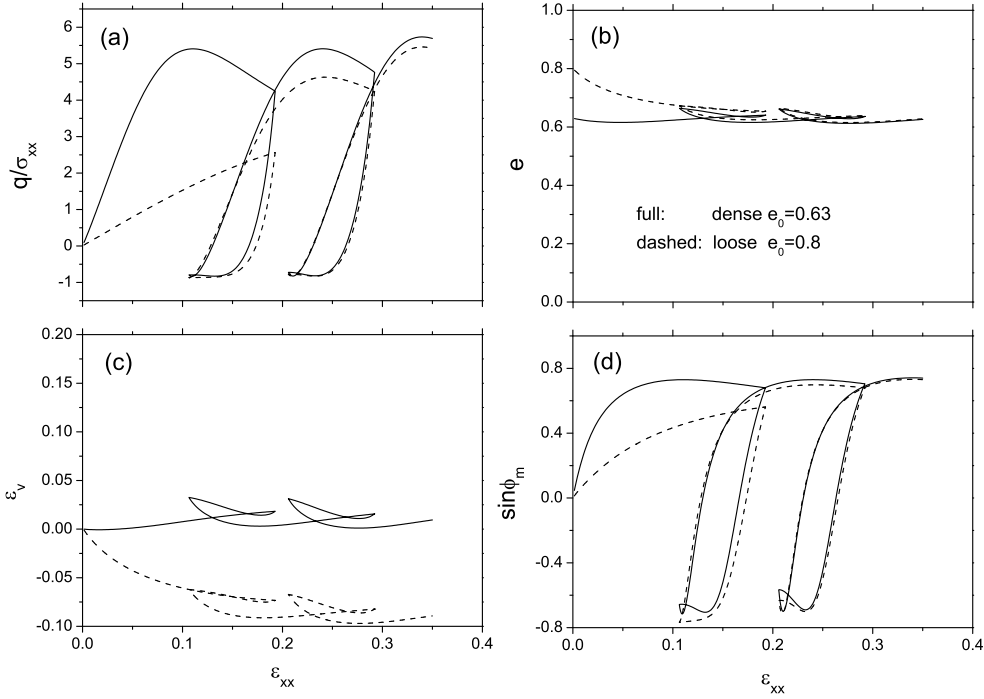}
\caption{In the geometry of triaxial tests, various quantities are  
computed employing {\sc gsh}, as functions of the strain $\varepsilon_{xx}$, holding $\sigma_{xx}=\sigma_{yy}$ constant.  (The axial direction is $z$.  The case with an initially higher density is rendered in solid lines, the looser one in dashed lines.) These are: (a)~deviatoric stress $q\equiv\sigma_{zz}-\sigma_{xx}$; (b)~void ratio $e$; (c)~volumetric strain $\varepsilon_v$; (d) the friction ange, $\sin \phi _{m}\equiv q/\left( 2\sigma_{xx}+q\right) $. We chose: $\alpha ,\alpha _{1},\lambda \sim \left( 1-\rho /\rho
_{cp}\right) ^{1.6}$ and $\eta_1,\gamma_1 \sim \left( 1-\rho /\rho _{cp}\right) 
^{-1}$.}
\label{triaxial}
\end{figure}

In~\cite{GSH&Barodesy,P&G2009}, the results of {\sc gsh} are compared to that of barodesy and hypoplasticity, with frequently quantitative agreement,  Some typical curves as produced by {\sc gsh} are given here, see Fig~\ref{butterfly1} and \ref{triaxial}, and the two papers for more details and the values for the parameters. 

\subsection{Stress-Controlled Experiments\label{aging}} 

Only rate-controlled experiments have been considered up to now. Employing Eqs~(\ref{Tg2}, \ref{eqD}, \ref{eqU}), we found that
the granular temperature quickly becomes a dependent quantity,  $T_g=T_c\equiv f|v_s|$, essentially reducing {\sc gsh} to the hypoplastic model,
with the exponential relaxation of $\Delta,u_s$ reproducing the approach to the critical state. In this section,  we examine what happens if we instead hold
the shear stress $\sigma_s=(1-\bar\alpha)\pi_s$  constant. (As discussed in the introductory sentences at the beginning of Sec.\ref{hypoplastic motion}, rate-independence is a misplaced concept here.)
Typical examples of experiments of given shear stresses includes relaxation of $T_g\propto v_s$ and shallow flows on an inclined plane or in rotating drums. In the second case, there is a delay between jamming ({\it angle of repose} $\varphi_{re}$) and fluidization ({\it angle of stability} $\varphi_{st}$), with $\varphi_{st}$ larger by a few degrees. All these are considered below.

\subsubsection{Diverging Strain and Long-Lived Temperature\label{TgRelaxation}}
If $T_g=0$, the system stays static, $\sigma_s=$ const, and there 
is no dynamics at all. If $T_g$ is initially elevated, $u_s$ relaxes, and with it also the stress $\sigma_s$. 
Maintaining a constant  $\sigma_s$ (or similarly, a constant $u_s$) therefore requires a compensating shear 
rate $v_s$. 
As long as $T_g$ is finite, $v_s(t)$ will accumulate, resulting in a growing shear strain 
$\varepsilon_s(t)=\int v_s{\rm d}t$. As we shall see, for $u_s$ close to its critical value $u_c$,  the characteristic time of $T_g$ is $\propto(1-u^2_s/u^2_c)^{-1}$ and long. Adding in the fact that the  relaxation of $T_g$ is algebraically slow rather than exponentially fast, the accumulated shear strain can be expected to be rather large. 

In a recent experiment, Nguyen et al.~\cite{aging} pushed the system 
to a certain shear stress at a given and fairly fast rate, producing an elevated $T_g$. Then, switching to maintaining the shear stress, they observed the accumulation of a large total strain  
$\varepsilon_s(t)$ that appears to diverge logarithmically. The authors referred to this phenomenon 
as {\it creeping}, and took it to be a {compelling evidence} that in spite of the very slow motion, their experiment contains a dynamics and was not quasi-static. we note that this conclusion sits well with a basic contention of {\sc 
gsh}, that what is usually taken as quasi-static motion is in fact hypoplastic, with an elevated $T_g$, as discussed above, see also Sec~\ref{quasi elastic motion} below.  

This experiment may in principle be accounted for by the equations of {\sc gsh}, though due to the highly nonuniform stress distribution, this would require solving a set of nonlinear partial differential equations with coefficients as yet uncertainly known. Hence we only consider a shear-stress controlled experiment in the hypoplastic regime with uniform variables. Also, we first assume that it is the elastic shear strain $u_s$ that is being kept constant, not the shear stress $\sigma_s\propto\sqrt\Delta\,u_s$, as both cases will turn out to be rather similar.
The relevant equations are still Eqs~(\ref{Tg2},\ref{eqD},\ref{eqU}). 
At the beginning, as the strain is being ramped 
up to $u_s$ employing a constant rate $v_1$, the granular temperature acquires the elevated initial value $T_0=fv_1$. Starting at $t=0$, $u_s$ is being held constant. From Eq~(\ref{eqU}), we therefore conclude
\begin{equation}\label{3b-7a}  
f|v_s|/T_g\equiv T_c/T_g={u_s}/{u_c},
\end{equation}
with $v_s$ the rate needed to compensate the stress relaxation.
Inserting this into Eqs~(\ref{Tg2},\ref{eqD}),
\begin{align}\label{3b-8a} 
\partial_t\Delta=-\lambda_1T_g[\Delta-(u_s/u_c)^2\Delta_c],
\\\label{3b-9} 
\partial_t T_g= -r_T\,T_g^2,\,\,\,r_T\equiv R_T[1-u_s^2/u_c^2],
\end{align}
we find the $T_g$-relaxation rate reduced from $R_T$ to $r_T$. Both equations may be solved analytically, if the coefficients are constant, which they are if the density is. The pressure $P(t)$ will then change with time, same as $\Delta(t)$. This is what we consider here.  (Keeping the pressure constant implies time-dependence of density and coefficients. Then, as with the critical state considered in Sec~\ref{pressure approach}, a general solution is possible only by numerical methods.)   
The first equation accounts for the relaxation of $\Delta$, from both below and above $(u_s/u_c)\Delta_c$. The relaxation is faster the more elevated $T_g$ is. 
Employing the initial condition $T_g=T_0$ at $t=0$, and  setting  $h=1$, the solution to the second equation is
\begin{equation}\label{3b-10} 
T_g={T_0}/({1+{r_T}{T_0}t}).
\end{equation}
Because of Eq~(\ref{3b-7a}),  the solution holds also for the shear rate, 
$v_s={v_0}/(1+ r_vv_0 t)$, with $v_0\equiv T_0/f$ and $r_v\equiv(fu_c/u_s){r_T}$.
This implies a slowly growing total shear strain
\begin{equation}
\varepsilon_s-\varepsilon_0\equiv\int v_s{\rm d} t=\ln(1+r_vv_0t)/r_v.
\end{equation}
However, $\varepsilon_s$ does not diverge, 
because as $T_g$ diminishes, it eventually enters the quasi-elastic regime, $\gamma_1h^2T_g^2\to \gamma_0T_g$, 
where its relaxation is exponential. More specifically, writing Eq.(\ref{3b-9}) as  $\partial_t T_g=-(r_0+r_TT_g)T_g$, with $r_0/\gamma_0=r_T/\gamma_1$, we have the general solution
\begin{equation}
T_g=r_0[(r_T+r_0/T_0)\exp(r_0t)-r_T]^{-1}.
\end{equation}

Assuming a large $T_0$ (implying large rate to ramp up the stress), $\Delta$ is quickly relaxed, $\Delta=(u_s/u_c)^2\Delta_c$. Fixing 
$u_s$ is then equal to fixing the shear stress, $\sigma_s\propto\pi_s\propto u_s\sqrt{\Delta} =(u_s^2/u_c)\sqrt{\Delta_c} $. With
$\pi_c\propto\sqrt{\Delta_c}\,u_c$, one may rewrite the factor in $r_T$ as
\begin{equation}\label{3b-12} 
1-u_s^2/u_c^2=1-\pi_s/\pi_c\approx1-\sigma_s/\sigma_c.
\end{equation}
The $T_g$-relaxation is slower the closer  $\pi_s$ is to $\pi_c$, infinitely so  for  $\pi_s=\pi_c$. Then we have $u_s=u_c$, $\Delta=\Delta_c$,   with $T_g(t)=T_0$ a constant, see Eqs.(\ref{3b-7a},\ref{3b-8a},\ref{3b-10}). 
This is indistinguishable from the rate-controlled critical state, which may be  maintained clearly also at given stress. 

If one chooses to keep $\sigma_s$ constant from the
beginning, irrespective how far $\Delta$ has relaxed, 
one needs to require
$\partial_tu_s=(u_s/2\Delta)\partial_t\Delta$, resulting in a different
proportionality $v_s\propto T_g$ to be inserted into the equations
of motion. The results are similar.

Next, we keep both the pressure and shear stress constant from the beginning. Though the general consideration does not appear analytically viable, one solution of a realistic situation exists: Keeping $\Delta, u_s=$  const in Eqs~(\ref{eqD},\ref{eqUs}), we have
\begin{equation}\label{3b-15}
\frac{u_s}{u_c}=\frac{T_c}{T_g},\quad \frac{\Delta}{\Delta_c}=\frac{T_c^2}{T_g^2}-\frac{v_{\ell\ell}}{T_g}\frac{1-\alpha}{\lambda_1\Delta_c}.
\end{equation}
For given $\Delta$, taking $u_s$ such that  ${\Delta}/{\Delta_c}={u_s^2}/{u_c^2}$, we have $v_{\ell\ell}=0$ and a constant density. Inserting ${u_s}/{u_c}={T_c}/{T_g}$ into 
the balance equation for $T_g$, Eq~(\ref{Tg2}), we again obtain Eq~(\ref{3b-9}) 
with (\ref{3b-12}). The only difference is that there is now a clear prescription for the 
experiment, because constant $\Delta,u_s,\rho$ means that pressure $P$ and shear stress $\sigma_s$ are kept constant. So one proceeds by applying an arbitrary pressure, then varying the shear stress until the density no longer changes. 
$T_g, v_s$ will then be as calculated.

Comparable calculation and analysis were carried out in~\cite{aging}, using two scalar equations 
that may roughly be mapped to the present ones. The quantities:  granular temperature $T_g$,  its relaxation and production rate, $R_T$ and 
$R_Tf^2$, were referred to as {\em fluidity, aging} and {\em rejuvenation parameter}. 
The above consideration is therefore not new, but does provide a  tensorial treatment that is embedded in {\sc gsh}, rendering it transparent, unified, and more realistic, also affording a better founded understanding. 
We also not that temporary, localized regions of strong deformation (called {\it hot spots}) were observed, with the fluidity (the averaged value of which is $T_g$) identified as their rate of occurrence.

As the stress distribution in the experiments of~\cite{aging} is rather nonuniform,  there will 
always be areas with a shear stress close to $\sigma_c$. And the system will tend to cave in there, resulting in a larger strain accumulation than what the average value for $\sigma_s$ 
would predict.   

In the experiment, a very soft spring was used to couple
the fan and the motor. This we believe is essential why this
experiment turned out as observed. Usually, triaxial apparatus with stiff walls are used. And the correcting rates employed by the feedback loop to keep the stress constant are of
hypoplastic magnitudes. As a result, much $T_g$ is excited, and we have the situation of consecutive constant rates, not that of constant stress. The soft spring, as discussed above, and in greater detail in Sec~\ref{soft springs}, enables quasi-static stress correction without exciting much $T_g$.

With an ambient temperature $T_a$, the  $T_g$ relaxes as $\partial_tT_g=r_T(T_g-\eta T_a)$, with  $\eta\equiv1/(1-u_s^2/u_c^2)$, see Eq~(\ref{Tg2}). This means, the values  $T_g$ and $v_s$ respectively relax to, ${\eta T_a}$ and ${\eta v_a}$, get strongly amplified close  to $u_s=u_c$. This is a large effect. 

\subsubsection{Stability above the Critical Shear Stress\label{scs}}
From the consideration of the last two sections we see that a granular assembly is, for an elevated $T_g$,   mechanically stable only up to the critical value for the elastic stress $\pi_c$. For $\pi_s<\pi_c$, $T_g$ grows, since $r_T$ is negative.  (As we shall see in Sec.\ref{nsb}, shear bands are formed as a result of this instability.) On the other hand,  for $T_g=0$, the system is stable at any static shear stresses exceeding $\pi_c$, as long as Eq.(\ref{sb12}) is not breached.
Now, since an infinitesimal $T_g$ is ubiquitous, and if it always grows, there is no stability for  static shear stresses exceeding $\pi_c$. It does not always grow: Only an initial $T_g$ of hypoplastic strength will explode, not an infinitesimal one, of quasi-elastic strength. This is because $h$ diverges for $T_g\to0$, and the critical stress diverges with $h$: Since $f\propto 1/h$, we have $u_c\propto h,\Delta_c\propto h^2$, and $\sigma_c\propto h^2$, see Eqs.(\ref{critV}). Therefore,  $r_T$ is always positive for very small $T_g$. In fact, what we have for strain values above  $u_c$ is a metastability, a stability that may be destroyed only by granular jiggling of sufficient strength. This fact is associated with familiar phenomena: A house on a cliff collapsing due to elastic waves from a distant earth quake, or a pneumatic hammer close by;  a gun shot initiating an avalanche. 

The elastic strain instability  for $u_s>u_c$ holds only for stress-controlled experiments, not rate-controlled ones, though this distinction  is not always clear-cut in experiments. For instance, if a step motor is used for a strain-controlled experiment, and one has a strain versus time curve such as given by Fig~\ref{StepMotor} below, than the stress is being hold constant at the plateaus, rendering  the stability of the uniform system precarious. This may well be the reason why shear band formation is so frequently observed in the cases where the initial density is high and the non-monotonic stress trajectory exceeds $u_c$, see Fig~\ref{fig3}.

Finally, we stress that these aspects of granular behavior are natural results of {\sc gsh}, not preconceived features planted in while constructing it. They stem from the interplay between yield and the critical state, or more precisely, between the instability of the elastic energy and the stationary solution of the elastic strain.

\subsubsection{Angle of Stability and Angle of Repose \label{sapc}}

Aranson and Tsimring were the first to construct a theory for these two angles~\cite{aranson,Aranson1}. Taking the stress $\sigma_{ij}$ as the sum of two parts, one solid, the other fluid-like, they define an 
order parameter  $\hat\varrho$ that is 1 for solid, and 0 for dense flow. They then postulate a free energy $f(\hat\varrho)$ such that it is stable with $\hat\varrho=1$ only for $\varphi< \varphi_{st}$, with $\hat\varrho=0$ only for $\varphi>\varphi_{re}$, and  $\varphi_{st}>\varphi>\varphi_{re}$ as the bi-stable region. The solid stress is taken as an input, assumed understood from some other theory. 
In comparison, the consideration below,  given within the context of {\sc gsh}, is somewhat more complete and less {\em ad hoc}. 

Fluidization, the collapse that occurs when one slowly tilts a plate supporting a layer of grains, is a process that happens at $T_g=0$, with no granular jiggling. Therefore, the Cauchy stress is given by the elastic one, $\sigma_{ij}=\pi_{ij}$. 
On a plane inclined by the angle $\varphi$,  with $y$ denoting the depth of the granular layer on the plane, and $x$ along the slope, we take the stress to be $\pi_{xx},\pi_{yy},\pi_{zz}=P_\Delta$, $\pi_{xy}=\pi_s/\sqrt2$, $\pi_{yz},\pi_{xz}=0$. Integrating $\nabla_j\pi_{ij}=g_i\rho$ assuming a variation only along $y$, we find $\pi_{xy}=g\sin\varphi\int\rho(y)dy$ and $\pi_{yy}= \pi_{xy}/\tan\varphi$. The angle of stability $\varphi_{st}$ is reached when the energetic instability of Eqs.(\ref{2b-3}) is breached. With  $\pi_s^{yield}\equiv P\sqrt{2{\cal A}/{\cal B}}$ denoting the yield shear stress, it is
\begin{equation}\label{sb12}
\tan\varphi_{st}=\pi_s^{yield}/\sqrt2 P=\sqrt{{\cal A}/{\cal B}}.
\end{equation} 
Effects derived from proximity to the wall or floor are considered in Sec.\ref{clogging}.

The angle of repose  $\varphi_{re}$ is related to the calculation of the last two sections. As long as the shear stress is held below the critical one, $\sigma_s<\sigma_c$, the $T_g$-relaxation will run its course, and the system is in a static, mechanically stable state  afterwards. At  $\sigma_s=\sigma_c$, however, the system becomes critical, and no longer comes to a standstill. Therefore, $\varphi_{re}$ is given by ${\sigma_c}$, 
\begin{equation}
\tan\varphi_{re}={\sigma_c}/\sqrt2\,P_c, \quad\text{with\,\,}\varphi_{re}<\varphi_{st}.
\end{equation}
The inequality holds because the critical state is an elastic solution, while $\varphi_{st}$ is the angle at which all elastic solutions become unstable.
That $\varphi_{re}$ and $\varphi_{st}$, material parameters, differ only slightly, is related to the microscopic fact that both account for the clearance with the profile of the underlying layer -- though one with granular jiggling and hence a little easier.


\subsection{The Visco-Elastic Behavior of Granular Media\label{visElaBeh} } 

All visco-elastic systems (such as polymer solutions) have a characteristic time $\tau$ that separates two frequency ranges:  fluid-like behavior for $\omega\tau\ll1$, and solid-like one for  $\omega\tau\gg1$.  Like granular media,  polymers are transiently elastic, though the transiency is constant and not variable, because $\tau$ is. The hydrodynamic theory of polymers, 
with a very similar elastic strain $u_{ij}$ that obeys the equation
$\partial_tu^*_{ij}-v^*_{ij}=-u_{ij}^*/\tau_{ve}$, 
%
is capable of accounting for many visco-elastic phenomena, including shear-thinning/thicken\-ing, elongational viscosity, the Cox-Merz rule, and the rod-climbing (or Wei\ss enberg) effect~\cite{polymer-1,polymer-2,polymer-3,polymer-4}. 

The main difference of the granular analogue, Eq~(\ref{2c-8}), is the fact that the relaxation time varies as $\tau\propto 1/T_g$ -- a granular system is fully elastic for $T_g\to0$, capable of sustaining a static shear stress. Moreover, rate-independence,  a  granular characteristics not observed in viscous elastic systems with a constant $\tau$, stems from the relation $1/\tau=\lambda T_g\propto v_s$.
However, when there is an ambient temperature in granular media, much larger than the temperature produced by the imposed shear rate,  $T_a\gg T_c\equiv f|v_s|$, polymers and granular media are very similar in their behavior, because $T_a$ is also a given quantity that does not depend on the local shear rate. The ambient temperature $T_a$ may be maintained by a standing sound wave, periodic tapping, or by diffusion from a region of great granular agitation. In all cases, the resultant $T_a$ enables the relaxation of the elastic strain and stress, implying  no static stress may be maintained, and the yield stress vanishes.

\subsubsection{The Creep Motion\label{creep motion}}
In granular media, one frequently observes shear bands, which borders on a  non-shearing, solid part. Careful experiments reveal that the shear rate is in fact continuous, with an exponentially decaying creep motion taking place in the solid, see Komatsu et al~\cite{komatsu}, Crassous et al~\cite{crassous}. 
We show here that this is a result of $T_g$ from the fluid region diffusing into the solid one, being present there 
as an ambient, spatially decaying temperature $T_a$ that enables stress relaxation. If the stress is to be maintained, there must be a compensating shear rate that also decays in space, along with $T_a$, and the velocity obtained from integrating the shear rate is the observed creep motion. 

Consider a ``liquid-solid boundary"  at $x=0$, with the shear rate being concentrated on one side, for $x>0$. (We shall return to consider the liquid side in Sec.\ref{sb}. Here, we only take  the fluid values at $x=0$ to provide the boundary conditions for $v_s,T_g$ in the solid part.) For a one-dimensional geometry, the pressure $P$, shear stress $\sigma_s$,  the shear rate $v_s$ and  $T_g$ are uniform, but $\rho$ need not be. We take $\rho$ to be discontinuous at $x=0$, but constant otherwise, with $v_{\ell\ell}=0$, and $T_g,v$ varying perpendicular to the boundary,  along $\hat x$. 
The circumstances are then quite similar to that of Sec~\ref{aging}, though variation is in space rather than time. First, with stationarity of  Eqs~(\ref{eqD},\ref{eqUs}) [see also Eq.(\ref{3b-15})], we have
\begin{equation}\label{props}
\frac\Delta{\Delta_c}=\frac{u_s^2}{u_c^2}=\frac{T_c^2}{T_g^2}=\frac{\pi_s}{\pi_c}=\frac{\sigma_s}{\sigma_c},\,\,\frac\Delta{u_s}=\frac{\Delta_c}{u_c}\frac{T_c}{T_g}.
\end{equation}
With $\Delta,u_s$ fixed, so are $P,\sigma_s$, where especially $P=P_c$ if $\sigma_s=\sigma_c$. Note also that since the stable branch of $P/\sigma_s=P_\Delta/\pi_s\equiv1/\mu$  increases monotonically with $\Delta/{u_s}$, see Eq.(\ref{2b-1}), the last above equation implies that the friction $\mu$ decreases for increasing $T_c/T_g$. 
The balance equation for $T_g$ [with $\partial_tT_g=0$ but including the diffusive current, see Eqs.(\ref{Tg2},\ref{3b-9})] reads
\begin{align}
\label{cr1}
\nabla^2T_g=T_g/\xi^2_{cr},\quad \xi_{cr}^2\equiv\xi_T^2/[1-\pi_s/\pi_c]
\\\label{cr2}
\text{implying}\quad v_s/v_s^{0}= T_g/T_g^{0}= \exp(-x/\xi_{cr}),
\end{align}
where $v_s^{0},T_g^{0}$ are the fluid values at $x=0$. 
That the decay length $\xi_{cr}\equiv \xi_T/\sqrt{1-\sigma_s/\sigma_c}$  diverges for $\sigma_s=\sigma_c$ is not surprising, because the solid region, turning critical, ceases to exist then. Although subcritical, $\sigma_s<\sigma_c$, the solid region sustains a finite rate $v_s\not=0$, because $T_g$ is being continually diffused from the fluid region.
Note $\sigma_s$ is a uniform quantity across the boundary, yet we necessarily have $\sigma_s<\sigma_c(\rho)$ on the solid side, $\sigma_s\geq\sigma_c(\rho)$ on the fluid side, implying a lower fluid density.  
Finally, the above exponential decay with the constant length $\xi_{cr}$ holds only in the hypoplastic regime. Once $T_g$ is sufficiently small, we have $h\to\infty$, and $\xi_T\propto h^{-1}$ vanishing quickly.

In two recent papers~\cite{kamrin,kamrin2}, Kamrin et al propose a nonlocal constitutive relation (KCR) well capable of accounting for steady flows in the split-bottom cell~\cite{fenistein}. A key ingredient is the fluidity $g\equiv v_s/\mu$. With  $\mu\equiv\sigma_s/P$, $\mu_s\equiv\sigma_c/P_c$, it is taken to obey
\begin{equation}
\xi^2_{cr}\nabla^2g=g-g_{loc},\quad \xi_{cr}\propto1/\sqrt{|\mu-\mu_s|}.
\end{equation}
Because $g_{loc}=0$ for $\mu<\mu_s$, this relations is rather similar to Eq.(\ref{cr1}), with $g$ assuming the role of $T_g$, and the two decay lengths diverging at the same stress values.

For $\mu\geq\mu_s$, the system is fluid, and $g=g_{loc}$ essentially constant. With $g_{loc}\propto\sqrt P(1-\mu_s/\mu)$, KCR is consistent with a first-order expansion of the MiDi relation, Eq.(\ref{poul}), in the inertial number. 
GSH is compared to MiDi in Sec.\ref{rdf}, showing broad agreement and some relevant disagreements. 
Here, we only discuss the additional differences of {\sc gsh} to KCR. 

First, KCR does not take the density as a variable, leading to inconsistencies: The stress is continuous at the solid-fluid interface and strictly constant in a one-dimensional geometry. As discussed below Eq.(\ref{cr2}), {we necessarily have $\sigma_s<\sigma_c(\rho)$ on the solid side, $\sigma_s\geq\sigma_c(\rho)$ on the fluid side, implying a lower fluid density.} Without the density, the same two conditions imply a discontinuity in  $\sigma_s$ or $\mu_s$ which violates momentum conservation. Another drawback is the fact that granular behavior depends sensitively 
on whether density or pressure is being held constant  see Sec.~\ref{pressure approach} above and \ref{udf} below. This cannot be reproduced employing KCR. 
Second, being {defined}  as $v_s/\mu$, the fluidity $g$ is not an independent variable like $T_g$, though it does possess a postulated, independent dynamics. If one eliminates $g$, rewrites its equation as $\mu\xi^2_{cr}\nabla^2(v_s/\mu)=v_s-v_s^{loc}$, a problem arises: This equation (in conjunction with $v_{\ell\ell}=0$) and the momentum conservation may both be used to calculate the velocity field for given density and stress. The results will in general be contradictory.     


\subsubsection{Nonlocal Fluidization\label{nonlocal fluidization}}
{\em Non-local fluidization} is an observation made (and named) by Nichol et al.~\cite{nichol2010}, see also Reddy et al.~\cite{reddy2011}. In a vessel of grains, after a shear band is turned on, the medium everywhere, even further away from the band, looses its yield stress, and the Archimedes law holds: A ball stuck at whatever height without the shear band starts to sink or elevate, until  its density is equal to the surrounding one. 
{\sc gsh}'s explanation for this behavior is quite simple: First, $T_g$ generated by the shear band diffuses through the solid phase, as accounted for by Eq.(\ref{cr1}), permeating the medium as a spatially decaying ambient temperature $T_a$. 
Second,  a medium such ``fluidized" obeys, as  observed earlier~\cite{huerta2005,caballero2009}), the Archimedes law, because the 
ball getting stuck in the sand deforms the grains around itself and  builds up an elastic shear stress.  Without an ambient temperature, $T_a=0$, this stress holds up the ball's weight if it is not too large,  and the ball is stationary. With $T_a\not=0$, the stress relaxes, requiring a compensating shear rate $v_s$ to maintain the stress balance, implying a
moving ball.  We note that  $T_a\not=0$ does not imply the grains need to jiggle violently. If the ball's descent takes an hour, a barely perceptible slip every minute would be quite sufficient. And $T_a$ is the spacial and temporal average of the changing energy contained in these slips.

More quantitatively, a solid object being  dragged by a constant force  $F^{ext}_i$ through a granular medium will quickly settle into a motion of  constant velocity $v_\infty$, implying a stationary stress and velocity distribution in the medium, in the rest frame of the object. So Eqs.(\ref{3b-15}) holds. This is remarkable, because the elastic stress $\pi_{ij}(\Delta,u_s)$ transforms, under the replacement $\Delta,u_s\to T_c\equiv f|v_s|,v_{\ell\ell}$, into a viscous stress. And this enables one to perform a calculation similar to that needed to arrive at the Stokes' law. 

The Stokes' law  $F^{drag}_i=6\pi R\eta v$ is derived assuming an incompressible (and infinitely extended) medium, with $v_{\ell\ell}=0$. The resulting velocity field, scaling with $v_\infty$, is a pure geometric quantity that does not depend on any parameters,  especially not the applied force  $F_{ext}$~\cite{LL6}.  In contrast, granular media possess sound velocities one to three times that of air and are rather compressible. As a result, both the velocity field and all parameters (that are functions of the density) will depend on  $F_{ext}$. 
In fact, that the viscosity seemingly depends on the mass of the steel ball ($\propto$ the gravitational force) was observed in~\cite{nichol2010}. 
Inserting Eqs.(\ref{3b-15}) into 
Eq~(\ref{2b-2}), we have, with $\sigma_{ij}=(1-\alpha)(P_\Delta\delta_{ij}+\pi_sv_{ij}^*/v_s)$, 
\begin{equation}\label{viscStress}
P_\Delta=\frac{{\cal A}u_c^2}{2\sqrt{\Delta_c}}\frac{T_c}{T_g},
\quad\pi_s=-2{\cal A}u_c\sqrt{\Delta_c}\frac{T_c^2}{T_g^2},
\end{equation}
where $P_\Delta$ contains only the of lowest order term in $v_s, v_{\ell\ell}$, while $\pi_s$ is valid assuming $v_{\ell\ell}=0$ (and appropriate for the steel plate below). Note
$T_g=\sqrt{T_c^2+T_a^2}$ has two contributions, $T_a$ from the remote shear band, and $T_c\equiv f|v_s|$ from the nonuniform local shear rate. For $T_a=0$, $P_\Delta$ and $\pi_s$ are rate-independent, and the system is in a (nonuniform) critical state. For $T_a\gg T_c$, 
the system is viscous, and one may define two effective viscosities, $P=\eta^{eff}_1v_s$, $\sigma_s=-\eta^{eff}_2v_s^2$, with $\eta^{eff}_1\propto1/T_g$, $\eta^{eff}_2\propto1/T_g^2$.   In~\cite{huerta2005}, faster ascent and  a smaller viscosity were  observed in regions of larger granular agitation (and attributed to ``pressure screening''). 

Given the form for the stress, one can calculate the velocity field depending on the geometry of the object. The drag force is then obtained by inserting the field into $\sigma_{ij}$, and integrating it over the surface of the object,
$F^{drag}_i=\oint \sigma_{ij}{\rm d}a_j$.  
The simplest case is that of a {\bf steel plate}, say perpendicular to $\hat x$ and being dragged along $\hat y$. The shear rate is a constant, $v_s=\frac12\nabla_xv_y$, with $v_{\ell\ell}=0$, 
and the force $F^{drag}$ per unit surface of the plate is $2\sigma_{xy}\propto v_s^2/T_a^2$. 
The velocity field for a {\bf ball of radius $\boldsymbol R$} is not as easily calculated, though it is clear that, for $T_c/T_a$ small, the drag force stems from the pressure and is linear (and not quadratic as with the plate):  $F^{drag}_i=\oint P{\rm d}a_i\propto v_s/T_g\propto v_\infty/T_g$, as observed in~\cite{caballero2009}.
Assuming incompressibility (as one does deriving the Stokes' law though inappropriately here), one finds 
 $F^{drag}_i=\oint P{\rm d}a_i=(9\pi^2/16)({\cal A}u_c^2f/\sqrt{2\Delta_c})\,(R v_\infty/T_a)$.


Any hydrodynamic theory starts from the basic assumption that its resolution is small compared to the system size, but much larger than any microscopic lengths -- in the present case, especially the grain diameter $d$. In~\cite{reddy2011}, the diameter of the probing rod, a system size, is only $2d$. Although averaging over time and runs usually retrieves the macroscopic behavior, this may not work quantitatively when the two scales are essentially the same.   

Summarizing, the dichotomy of the elastic stress and a $T_g$-dependent viscosity is the basic {\sc gsh}-explanation for granular visco-elasticity. In this more general picture, creep motion may equally well be understood as the viscous motion under constant moment of inertia. 
                                

\subsection{Narrow Shear Bands\label{nsb}}

Typical constitutive models such as hypoplasticicty or barodesy do not properly account for shear bands, and the reason is the lacks of a length scale. There are 
various approaches to overcome this short coming, by introducing gradient terms~\cite{wu2} or adding state variables to account for the couple stress and the Crosserat rotation~\cite{wu1}.
Especially the Crosserat method works well, but it leads to a  far more complex theory, constructed for the sole purpose of solving the shear band problem. Moreover, it throws up the question about the underlying physics: If couple stress and rotational motion are important in the shear band, because it is fluid, why then are they not important in the uniformly fluid and gaseous state of granular media, see Sec.\ref{rdf}, or more generally, in nematic liquid crystals~\cite{deGennes}?


The purpose of this section is to point out that {\sc gsh} is well capable of accounting for the shear band without any modification. We consider a system of uniform density and stress, with all variables stationary, such that Eqs.(\ref{props}) hold.
The balance equation (\ref{3b-9}) for $T_g$, accounting for $T_g$'s relaxation to 0 if $\pi_s<\pi_c$, implies $T_g\equiv0$ is the uniform stationary solution, see Sec.\ref{TgRelaxation}. For $\pi_s=\pi_c$, the system is in the critical state, $T_g$ does not relax  and the strain rate is indeterminate. For $\pi_s>\pi_c$, no uniform solution is stable, but a localized one is, with $T_g\equiv0$  for $x\le0$ or $x\ge\xi_{sb}$, and
\begin{align}\nonumber
\nabla^2 T_g=-T_g/\xi^2_{sb},\quad{\xi}^2_{sb}\equiv\xi_T^2/[\pi_s/\pi_c-1],
\\
 v_s/v_s^0=T_g/T_g^0=\sin(\pi x/\xi_{sb})\label{nsb2}
\end{align}			
in between. [Note the similarity to Eq.(\ref{cr1}). Allowing $\rho$ to vary will render $T_g$ differentiable at $0,\xi_{sb}$.] The  velocity difference from 0 to $\xi_{sb}$ is $\Delta v=\int v_s{\rm d}x=\int (T_g/f)\sqrt{\pi_s/\pi_c}{\rm d}x$, hence 
\begin{equation}
T_g^0/f=\sqrt{\pi_c/\pi_s}\, v_s^0 =\sqrt{\pi_c/\pi_s}\,\Delta v/(2\xi_{sb}).
\end{equation}
The critical state and the narrow shear band are the same rate-independent solution, behaving differently depending on how large $\pi_s$ is. That the correlation length $\xi_{sb}$ diverges for $\pi_s=\pi_c$ gives a retrospective justification of the term {\it critical.}
For increasing $\Delta v$, the variables $v_s,u_s,\Delta/u_s$ also grow, and the system will eventually leave the rate-independent, hypoplastic regime. Shear bands become wider then, and have to be treated as in Sec.\ref{sb}. 

The above is an idealized and simplified consideration of narrow shear bands, assuming uniform density and stress, and employing {\sc gsh} expressions that have been linearized and simplified. (Neither did we invoke the higher order strains terms of Sec.\ref{clogging}, implying in essence $\xi_{sb}\gg \theta,\theta_1$.) The qualitative and structurally stable part of the results is a localized shear band solution of {\sc gsh}, for overcritical stress values, with a characteristic length that decreases with increasing $\pi_s$. When approaching the critical state  non-monotonically, starting from a dense initial state, with $\pi_s>\pi_c$ for part of the path, there is a high probability for the $T_g$-instability discussed in Sec.\ref{scs} to occur and shear bands to form. 

Details such as the spacial distribution of $T_g(x)\propto v_s$, or that the friction angle $\sigma_s/P=\pi_s/P_\Delta$ decreases with increasing $T_c/T_g\propto\Delta{u_s}$, however, should be taken with a grain of salt, as these depend on the details and may change with the starting assumptions and  {\sc gsh} expressions. For instance, the original $T_g$ equation and the associated solution are
\begin{equation}
\label{Tg3}
\nabla_i (T_g\nabla_i T_g)=-T_g^2/\xi^2_{cb},\quad T_g=T_0\sqrt{\sin(\sqrt2\,x/\xi_{sb}}),
\end{equation}
see the discussion preceding Eq.(\ref{Tg1}), leading to the neglect of the nonlinear term $(\nabla_iT_g)^2$, small for slow variations, in both  Eq.(\ref{Tg1}) and (\ref{nsb2}). 
Same holds for Eq.(\ref{cr1}). 
Finally, if the density is nonuniform, say due to an aggregation of macropores, we will have $\pi_s>\pi_c(\rho)$ only in some regions. $T_g$ will be larger there, diffusing away, making the situation less clear-cut.

\subsection{Clogging and the Proximity Effect\label{clogging}}
The phenomenon of clogging implies that a free surface, if several grain diameter wide, may be stable even when facing downward, implying an angle of stability of $180^\circ$, 
much larger than than the  usual $30^\circ$ or $40^\circ$, as discussed around Eqs.(\ref{sb12}), valid only if the surface area is sufficiently large. 
Although {\sc gsh} in its present form, as given in Sec.\ref{GSH}, does not account for clogging, there is a tried and proven method of amending it. One example is the Ginzburg-Landau description of the superfluid  transition~\cite{Khal}, which includes gradients of the order parameter's magnitude in the energy. In the present case, we need to include gradients of the
elastic strain that express the extra energetic cost of a nonuniform strain field. Without these terms, unclogging occurs accompanied by a discontinuity in $\Delta,u_s$. 
With them, divergent gradients are forbidden by the infinite energy. A length scale on which elastic strains will change is thus introduced. 
%
With $w_\Delta=w_\Delta(u_{ij},\nabla_ku_{ij})$, $-\pi_{ij}\equiv\partial w_\Delta/\partial u_{ij}$, $\phi_{ijk}\equiv\partial w_\Delta/\partial\nabla_ku_{ij}$, the elastic and total stress are, respectively
\begin{equation}\label{GSE3}
\hat\pi_i\equiv\pi_{ij}+\nabla_k\phi_{ijk}, \quad \sigma_{ij}=[1-\alpha(T_g)]\hat\pi_i.
\end{equation}
Denoting the  two characteristic lengths as $\theta,\theta_1$, a simple example for such an energy is 
\begin{align}\label{GSE8}
w_\Delta=\sqrt{\Delta }[2{\mathcal B}\Delta ^2/5+
{\mathcal A}u_s^2] +{\mathcal A}(\theta\nabla_k
u_s)^2+{\mathcal B}(\theta_1\nabla_k
\Delta)^2],
\\\label{GSE9a}\text{implying}\quad
\hat P=P-2{\cal B}\theta_1^2\nabla_k^2\Delta,\quad 
\hat\pi_s=\pi_s+2{\cal A}\theta^2\nabla_k^2u_s,
\end{align}
with $P,\pi_s$ the uniform contributions, assuming $u^*_{ij}/|u_s|=$ const. Note that with this energy, the convexity transition,  $u_s/\Delta\le\sqrt{2{\cal B}/{\cal A}}$ of Eq.(\ref{2b-3}) is unchanged (though $\pi_s/P_\Delta\le\sqrt{2{\cal A}/{\cal B}}$ does change), because with $w=w_1(a)+w_2(\nabla a)$ and 
\begin{align*}
\delta^2w=\delta(\delta w)=\delta\left(\frac{\partial w}{\partial a}-\nabla\frac{\partial w}{\partial \nabla a}\right)\delta a=\delta\left(\frac{\partial w_1}{\partial a}-\nabla\frac{\partial w_2}{\partial \nabla a}\right)\delta a\qquad
\\
=\left(\frac{\partial^2 w_1}{\partial a^2}\delta a-\nabla\frac{\partial^2 w_2}{\partial (\nabla a)^2}\delta\nabla a\right)\delta a=\left(\frac{\partial^2 w_1}{\partial a^2}+\frac12\nabla^2\frac{\partial^2 w_2}{\partial (\nabla a)^2}\right)(\delta a)^2,
\end{align*}
$a$ standing for $\Delta$ or $u_s$, we have $\delta^2w/\delta a^2={\partial^2 w_1}/{\partial a^2}$.  (Note $\int\nabla[{\partial^2 w_2}/{\partial (\nabla a)^2}]\delta\nabla a\delta a = 
-\int\nabla^2 [{\partial^2 w_2}/{\partial (\nabla a)^2}]\delta a^2-\int\nabla[{\partial^2 w_2}/{\partial (\nabla a)^2}]\delta a\delta\nabla a$ if the surface integral vanishes.)

We employ this result and the model energy Eq.(\ref{GSE8}) to consider,  qualitatively, clogging and the proximity effect. More quantitative treatment will be provided in a 
separate work. First the effect that the angle of stability $\varphi_{st}$ is, for a few layers of grains, much larger than given in Eq.(\ref{sb12}). Simpler, that $\pi_s/P_\Delta$ can 
be larger than $\sqrt{2{\cal A}/{\cal B}}$ in a one-dimensional, simple shear geometry of the width $L$. 
We consider the strain fields for $-L<y<L$: $\Delta=\Delta_0$, $u_s=u_0+\alpha y^2/3L^2$ [ie. displasement $U_x=u_0y+\alpha y^3/12L^2$], with $\Delta_0,u_0=$ const, $u_0/\Delta_0\le\sqrt{2{\cal B}/{\cal A}}$  , and 
$\alpha\ll u_0$ such that the direct contribution to $\pi_s$ is negligible. Then $\hat P=P$, $\hat\pi_s=\pi_s+{\cal A}\alpha\theta^2/L^2$, and the uniform correction is considerable for  $\theta\gg L$. The fact that crushing is most efficient when the shearing walls are only a few grain diameters apart is clearly related to the above consideration that elastic solutions remains stable at large $\pi_s$. The grains remain static until they are crushed in narrow geometries, while transitioning into sliding, rotating, critical states in wider ones.    

Next, a crude model for  clogging. Since the stress vanishes for any free surfaces, we examine the 1D-situation in which it is zero for  $-L<x<L$, but finite at $x=\pm L$ and beyond. Taking
$\hat P_\Delta,\hat \pi_s=0$ as the differential equations, we solve them for $-L<x<L$ subject to the boundary conditions $\Delta=\Delta_0, u_s=u_0$ for  $x=\pm L$. Assuming for simplicity that $\theta_1\ll L$, we take  $\Delta\equiv\Delta_0$, implying $(1-\bar\theta^2\nabla_k^2)u_s=0$ with $\bar\theta={\theta}/{\sqrt[4]{\Delta_0}}$, or
\begin{equation}
\frac{u_s(x)}{u_0}\left[1+\exp\left(\frac{2L}{-\bar\theta}\right)\right]=\exp\left[\frac{x+L}{-\bar\theta}\right]+\exp\left[\frac{x-L}{\bar\theta}\right].
\end{equation}
If  $u_0$ satisfies the stability condition, $u_0/\Delta_0\le\sqrt{2{\cal B}/{\cal A}}$, the solution  $u_s(x)$ also does, and therefore represents a stable elastic situation.


\section{Rapid Dense Flow\label{rdf}} 
\subsection{The $\mu-$Rheology versus GSH\label{udf}}
When considering hypoplastic motion in the last section, \ref{hypoplastic motion}, we neglected the kinetic pressure $P_T$ and the viscous shear stress $\propto\eta_g$, see Eqs.(\ref{2c-2},\ref{2c-2a},\ref{2c-2b}). Here, we consider faster flows in which they are important, some times even dominant. Including them, we are leaving the rate-independent, hypoplastic regime. 
Being  quadratic in the shear rate, the correction come on slowly, leaving a large rate regime in which rate-independence holds.

How the stress of a system, in it stationary state, depends on the density $\rho$ and shear rate $v_s$, is called its {\it rheology}. Probing it over a wide range of shear rates is 
a useful inquiry for coming to terms with complex fluids including granular media.  Granular rheology has many facets, and typically, the shear rate $v_s$ is given. 
If it is low, the system executes complex elasto-plastic motion with a rate-independent stress, converging onto the critical state at constant rates, with a universal shear stress $\sigma_c$ that depends only on the density, not the rate or the initial stress, as considered in Sec~\ref{hypoplastic motion}. 
If  $v_s$ is high and the density sufficiently low, the system is in the Bagnold regime, with all components of the stress proportional to shear rate squared~\cite{Bagnold}. We consider the whole regime below. 
If the shear stress is given instead of $v_s$, circumstances are yet different. Examples are flows on an inclined plane or in a rotating drum, with a delay between jamming ({\it angle of repose} $\varphi_{re}$) and liquefaction ({\it angle of stability} $\varphi_{st}$), see Sec~\ref{sapc}. Part of the results of this section is in~\cite{P&G2013}.


\subsubsection{The $\mu-$Rheology\label{mu-rheo}} Sixty years ago, Bagnold examined how a granular system behaves at high rates and low densities, finding the pressure $P$ and shear stress $\sigma_s$ given as 
\begin{equation}\label{intro1}
P=e_p(\rho)v_s^2, \quad \sigma_s=e_s(\rho)v_s^2,  
\end{equation}
with $\mu_2\equiv\sigma_s/P=e_s/e_p$ a constant~\cite{Bagnold}. This result has been variously verified employing the kinetic theory to consider binary collisions among rarefied,  dissipative grains~\cite{kin1,kin2,kin3,kin4}.  

 A decade later, granular rheology at low rates and high densities was studied. Again, a surprisingly universal so-called {\it critical state} was observed~\cite{schofield,wood1990,nedderman,gudehus2010}. 
Starting from any initial stress, the system will, at constant densities and shear rates,  acquire values for the pressure and shear stress that depend on $\rho$ but not the rate, with the friction $\mu_1\equiv\sigma_s^c/P^c=$ const. 

Faced with these results, many find it plausible to account for the intermediate behavior by interpolating between the two rate- and density-independent plateaus~\cite{interpolate2,interpolate3,interpolate4,interpolate5},
\begin{equation}\label{intro2}
P=P^c+e_p(\rho)v_s^2, \,\, \sigma_s=\mu_1P^c+\mu_2e_pv_s^2,
\,\, \mu\equiv\sigma_s/P,
\end{equation}
implying $\mu\to\mu_1$ for $v_s\to0$ and  $\mu\to\mu_2$ for $v_s\to\infty$.

Embarking on an approach  independent from the above  and stressing first principles, the French research group GDR MiDi consider infinitely rigid grains~\cite{midi, pouliquen1}, and  point out that its rheology has only three independent  numbers: the friction $\mu$, the packing fraction $\phi\equiv \rho/\rho_g$ and the inertial number $I\equiv d \sqrt{\rho_g}( v_s/\sqrt P)$, with $\rho_g$ the bulk density, $d$ the granular diameter. 
Taking two as functions of the third, $\mu=\mu(I)$, $\phi=\phi(I)$, Forterre and  Pouliquen~\cite{pouliquen2}  take granular rheology to be accounted for by
\begin{equation}\label{poul}
\mu=\mu_1+({\mu_2-\mu_1})I/({I + I_0}),
\end{equation}
with $\mu_1\approx\sqrt2\tan 21^\circ$, $\mu_2\approx\sqrt2\tan 33^\circ$, $I_0\approx0.3$. Containing two plateaus, same as Eqs.(\ref{intro2}), this formula is shown capable of accommodating many experiments and simulations, and has recently also been successfully applied to dense suspensions~\cite{pouliquen4}.  

However, there is a fundamental problem. The relations $\mu=\mu(I)$, $\phi=\phi(I)$ are (irrespective of their functional dependences) equivalent to Eqs.(\ref{intro1}), implying  $P^c, \sigma_s^c=0$: First, $\phi=f(I)$ is clearly equivalent to $P=v_s^2/f^{-1}(\phi)$; second, $\mu(I)=\mu[f^{-1}(\phi)]$ is a function of  $\phi$ alone, and the two plateaus are for large and small packing fractions, respectively, impßlying  $P,\sigma_s\propto v_s^2$. However, this contradicts half a century worth of research in  soil mechanics, unambiguously showing {\it rate-independent stresses} for elasto-plastic motion, $v_s\to0$. 

The validity of Eqs.(\ref{intro1}) for infinitely rigid grains has been rigorously proven by Lois et al.~\cite{lois}, for any rates and densities, not only where the kinetic theory holds. 
Yet the speed of elastic waves in glass beads is between 350 and 800 m/s~\cite{jia2009}, which in comparison to air, water, bulk glass (with velocities of 300, 1500, 4000~m/s,  respectively) indicates  a very soft medium.  The difference between glass beads and bulk glass stems from the geometry of the Hertz contact. When considering binary collisions, assuming infinitely rigid grains reduces the collision time to zero, but does not change the physics qualitatively. Assuming incompressibility in dense media eliminates elastic waves and the critical state.

When arguing that one may treat grains as infinitely rigid, the authors of~\cite{lois}, citing a paper by Campbell~\cite{campbell}, assert that as long as $M\equiv dv_s/c_s$ (with $c_s$ the sound velocity) is small (typically for $ v_s\ll10^3$/s), grains behave as if they were perfectly stiff. This is oddly reversed, because quasi-static 
deformations occur at small rates, and are disrupted at higher ones. Indeed, 
perusing~\cite{campbell}, one finds Campbell stating clearly: (1)~It is the {\it inertially induced} contact deformation that vanishes with $M$. (2)~Stresses are generated by elastic deformations in the rate-independent, ``elastic-quasi-static''  regime. 

\subsubsection{The Dense Flow Results of GSH\label{vsGSH}} Treating dense granular media as compressible, {\sc gsh} shows the appropriateness of Eqs.(\ref{intro2}). Starting from Eqs.(\ref{2c-2},\ref{2c-2a}), we substitute the elastic contributions with the critical state expressions,  Eqs~(\ref{3b-4a}), appropriate for constant shear rates, while noting  $P_T=g_pT_g^2=g_pf^2v_s^2\equiv e_p v_s^2$, see Eq.(\ref{2b-5},\ref{Tcf}), also $\eta_1T_gv_s=\eta_1f v_s^2\equiv e_sv_s^2$, to obtain
\begin{equation}\label{3b-17}
P=P_c+e_p{v_s^2},
\,\,\sigma_{s}=\sigma_c+e_sv_s^2.
\end{equation}
%
%
The observed density-independence of $\mu_1,\mu_2$ implies the constancy of  $P_c(\rho)/\sigma_c(\rho)=\mu_1$ and $e_s(\rho)/e_p(\rho)=\mu_2$,  an experimental input.
(A viscous stress linear in $v_s$, as observed in~\cite{campbell} at high densities, has not been included above but is a possibility, see~\cite{granR2,granR3}.  It appears if a macroscopic shear flow not only heats up $T_g$, but $T$ as well. This is the case for instance when the sand is saturated with water. The pressure $P$ would receive a term liner in $T_g$ if  one modify the energy, $w=w_T+w_\Delta+w_x$, by adding a cross term such as $w_x=cs_g\Delta^{1.5}$. The total pressure $P$ and $T_g$ then obtain the respective additive term, $cs_g\sqrt\Delta$ and $c\Delta^{1.5}$, implying that the linear term  $\propto T_g$ exists only for $\rho>\rho_{\ell p}$ and $\Delta\ne0$.)

If the density is low, $\rho<\rho_{\ell p}$, the grains lack enduring contacts and no elastic solution is stable, $P_\Delta,\pi_s=0$, see Eq.(\ref{2b-4}). Then Eqs.(\ref{intro1}) are the appropriate formulas.
When studying granular rheology by varying the shear rate $ v_s$, one can either keep the density or the pressure constant. For any realistic shear rates, we have $P^c\gg e_p(\rho) v_s^2$, $\sigma_s^c\gg e_s(\rho) v_s^2$, and it is hard to arrive at the $\mu_2$-limit for given $\rho$. Not so for given pressure, because the density decreases for increasing $v_s$. A discontinuous transition from  Eqs.(\ref{intro2}) to (\ref{intro1}) takes place at $\rho=\rho_{\ell p}$,  when $\mu$ jumps from $\mu_1$ to $\mu_2$, while $P, \sigma_s$ decrease dramatically, by around three orders of magnitude~\cite{campbell}. 

Two important points remain to be discussed, first when and why there is, as observed~\cite{Brodsky1,Brodsky2}, a minimum in the shear stress as a function of the rate; and second, why the MiDi relation is, in spite of its shortcomings, so successful.
Keeping the density constant, $P^c(\rho)$,   $\sigma_s^c(\rho)$, $e_p(\rho)$, $e_s(\rho)$ also are, implying $P,\sigma_s$ increase monotonically with  $v_s^2$.  Keeping $P=$ const, the circumstances, though still given by  Eqs.(\ref{intro2}), are different. 
In the hypoplastic regime, the shear stress at given pressure,  $\sigma_s^c=\mu_1P$, is simply a constant, similarly in the Bagnold regime, $\sigma_s=\mu_2P$. 
In between, both $\rho,\sigma_s$ are rate- and density-dependent, given by  $\sigma_s=\mu_1P^c(\rho)+\mu_2e_p(\rho) v_s^2$, with $\rho(P, v_s)$ from $P=P^c(\rho)+e_p(\rho) v_s^2$ plugged in.  Then there is no reason for  $\sigma_s(P, v_s)$ to be monotonic, see~\cite{StressDip} for more details.

Eqs.(\ref{intro2}) are algebraic relations that hold for uniform systems. To account for nonuniform ones, gradient terms from {\sc gsh} become important, and the large set of
nonlinear, partial differential equations that  {\sc gsh} is needs to be solved. Even disregarding this, there are still complications that one needs to heed. 
For instance, enforcing a constant total volume does not prevent the local density to vary, 
and a stress dip may still occur. 

To understand this better, consider two uniform volumes $V_1,V_2$, with $V_1+V_2=$ const. Being in contact via a flexible membrane, they  may serve as a simple model for the continuous non-uniformity of a constant volume 
experiment. Initially, the
total system is uniform, with both densities equal,
$\rho_1=\rho_2$, and both shear rates vanishing,
$\dot\gamma_1, \dot\gamma_2=0$. Now, if $\dot\gamma_2$ is
cranked up, but $ \dot\gamma_1$ remains zero, because of
pressure equality,
$P_1(\rho_1, \dot\gamma_1)=P_2(\rho_2, \dot\gamma_2)$, the
density must change and the membrane will stretch, with $\rho_2$ decreasing and
$\rho_1$ increasing. If system~1 is much larger than 2,
the stretching of the membrane will not change $\rho_1$
much, and $P_1(\rho_1, \dot\gamma_1)$ will remain
essentially constant as a result. So will $P_2=P_1$, and 
the pressure-controlled limit holds in system 2. Otherwise, we have an
intermediate case between the pressure- and
density-controlled limits. In both cases,  a stress dip may appear.


Finally, we give  three reasons for the undeniable success of the MiDi relation: First, $\phi=\phi(I)$ is correct for $\rho<\rho_{\ell p}$, while $\mu=\mu(I)$ as given by Eq.(\ref{poul}) is right for  $\rho>\rho_{\ell p}$. Very few papers span both limits and employ both relations simultaneously. 
Second, many experiments are nonuniform, lying between the density- and pressure-controlled limits. An unreflective comparison of the relation to a subset of data such as the average density or stress is then neither accurate nor discriminating. In fact, by employing Eqs.(\ref{intro1},\ref{intro2}),  Berzi et al.~\cite{interpolate5} were able to achieve  quantitative agreement with both the simulation on simple shear in~\cite{daCruz} and the experiment on incline flows in~\cite{Opouli}. Both were deemed strong support for  the MiDi relation. 
Third, the frequently observed collapse of different curves, when $\mu$ is plotted as a function of $I$, may be understood because $\mu$ depends on $\hat I\equiv e_p(\rho) v_s^2/P$ alone, and $\hat I$ is close to $I^2$. [One writes $\mu=({\sigma_c}/{P})({P-e_p v_s^2})/{P_c}+{e_s v_s^2}/{P}$ $=\mu_1(1-e_pv_s^2/P)+\mu_2e_p/P$ $=\mu_1+(\mu_2-\mu_1)\hat I$.
Generally speaking, granular rheology is given by $P,\sigma_s=f(\rho,v_s)$. One may switch to $\rho,\phi=f(P,v_s)$ or $\mu,\phi=f(I,v_s)$, two variables remain and there is no collapse. $\mu=f(\hat I)$, $\hat I\stackrel{v_s\to\infty}{\longrightarrow}1$ is an exception.] Note depending whether $\rho$ or $P$ is being held constant, one must take, respectively,  $\hat I=e_p(P,v_s^2)v_s^2/P$ and  $\hat I=e_p(\rho)v_s^2/P(\rho,v_s^2)$.

\subsection{Wide Shear Bands\label{sb}} 

The narrow shear band has already been considered in Sec.\ref{nsb}. Here, we consider a wide shear band, which is in essence the coexistence of static granular solid and uniform dense flow. In the first, the grains are deformed and at rest, $T_g=0$, with all energy being elastic. In the 
second, the grains jiggle, rattle, move macroscopic distances,  with $T_g\propto v_s$ and a  portion of the energy in $T_g$. 
Increasing the shear rate, the transition from the rate-independent critical state to the 
Bagnold regime of dense flow is, as discussed in Sec~\ref{udf}, continuous at given density and discontinuous at given pressure, but always uniform. Here, we consider a nonuniform path,  a narrow shear band that suddenly appears, as the result of an instability, see Sec.\ref{nsb}, then continuously widens as  the externally applied velocity difference increases, until the band covers the whole system, and uniformity is restored.  

Approaching the critical state with a high initial density, the evolution of the shear stress 
$\sigma_s$ is non-monotonic, assuming overcritical values part of the path. This is 
where the system has a high probability of breaching an instability, either of the elastic energy at a point on the yield surface, as discussed  around Eq.(\ref{2b-3}), or that of $T_g$, as discussed in Sec~\ref{scs}. 
The transition is difficult to account for, but the stable shear band is again simple.

As we have seen, the narrow shear band of low shearing velocity $v$ has a rate-independent width. If $v$ is higher, the system's behavior depends on the setup. For given pressure, the width $\ell$ grows linearly with $v$, implying a constant rate $v/\ell$ in the liquid phase. As a result, the shear stress, a function of the rate, remains independent of $v$. This {\em faux rate-independence} goes on until the band covers the whole system, at which point the quadratic rate dependence of uniform dense flow sets in. 
For given volume, the band width remains independent of $v$, but the shear stress grows quadratically with it. The transition to uniform dense flow is for given volume  discontinuous. It happens when the shear stress exceeds the critical value of the solid density, at which point the solid phase is no longer stable.

To understand wide shear bands, we study the simple case of uniform fluid and solid regions connected via a flat surface. (Separately, they are already understood.) Denoting the solid and fluid parts with the superscripts $^S$ and $^F$,  respectively, these two regions have equal pressure, shear stress, and chemical potential,
\begin{equation}\label{3b-21}
P^S=P^F,\quad \sigma_s^S=\sigma_s^F, \quad\mu^S=\mu^F.
\end{equation}
[The chemical potential is defined as $\mu\equiv\partial w/\partial\rho$, see Eq~(\ref{2-2}). The equality holds because otherwise a particle current would flow across the phase boundary.]  All three fields have an elastic and a seismic contribution:  With $P=({1-\alpha})P_\Delta+P_T$, ${\sigma_s}=({1-\alpha})\pi_s+\eta_1T_g v_s$, see Eqs~(\ref{2c-2},\ref{2c-2a}), and $\mu=\mu_\Delta+\mu_T$,  where
\begin{align}
\label{5-7}
\mu_T\equiv T_g^2\,\frac {b_0\rho}2\left[1-\frac{\rho}{\rho_{cp}}\right]^a\frac{(1+a)\rho-\rho_{cp}}{\rho_{cp}-\rho},
\\\mu_\Delta\equiv{0.15w_\Delta
}{(\rho_{cp}-\bar\rho)}/[{(\rho_{cp}-\rho)}{(\rho-\bar\rho)}].\end{align}
Denoting  the width of the shear band as $\ell$, and the velocity difference across the shear band as $v$,  we take
\begin{flalign}
\text{in fluid:}&\quad v_s=v/\ell\propto T_g,\,\, \Delta^F=\Delta_c,\,\, u_s^F=u_c,
\\
\text{in solid:}&\quad \alpha, T_g, v_s=0.
\end{flalign}
In other words, the elastic strain $\Delta$ and $u_s$ have critical values in the $F$-phase, and appropriate static values in the $S$-phase.  Strictly speaking, the discontinuities at the $S-F$ boundary are in $\rho,\Delta, u_s$, but not in  $T_g$ and $v_s$, as both  diffuse into the solid, decaying exponentially there, see Sec~\ref{creep motion}. We neglect this detail, approximating the decay with a discontinuity to keep the formulas simple, and to work  at the qualitative understanding first. The price we pay is a slightly fuzzy $\ell$ that includes the two decay zones in the solid.

\subsubsection{The Fluid Region}
The elastic contribution $\mu_\Delta$ is a very small quantity: In $P_\Delta\propto{\cal B}\Delta^{1.5}$, a large ${\cal B}$ compensates a small $\Delta^{1.5}$, such that $P_\Delta$ is either much larger than, or comparable to, $P_T\propto T_g^2$. Now,  $\mu_T$ is of the order of $P_T/\rho$, but $\mu_\Delta\propto{\cal B}\Delta^{2.5}\propto \Delta P_\Delta$ is smaller by the factor $\Delta$, around $10^{-3}-10^{-4}$. Therefore, as long as $P_T\gg\Delta P_\Delta$, we have  $\mu_T\gg\mu_\Delta$, and $\mu^S=\mu^F$ reduces to $\mu_T=0$, implying  the density in the shear band is (in dry sand) fixed as
\begin{equation}\label{5-8}
\rho^{F}=\rho_{cp}/(1+a).
\end{equation}
Measuring $\rho^{F}$ therefore yields the value of $a$, see Eq~(\ref{2b-5}). In what follows, we need to assume a sufficiently small $a$, such that $\rho^F>\rho_{\ell p}$. Because $\Delta^F=\Delta_c (\rho^F), u_s^F=u_c(\rho^F)$, the elastic pressure $P_\Delta(\rho,\Delta,u_s)$ in the fluid is also known. 

\begin{figure}[t]
\begin{center}
\includegraphics[scale=.45]{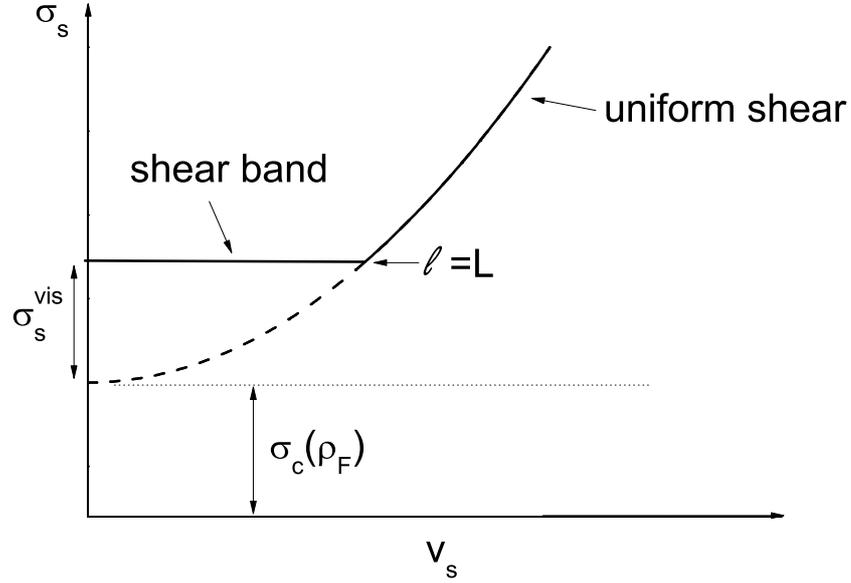}
\end{center}\caption{Faux rate-independence: Shear stress $\sigma_s$ as a function of the velocity difference $v$, or of the apparent shear rate $v_s\equiv v/L$,  for given pressure, in a simple-shear geometry. The offset gives the elastic contribution, $\sigma_c(\rho_F)$; the parabola is the case without a shear band. The thick horizontal line depicts the situation with a shear band, of width $\ell$, which is smaller towards left, and equal to the system's width $L$ at the right end. The rate-independence of $\sigma_s$ derived from  $\ell$ adjusting itself such that $T_g\propto v/\ell$ remains constant for given pressure.  \label{fig4}} 
\end{figure}

{\bf Given Pressure\quad}
Next, we consider the case of given velocity difference $v$ across the shear band, and given external pressure, $P^{ex}=P^S=P^F$,
\begin{align}\label{sb-1}
P=P_c(\rho^F)+\frac{T_g^2}{2}\,\frac{(\rho^F)^2\,a\,b/\rho_{cp}}{(1-\rho^F/\rho_{cp})},
\\\label{sb-2}
\sigma_s=\sigma_c(\rho^F) -\eta_1T_g\,v/\ell.
\end{align}
Since $P^F,\rho^F$ fix $T_g,v_s=T_g/f$, and for given $v$, the width of the shear band  $\ell=v/v_s$ is also fixed, we have all there is to know about the fluid region. 
Remarkably, the system now displays a faux rate-independence: $\ell$ adjusts itself such that $T_g\propto v/\ell$ remains constant for given pressure,
independent what $v$ is. The parabola of Fig~\ref{fig4} depicts  $\sigma_s$. The offset gives the elastic contributions, $\sigma_c$. The horizontal line is a result of  $\ell$ adjusting.   

Increasing the velocity $v$ at given pressure alters the width $\ell$, as long as it is smaller than the width of the total system $L$. For larger velocities, the system is again uniform, without a solid region. And the consideration of Sec~\ref{udf} holds. Until this point, the stress is rate-independent, much longer than without a shear band. 

Given the solid density $\rho^S$ (see Sec.\ref{sr}), and the mass per unit surface $M$, mass conservation $\rho^S(L-\ell)+\rho^F\ell=M$ determines the total width $L$ for given pressure $P$. 

{\bf Given Total Volume\quad}
At given total volume $L$, because of mass conservation
and because $\rho^S,\rho^F$ are given in addition to $L$, the band width $\ell$ is fixed, irrespective what the velocity $v$ is. As a result, both the shear stress and pressure grow as $(v/\ell)^2\propto v^2$, not at all rate-independent. The transition to uniform dense flow happens discontinuously, when Eq~(\ref{94}) is violated, for $\sigma_c(\rho^S)=\sigma^S$.

\subsubsection{The Solid Region\label{sr}}
Because we have terms of such different magnitudes in the connecting condition $\mu^S=\mu^F$, it fixes $\rho^F$ instead of 
giving a relation between $\rho^F$ and  $\rho^S$. Therefore, the condition is always satisfied, 
irrespective what value $\rho^S$ assumes. So $\rho^S$ can only be  a result of the dynamics: When an instability is breached, the density is changed until it gets stuck at 
some value for $\rho^S$, at which the system is again stable. Then of course, $\Delta^S, u_s^S$ may be determined for given pressure and shear stress.
Nevertheless, we do know  
\begin{equation}\label{92}
\rho^F<\rho^S \quad\text{and}\quad \rho^F\le\rho_c(P)
\end{equation}
must hold. The first inequality can be seen from 
\begin{equation}\label{94}
\sigma_c(\rho^S)>\sigma^S=\sigma_c(\rho^F)+\eta_1T_gv_s\ge\sigma_c(\rho^F),
\end{equation}
where the first greater sign is related to the discussion in Sec~\ref{scs}; the equal sign is a connecting condition; and the second greater sign is a result of $\eta_1T_gv_s$ being positive, in addition to the fact that $\sigma_c$ is a monotonically increasing function of the density, cf. the discussion in Sec.\ref{Stationary Elastic Solution}. The second inequality, $\rho^F\le\rho_c(P)$, holds because of two reasons: First,  
in the critical state, there is only one free parameter. Once $\rho$ is given, $\Delta_c,u_c,P_c,
\sigma_c$ also are. Alternatively, one may fix the external pressure $P$, then $\rho_c(P)$ is a dependent quantity. Second, given  $P=P_c(\rho^F)+P_T$, we have $\rho_F=\rho_c$ for $P_T=0$. $\rho^F$ may be smaller, but if it were larger,  shear band will not exist, and the flow is uniform. 
For $\rho^F<\rho_{\ell p}$, there is no elastic contribution in the shear band, $P_c,\sigma_c=0$ in Eqs.(\ref{sb-1},\ref{sb-2}), and Eq.(\ref{92}) holds trivially. All other conclusions remain valid, also Fig.\ref{fig4}, though without the offset $\sigma_c$.  

When the velocity $v$ decreases, the above consideration stops to be valid at some point. For instance, $\rho^F$ is no longer given if $P_T\gg\Delta P_\Delta$ does not hold. But before this happens, the narrow band solution should already have taken over.


\section{Velocity and Damping of Elastic Waves\label{elastic waves}}
That elastic waves propagate in granular media~\cite{jia1999,jia2004} is an important fact, because it is an unambiguous proof that granular media possess an 
elastic regime. In this section, we consider elastic waves and propose to employ them as a tool to detect the elastic to plastic transition.
There is a wide-spread believe in the granular community that small, quasi-static increments from any 
equilibrium stress state is elastic, but large ones are plastic.  As discussed in 
Sec~\ref{quasi elastic motion}, this assumption appears illogical, because any large increment can always be taken as the sum of small ones. In {\sc gsh}, the parameter that sets the 
boundary between elastic and plastic regime is the granular temperature $T_g$. We have quasi-elastic regime for vanishing $T_g\propto v_s^2$, and the 
hypoplastic one for elevated $T_g\propto v_s$. 

A perturbation in the elastic strain or stress propagate as a wave only in the quasi-elastic regime, while it diffuses in the hypoplastic one. 
More specifically, we derive a telegraph equation from {\sc gsh}, with a quantity $\propto T_g$ taking on the role of the electric resistance~\cite{zhang2012}. It defines a 
characteristic  frequency $\omega_0=\lambda T_g$, such that elastic perturbations of the frequency $\omega$ diffuse for $\omega\ll\omega_0$, and propagate for 
$\omega\gg\omega_0$.  We have $\omega_0\to0$ in the quasi-elastic regime, so all perturbations propagate. In the hypoplastic regime, when $T_g$ is elevated, so is 
$\omega_0$, pushing the propagating range to ever higher frequencies. Eventually, the associated wave length become comparable to the granular diameter,  exceeding  {\sc gsh}'s range of validity.  

To derive the telegraph equation, we start with two basic equations of {\sc gsh}, Eqs~(\ref{2c-8},\ref{2c-2a}),
\begin{align}  \label{ew1}
\rho\partial_tv_i-(1-\alpha)\nabla_mK_{imkl}u^*_{kl}=0, \\
\partial_tu^*_{ij}-(1-\alpha)v^0_{ij}=-\lambda T_gu^*_{ij},  \label{ew2}
\end{align}
with $K_{imkl}\equiv\partial^2w/\partial u_{im}\partial u_{kl}$. (For simplicity, we concentrate on shear waves, assuming $v_{\ell\ell}\equiv0 $.) 
For $T_g\to0$, the plastic terms $\lambda T_gu^*_{ij}$ and $\alpha\propto T_g$ are  negligibly small, and these two equations represent conventional elasticity theory. 
The  wave velocity $c$ [given by the eigenvalues of $K_{imnj}q_mq_n/(\rho q^2)$ with $q_m$ the wave vector], as a function of stress and density, is then easily calculated. The results~\cite{ge4} agree well with observations~\cite{jia2009}.

There are two ways to crank up $T_g$ and the plasticity, either by introducing external perturbations $T_a$, or by increasing the amplitude 
of the wave mode, because its own shear rate also creates $T_g$. The characteristic time of $T_g$ is $1/R_T\lesssim10^{-3}$~s in dense media, see Eq~(\ref{Tg2}). Therefore, we assume that the  wave mode's frequency is much larger than $R_T$ , such that $T_g$ and $\alpha(T_g)$ are essentially constant, or
\begin{align}\label{ew3}
2(\partial^2_t+\lambda T_g\partial_t)\,u^*_{ij}=(1-\alpha)^2\times\qquad
\\
\nabla_m[K_{imkl}\nabla_ju^*_{kl}+K_{jmkl}\nabla_iu^*_{kl}].
\nonumber
\end{align}
Concentrating on one wave mode along $x$, with $c_{\rm qs}$ the quasi-elastic, $c\equiv(1-\alpha)c_{\rm qs}$ the actual velocity, and $\bar u\propto e^{iqx-i\omega t}$ the eigenvector's amplitude, we have the telegraph equation, 
\begin{equation}\label{ew4}
(\partial^2_t+\lambda T_g\partial_t)\,\bar u= (1-\alpha)^2c_{\rm qs}^2\nabla_x^2\,\bar u
\equiv c^2\nabla_x^2\,\bar u.
\end{equation}
The coefficient  $(1-\alpha)^{2}$, accounting for granular contacts softening  and the effective elastic stiffness decreasing, is, in the language of electromagnetism,  the inverse dielectric permeability.
Inserting $\bar u\propto e^{iqx-i\omega t}$ into Eq~(\ref{ew4}), we find
$c^2 q^2={\omega^2+i\omega\lambda T_g}$,
implying diffusion for the low frequency limit, $\omega\ll\lambda T_g$,
\begin{equation}
q\approx\pm\frac{\sqrt{\omega\lambda T_g}}c \, \frac{1+i}{\sqrt2},
\end{equation}
and propagation for the high-frequency limit, $\omega\gg\lambda T_g$,
\begin{align}
cq\approx\pm\omega \left(1+i\,{\lambda T_g}/{2\omega}
\right),
\\
\bar u\propto\exp{\left[-i\omega\left(t\mp x/c\right) \mp x({\lambda T_g}/{2 c})\right]}.
\end{align}
The first term in the square bracket accounts for wave propagation, the second a decay length $2c/\lambda T_g$, independent of the  frequency if $T_g=T_a$ is an ambient temperature. It is strongly frequency and  amplitude dependent if  $T_g=f|v_s|\propto\omega q\bar u\propto\omega^2\bar u$ is produced by the elastic wave itself, because the inverse length varies with $T_g$, going from $T_g\propto v_s^2$ to $T_g\propto v_s$. 
     
A brief wave pulse, arbitrarily strong, can always propagate through granular media if its duration is too brief to excite sufficient $T_g$ for the system to enter the hypoplastic regime. The duration must be much smaller than $T_g$'s characteristic time  $1/R_T$. 
        
\section{Compaction\label{compaction}}
The present understanding of {\it compaction under tapping} takes it to be a rather insular phenomenon, in need of an special entropy not useful for any of the other granular phenomena. We shall return to the so-called {\it Edwards entropy} in Sec.\ref{tapping}, after having pondered whether tapping may be related to a  ubiquitous variety of compaction that has been known to engineers for a long time, the slow increase of the density at given pressure under shear, or in the presence of an ambient temperature $T_a$. This more typical phenomenon is easily  understood to be a result of the fact that $\Delta$ relaxes, as accounted for by Eq.(\ref{eqD}). Keeping the pressure $P_\Delta={\cal B}(\rho)\Delta^{1.5}$ constant, the density increases to compensate. (Note that Approaching the critical state under a constant shear, the circumstances are more general, because $\Delta$ relaxes and is being increased by a shear rate at the same time. It may increase, leading to dilation, or decrease, to  contraction, as considered in Sec~\ref{pressure approach}.) 

\subsection{Reversible and Irreversible Compaction}
Consider the pressure $P=(1-\alpha)P_\Delta+P_T$ assuming vanishing shear strain and rate, $u_s,v_s=0$, with $P_\Delta$ the elastic, and $P_T$ the seismic, contribution, see Eqs~(\ref{2b-2b},\ref{2b-5},\ref{2b-4}),
\begin{equation}\label{gc5}
P_\Delta={}{\cal B}(\rho)\Delta^{1.5},\quad P_T= g_p(\rho)T_g^2
\end{equation}
where both $\cal B$ and $g_p$ are, for dense media, monotonically increasing functions of $\rho$. 
At small $T_g$, the seismic pressure $P_T$ may be neglected, so $\rho$ must increase when $\Delta$ relaxes,  for $P=P_\Delta=$ const. The increase is irreversible because the relaxation is. This is the limit most soil mechanical experiments are in. Only irreversible compaction is observed. 

For $T_g$  larger, the seismic pressure $P_T$ needs to be included. Because the density change in $g_p$ is faster than in $\cal B$, the relaxation of $\Delta$  increases $P_T$ and decreases $P_\Delta$, with $P_\Delta+P_T=$ const. After the relaxation has run its course, $\Delta,P_\Delta\to0$, if one modifies 
$T_g$ (ie. the amplitude of the perturbation) but maintains $P=P_T$, the density will change in response, in both direction and reversibly. Since $P_T(\rho,T_g)\equiv\partial(w/\rho)/\partial(1/\rho)$ is a thermodynamic derivative,  the change is also thermodynamic. 

\subsection{History Dependence versus Hidden Variables\label{hdvhv}}

Changing $T_g$ midway at constant $P$, with $\Delta$ still finite,
will mainly lead to a change in $\Delta$,  because the density responds much more slowly. 
It disrupts the relaxation of $\Delta$, in essence resetting its initial
condition. This phenomenon was observed in~\cite{mem} and interpreted as a
memory effect. Generally speaking, ``memory" is
usually a result of hidden variables: When the system
behaves differently in two cases, although all state
variables appear to have the same values, we speak of memory-, or
history-dependence. But an overlooked variable that
has different values for the two cases will naturally
explain the difference. In the case of compaction, the manifest and hidden
variables are  $\rho$ and $\Delta$, respectively.

\subsection{Tapping and the Edwards Entropy\label{tapping}}

Numerous experiments have shown that tapping leads to  reversible and irreversible compaction, see the review article~\cite{1nico}. It is usually accounted for by the 
specifically tailored {\it granular statistical mechanics}~\cite{Edw} and the Edwards 
entropy $S_{Ed}$, or some generalization of it. Substituting the volume
$V$ for the energy $E$, and compactivity $X$ for the
temperature $T$, this theory employs ${\rm d}V=X{\rm
d}S_{Ed}$ as the basic thermodynamic relation for a
{\it``mechanically stable agglomerate of infinitely
rigid grains at rest"}~\cite{Edw}. The entropy
$S_{Ed}$ is obtained by counting the 
possibilities to package grains stably for a given volume,
equating it to $e^{S_{Ed}}$. 

Two reasons prompt us to doubt  its appropriateness. 
First, the number of possibilities to arrange
grains concerns inter-granular degrees of freedom. These are vastly overwhelmed by the much more numerous configurations of the inner-granular degrees of freedom. In other words, the Edwards entropy $S_{Ed}$ is a special case of the granular entropy $S_g$, and as discussed in the introduction, we always have $S_g\ll S$. 
One would be able to neglect $S$ and concentrate on $S_g$ if these two were only weakly coupled, if the energy decay from $S_g$ to $S$ were exceedingly slow. This is not the case, the relaxation of  $T_g\propto s_g$ is fast.

Second, even assuming a weak coupling, 
$S_{Ed}$ would still be a overwhelmed measure. The starting point of the Edwards entropy is the fact that the energy $E$ is always zero for infinitely rigid,  non-interacting grains at rest, however they are packaged. Taking $S_g$ generally as a function of energy and volume, $S_g(E,V)$, we have, 
\[{\rm d}S_g=\frac{\partial S_g}{\partial E}{\rm d}E+\frac{\partial
S_g}{\partial V}{\rm d}V \equiv\frac1{T_g}{\rm d}E+\frac P{T_g}{\rm d}V.\] 
Usually, one keeps the volume constant, and consider ${\rm d}S_g=(1/T){\rm d}E$. Taking instead $E\equiv0$, we have ${\rm d}S_g=(P/T)
{\rm d}V$, equivalent to the Edwards expression ${\rm d}V=(T/P){\rm d}S_g\equiv X{\rm d}S_{Ed}$. 

This derivation ignores three essential points: First, perturbing the system, allowing it to explore the phase space, introduces kinetic energy that one must include. But then $E\not\equiv0$. Second, because of the Hertz-like contact 
between grains, little material is deformed at first contact, and the compressibility diverges at 
vanishing compression. This is a geometric fact independent of how rigid the bulk material is. 
Therefore, infinite rigidity is never a realistic limit in granular media, and there is always considerable 
elastic energy stored among grains in mechanically stable agglomerates.  Third, $S_{Ed}$ as defined is the granular entropy at vanishing granular motion and compression. Its phase space is therefore severely constrained. 
Generally speaking, each classical particle has states in a 6D space, three for the position and three for the velocity. The Edwards entropy only includes states in a 3D space. So $\exp(S)$ is the number of states times the Loschmidt's number; $\exp(S_g)$ is the number of states in 6D space times the number of grains, and $\exp(S_{Ed})$ is the number of states in 3D space (no velocities)  times the number of grains. Therefore  
\begin{equation}
S_{Ed}\ll S_g\ll S.
\end{equation}
Going toward equilibrium, a system searches for the greatest number of states to equally redistribute its energy. One bears the burden of proof for the claim that it is sensible for the system to neglect $S,S_g$ and concentrate on $S_{Ed}$. In contrast, {\sc gsh} identifies compaction as a process taking place at finite $T_g$ and compares the true entropy $S$ of macrostates at that $T_g$. It also accounts for entropy increase, by detailing how macroscopic energy decays into granular heat, and how this is converted to true heat. 

More specifically, maximizing the true entropy $S$, {\sc gsh} obtains two sets of equilibrium conditions, one for the solid and another  for the fluid state~\cite{granR2,granR3}, 
\begin{align}
\nabla_i\pi_{ij}=\rho\, {g}_i,\quad T_g=0;\\
\pi_{ij}=0, \,\,\,
\nabla_iP_T=\rho\, {g}_i.
\end{align}
The first is the result of $T_g$ vanishing quickly, leaving a jammed, elastically deformed system. The second (implying $\Delta, P_\Delta=0$)  holds, when $T_g=T_a$ is being maintained externally. This is the limit of reversible, thermodynamic compaction, for $\Delta=0$.

Reversible and irreversible compaction as accounted for by {\sc gsh} is a universal granular 
phenomenon. It occurs at given pressure and  $T_a$, however $T_a$ is created. This corresponds well to the observation that tapping, though especially efficient,  is but one way to achieve compaction, leading to results vary similar to that of many other methods~\cite{1nico}.  So it is natural to take the consideration of the last section to hold for tapping as well. This rings true for gentle tapping, but stronger one warrants further scrutiny.  

Gentle tapping leads to granular jiggling and a small $T_g$,  
though one that fluctuates in time, with periodic flare-ups. As long as  $P_T$ may be neglected, 
$\Delta$ will relax according to the momentary value of $T_g$, haltingly but monotonically. 
Since the relaxation is a slow process, one could average over many taps to yield a coarse-grained 
account. Given a granular column with a free upper surface in the 
gravitational field, because a given layer is subject to a constant pressure, the density will increase to  compensate for the diminishing $\Delta$. The characteristic time of $\Delta$-relaxation diverges towards the end, and is not a constant, see~\cite{compaction}.  

Stronger tapping leads to a higher $T_g$, with $\Delta$ relaxing more quickly.  
$P_T$ must now be included. Periodically, when all grains are at rest, $P_T$ vanishes, and  $\Delta$ is necessarily increased to maintain the given pressure. This introduces a non-monotonicity into $\Delta(t)$, and raises the question, whether the system, when being tapped again to arrive at an elevated $T_g$, will pick up the relaxation of $\Delta$ where it was left when the system last crushed to a stop, and also why it should do so. If it does, we can again take tapping as coarse-grainable, intermittent compaction. Then {\sc gsh} indeed provides a complete picture for compaction, with an 
understanding  that is transparent, conventional and demystified. 

\section{The Quasi-Elastic Regime\label{quasi elastic motion}}
Textbooks on soil mechanics take granular motion in the hypoplastic  regime -- say the approach to the critical state -- to be quasi-static,  because it is slow and rate-independent. Yet since it is also strongly dissipative and irreversible, we do not believe this is right: Quasi-static motion is not dissipative. 

Consider sound propagation in any system including Newtonian fluid, elastic medium or liquid crystals. The sound velocity is always an order in the frequency lower than the damping.
This is a general feature: Changing a state variable $A$ slowly, dissipation is $\propto\partial_tA$. For $\omega\to0$, the motion is free of dissipation and rate-independent. One calls it  {\em quasi-static} because the system is at this frequency visiting static states consecutively.  

In the hypoplastic regime, reactive and dissipative terms in {\sc gsh} are of the same order in  the frequency, and comparable in size -- they are exactly equal in the critical state --  and elastic waves are over-damped. 
So there must be  a true quasi-static regime at even lower frequencies. 
A more convoluted explanation, popular in the geotechnical community, is to assume that a small incremental strain is elastic and free of dissipation, but a large one is elasto-plastic and dissipative. Unfortunately, this is incompatible with the basic notion of  quasi-static motions: Starting from a static state of given stress, and applying a small incremental strain that is elastic, the system is again in a static state and an equally valid starting point. The next small increment must therefore again be elastic. Many consecutive small increments yield a large change in strain, and if the small ones are not dissipative, neither can their sum be. This cannot go on for ever, and the limit is
the elastic convexity transition of Eq.(\ref{2b-3}), at which no elastic state is stable. So in the quasi-elastic regime, granular media behave in accordance to the simplest elasto-plastic theory: completely elastic for small shear stresses, and ideally plastic when the yield stress is breached.  

Together, these reasons let us believe that it is $T_g$, rather than strain amplitude that decides whether the system is elastic or elasto-plastic.  Of course,  small strain increments achieved with a high but short lasting shear rate will indeed provoke elastic responses, if $T_g$ does not have time to get large and produce plastic responses. 
To be specific, we quote a few numbers, well aware that these are at best educated guesses for the case of dry sand: 
The Bagnold regime starts at rates of one or two hundred Hz, the hypoplastic regime is say between $10^{-3}-1$Hz, and quasi-elastic regime lies possibly below $10^{-5}$Hz.

{\sc gsh} accounts well both for static stress distribution and the hypoplastic regime. 
Its prediction of what should happen in between, in the quasi-elastic regime, derives from a continuous connection of these two behavior, and is not yet verified 
experimentally.  Granular media are taken to be completely elastic, with the elastic energy given by Eq.(\ref{2b-2}) for the static case of identically vanishing shear rate, 
$v_s\equiv0$. Many known static stress distributions have thus been successfully reproduced, including silos, sand piles and point load on a granular sheet, 
see~\cite{ge1,ge2,granR1}. Also, Incremental stress-strain relation starting from varying static stress points~\cite{ge3}, and the propagation of anisotropic elastic waves at varying static stresses~\cite{ge4} are well accounted for. 
The elasto-plastic motion that are on display for hypoplastic shear rates and elevated $T_g$ is also in full agreement with experiments and state-of-the-art engineering theories such as hypoplasticity and barodesy, see Sec.\ref{Constitutive Relations}. 

Given the two limits, there is only little leeway of how to connect both. {\sc gsh} employs  $h$ of Eq.(\ref{2c-6}) as the switch, such that $h=1$ and $T_g\propto v_s$ in the rate-independent hypoplastic regime, while $h\to\infty$ and $T_g\propto v_s^2$ quadratically small in the quasi-static one. Since deviations from  elasticity of all expressions vanish with $T_g\to0$, the transition is smooth.

For experiments at given shear rates, the key difference between the hypoplastic and quasi-elastic regime lies in whether the system retrace the stress-strain curve when the rate is reversed, see next section. For experiments at given shear stresses (employing a soft spring, see Sec.\ref{soft springs} below) in the hypoplastic regime, an initially elevated $T_g$ will relax sufficiently slowly to give rise to an apparently diverging creep, see Sec.\ref{aging}. This does not happen in the quasi-elastic regime. The first was observed in~\cite{aging}, and the authors concluded reasonably that the system harbors a slow dynamics and is not quasi-static.

\subsection{The Steep Stress-Strain Trajectory\label{A Steep Stress Trajectory}}
As discussed above, in the quadratic regime of very slow shear rates,  $T_g\propto |v_s|^2\to0$, the granular temperature is so small that the system is essentially elastic, moving from one elastic equilibrium state to a slightly different elastic one. This is the reason we call it {\em quasi-elastic}. Because $\sigma_s=\pi_s$ and $\partial_t u_s=\partial_t\varepsilon_s=v_s$, the change of the the shear stress $\sigma_s$ is well approximated by the (hyper-) elastic relation, 
\begin{equation}\label{3a-1}
\partial_t\sigma_s=\frac{\partial\sigma_s}{\partial
u_s}\partial_t u_s =\frac{\partial\pi_s}{\partial
u_s}\partial_t\varepsilon_s=-\frac{\partial^2 w}{\partial
u_s^2}v_s. 
\end{equation} 
Shearing a granular medium at quasi-elastic rates, the result will be a trajectory $\sigma_s(\varepsilon_s)$ that is much steeper than in experiments at hypoplastic rates, such as observed during an approaching to the critical state. The gradient is given directly by the stiffness constant ${\partial^2 w}/{\partial u^2_s}$, and possibly three to four times as large as the average between loading and unloading at hypoplastic rates [because Eq~(\ref{2c-8}) lacks the factor of $(1-\alpha)$]. This goes on until the system reaches a yield surface of the elastic energy, say Eq.(\ref{2b-3}). We expect the system to form shear bands at this point, see Sec~\ref{nsb},\ref{sb}. The critical state will not be reached. Reversing the shear rate in between will retrace the function $\sigma_s(\varepsilon_s)$. 

\subsection{Soft Springs versus Step Motors\label{soft springs}}
\begin{figure}[t] 
\hspace{-1.5cm}
\includegraphics[scale=0.5]{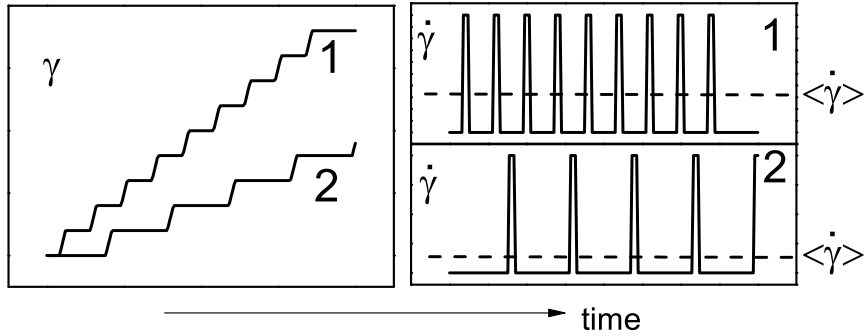}
\caption{Why it is hard to observe the quasi-elastic regime if step motors are used, see text.\label{StepMotor}} 
\end{figure} 

Quasi-elastic behavior has not been observed in triaxial apparatus, even at the lowest rates. This maybe because they are simply not slow enough. But we suspect other 
reasons: First, triaxial experiments are frequently performed with sand saturated in water, and squeezing water through the narrow gaps between grains is an efficient mean 
of producing $T_g$. This may push the transition from the elastic to hypoplastic regime to much lower rates than in dry grains. second,  the wide usage of step motors in 
the triaxial appliances may have contributed to a wrong perception. Plotting the shear rate versus time, $ v_s(t)$, different shear rates are approximately given as depicted by 
the two curves of Fig~\ref{StepMotor}. Although the curves have different average rates $\langle v_s\rangle$, the time-resolved, maximal rates  $ v_s^{Max}$ are identical. And if 
the time span of  $ v_s^{Max}$ is long enough for $T_g$ to respond, and $ v_s^{Max}$ is high enough for the system to be in the linear regime, $T_g\propto  v_s^{Max}$, the 
system will display consecutive hypoplastic behavior in both cases, irrespective of  the average rate $\langle v_s\rangle$. 

We suggest two ways here to enter the quasi-elastic regime. Since a given slow stress rate has a high shear rate at elevated $T_g$ and a low one at vanishing $T_g$, the idea is to find the latter.  One method is to slowly tilt an inclined plane supporting a layer of grains. In such a situation, the shear rate remains very small, and the system starts flowing only when a yield surface is breached. In contrast, employing a feedback loop in a triaxial apparatus to maintain a stress rate would not work well, because the correcting motion itself typically has strain rates that are too high.
A second method is to insert a very {soft spring}, even a rubber band, between the 
granular medium and the device moving at a given velocity $v$ to deform it. If the spring is softer by a large factor $a$ than the granular medium, it will absorb most of the displacement, leaving the granular medium deforming at a rate smaller by
the same factor $a$ than without the spring. In other words, the soft spring serves as a ``stress reservoir'' for the granular medium. The same physics applies when the feedback loop is connected via a soft spring. Little $T_g$ is then excited, as in the experiment~\cite{aging}, see Sec~\ref{aging}.

\section{Conclusions}
This paper represents half a decade worth of attempts to come to terms, at least qualitatively, with the  many observations of granular dynamics, by employing {\sc gsh} as the description and unifying framework. We are happy to report that it has not failed us once, although the outcome was rarely obvious when we started to examine a new experiment. Retrospectively, of course,  circumstances appear much clearer and naturally systematic, and this is also how we present them above. The range of phenomena considered is clearly considerable, much wider than any macro-theory to date. Necessarily, a number of corollary predictions have also been made, especially in the context of wide shear bands and the quasi-elastic regime. They cry out for verification. Also, an observation of the difference between yield (or elastic instability) and the critical state would be highly desirable.


\begin{thebibliography}{99} 
\bibitem{schofield}
P.~Wroth A.~Schofield.
\newblock {\em Critical State Soil Mechanics}.
\newblock McGraw-Hill, London, 1968.

\bibitem{nedderman}
R.M. Nedderman.
\newblock {\em Statics and Kinematics of Granular Materials}.
\newblock Cambridge University Press, 1992.

\bibitem{wood1990}
D.~M. Wood.
\newblock {\em Soil Behaviour and Critical State Soil Mechanics}.
\newblock Cambridge University Press, 1990.

\bibitem{kolymbas1}
D.~Kolymbas.
\newblock {\em Introduction to Hypoplasticity}.
\newblock Balkema, Rotterdam, 2000.

\bibitem{kolymbas2}
W.~Wu and D.~Kolymbas.
\newblock {\em Constitutive Modelling of Granular Materials}.
\newblock Springer, Berlin, 2000.

\bibitem{gudehus2010}
G.~Gudehus.
\newblock {\em Physical Soil Mechanics}.
\newblock Springer SPIN, 2010.

\bibitem{goddard2013}
J. Goddard.
\newblock {\em Revs. Appl. Mech.}
\newblock (to be published, 2013)


\bibitem{hutter2007}
S.P. Pudasaini and K.~Hutter.
\newblock {\em Avalanche Dynamics}.
\newblock Springer, 2007.

\bibitem{LL6}
L.~D. Landau and E.~M. Lifshitz.
\newblock {\em Fluid Mechanics}.
\newblock Butterworth-Heinemann, 1987.

\bibitem{Khal}
I.~M. Khalatnikov.
\newblock {\em Introduction to the Theory of Superfluidity}.
\newblock Benjamin, New York, 1965.

\bibitem{deGennes}
P.G. de~Gennes and J.~Prost.
\newblock {\em The Physics of Liquid Crystals}.
\newblock Clarendon Press, Oxford, 1993.

\bibitem{microMech}  See eg. F. Nicot and F. Darve, {\it Mechanics of Materials}
{\bf 37}-9, 980 (2005);
and {\it Second International Symposium on Computational Geomechanics (ComGeo II)}, 27-29/04/2011, Cavtat-Dubrovnik, HRV; and references therein.

\bibitem{hydro-1}S. R. de Groot and P. Masur,
{\it Non-Equilibrium Thermodynamics}, (Dover, New York 1984).

\bibitem{hydro-2}D. Forster, Hydrodynamic Fluctuations, Broken Symmetry and
Correlation Functions (Benjamin, New York, 1975).

\bibitem{liqCryst-1} P.G. de Gennes and J. Prost, {\em The Physics of
Liquid Crystals} (Clarendon Press, Oxford 1993). 

\bibitem{liqCryst-2} P.C.
Martin, O. Parodi, and P.S. Pershan, {\it Unified Hydrodynamic Theory for
Crystals, Liquid Crystals, and Normal Fluids}, Phys. Rev. A 6, 2401 (1972).

\bibitem{liqCryst-3} T.C. Lubensky, {\it Hydrodynamics of Cholesteric
Liquid Crystals}, Phys. Rev. A 6, 452 (1972). 

\bibitem{liqCryst-4} M. Liu,
{\it Hydrodynamic Theory near the Nematic Smectic-A Transition}, Phys. Rev.
{\bf A 19}, 2090 (1979); 

\bibitem{liqCryst-5} M. Liu, {\it Hydrodynamic
theory of biaxial nematics}, Phys. Rev. {\bf A 24}, 2720 (1981).
\bibitem{liqCryst-6} M. Liu, {\it Maxwell equations in nematic liquid
crystals}, Phys. Rev. {\bf E 50}, 2925, (1994). 

\bibitem{liqCryst-7} H.
Pleiner and H.R. Brand, in {\it Pattern Formation in Liquid Crystals},
edited by A. Buka and L. Kramer (Springer, New York, 1996).

\bibitem{he3-1} R. Graham, {\it Hydrodynamics of 3He in Anisotropic A
Phase}, Phys. Rev. Lett. {\bf 33}, 1431 (1974). 

\bibitem{he3-2} R. Graham
and H. Pleiner, {\it Spin Hydrodynamics of 3He in the Anisotropic A Phase},
Phys. Rev. Lett. {\bf 34}, 792 (1975). 

\bibitem{he3-3} M. Liu, {\it
Hydrodynamics of $^3$He near the A-Transition,} Phys. Rev. Lett. {\bf 35},
1577 (1975). 

\bibitem{he3-4} M. Liu and M.C. Cross, {\it Broken Spin-Orbit
Symmetry in Superfluid $^3$He and the B-Phase Dynamics,} Phys. Rev. Lett.
{\bf 41}, 250 (1978). 

\bibitem{he3-5} M. Liu and M.C. Cross, {\it Gauge
Wheel of Superfluid $^3$He,} Phys. Rev. Lett. {\bf 43}, 296 (1979).

\bibitem{he3-6} M. Liu, {\it Relative Broken Symmetry and the Dynamics of
the $A_1$-Phase,} Phys. Rev. Lett. {\bf 43}, 1740 (1979).

\bibitem{SC-1} M. Liu, {\it Rotating Superconductors and the
Frame-independent London Equations,} Phys. Rev. Lett. {\bf 81}, 3223,
(1998). 

\bibitem{SC-2} Jiang Y.M. and M. Liu, {\it Rotating Superconductors
and the London Moment: Thermodynamics versus Microscopics,} Phys. Rev. {\bf
B 6}, 184506, (2001). 

\bibitem{SC-3} M.~Liu, {\em Superconducting
Hydrodynamics and the Higgs Analogy,} J. Low Temp. Phys. 126, 911, (2002)

\bibitem{hymax-1} K. Henjes and M. Liu, {\it Hydrodynamics of Polarizable
Liquids,} Ann. Phys. {\bf 223}, 243 (1993). 

\bibitem{hymax-2} M. Liu, {\it
Hydrodynamic Theory of Electromagnetic Fields in Continuous Media,} Phys.
Rev. Lett. {\bf 70}, 3580 (1993). 

\bibitem{hymax-3} {\it Mario Liu
replies,} Phys. Rev. Lett. {\bf 74}, 1884, (1995). 

\bibitem{hymax-4} Y.M.
Jiang and M. Liu, {\it Dynamics of Dispersive and Nonlinear Media,} Phys.
Rev. Lett. {\bf 77}, 1043, (1996).

\bibitem{FF-1} M.I. Shliomis, {\em Magnetic Fluids}, Sov. Phys. Usp. 17,
153 (1974). 

\bibitem{FF-2} R.E. Rosensweig, {\em Ferrohydrodynamics},
(Dover, New York 1997). 

\bibitem{FF-3} M. Liu, {\it Fluiddynamics of
Colloidal Magnetic and Electric Liquid,} Phys. Rev. Lett. {\bf 74}, 4535
(1995). 

\bibitem{FF-4} M. Liu, {\it Off-Equilibrium, Static Fields in
Dielectric Ferrofluids,} Phys. Rev. Lett. {\bf 80}, 2937, (1998).

\bibitem{FF-5} M. Liu, {\it Electromagnetic Fields in Ferrofluids}, Phys.
Rev. {\bf E 59}, 3669, (1999). 

\bibitem{FF-6} H.W.~M\"{u}ller and M.~Liu,
{\it Structure of Ferro-Fluiddynamics,} Phys. Rev. {\bf E 64}, 061405
(2001). 

\bibitem{FF-7} H.W. M\"{u}ller and M. Liu, {\em Shear Excited Sound
in Magnetic Fluid}, Phys. Rev. Lett. {\bf 89}, 67201, (2002).

\bibitem{FF-8} O. M\"{u}ller, D. Hahn and M. Liu, {\em Non-Newtonian
behaviour in ferrofluids and magnetization relaxation,} J. Phys.: Condens.
Matter 18, 2623, (2006). 

\bibitem{FF-9} S. Mahle, P. Ilg and M. Liu, {\em
Hydrodynamic theory of polydisperse chain-forming ferrofluids,} Phys. Rev.
{\bf E 77}, 016305 (2008).

\bibitem{FF-10} M. Liu and K. Stierstadt, ``Thermodynamics, Electrodynamics,
and Ferrofluid Dynamics,'' in {\it Colloidal Magnetic Fluids: Basics, Development and Application of Ferrofluids}, Lect. Notes Phys. 763, edited by S. Odenbach, (Springer, Berlin Heidelberg 2009), DOI 10.1007/978-3-540-85387-9

\bibitem{polymer-1} H. Temmen, H. Pleiner, M. Liu and H.R. Brand, {\it
Convective Nonlinearity in Non-Newtonian Fluids,} Phys. Rev. Lett. {\bf
84}, 3228 (2000). 

\bibitem{polymer-2}H. Temmen, H. Pleiner, M. Liu and H.R.
Brand,{\it Temmen et al. reply}, Phys. Rev. Lett. {\bf 86}, 745 (2001).

\bibitem{polymer-3} H. Pleiner, M. Liu and H.R. Brand, {\it Nonlinear Fluid
Dynamics Description of non-Newtonian Fluids}, {Rheologica Acta} {\bf 43},
502 (2004). 

\bibitem{polymer-4}O. M\"{u}ller, {\em Die Hydrodynamische
Theorie Polymerer Fluide}, PhD Thesis University T\"{u}bingen (2006).

\bibitem{granR2}
Y.M.~Jiang and M.~Liu.
\newblock Granular solid hydrodynamics.
\newblock {\em Granular Matter}, 11:139, May 2009.
Free download: http://www.springerlink.com/content/a8016874j8868u8r/fulltext 

\bibitem{granR3}
Y.M.~Jiang and M.~Liu.
\newblock The physics of granular mechanics.
\newblock In D.~Kolymbas and G.~Viggiani, editors, {\em Mechanics of Natural
  Solids}, pages 27--46. Springer, 2009.
\bibitem{granRgudehus}
G.~Gudehus, Y.M. Jiang, and M.~Liu.
\newblock Seismo- and thermodynnamics of granular solids.
\newblock {\em Granular Matter}, 1304:319--340, 2011.

\bibitem{granR4}  
  Y.M.~Jiang and M.~Liu.
  \newblock Granular Solid Hydrodynamics  (GSH): a broad-ranged macroscopic theory of granular media.
\newblock {\em Acta Mech.}, Acta Mech. (2014)
10.1007/s00707-014-1131-3.
  
 

\bibitem{GGas}Y.P. Chen, M.Y. Hou, Y.M. Jiang, and M. Liu
Hydrodynamics of granular gases with a two-peak distribution
{\it Phys. Rev.} {\bf E88}, 052204 (2013)
\bibitem{luding2011}
V. Magnanimo, S. Luding
A local constitutive model with anisotropy for ratcheting under 2D
axial-symmetric isobaric deformation.
\newblock {\em Granular Matter}, 13:225-232, 2011.
\bibitem{ge4}
M.~Mayer and M.~Liu.
\newblock Propagation of elastic waves in granular solid hydrodynamics.
\newblock {\em Phys. Rev. E}, 82:042301, 2010.
\bibitem{luding2009}
Stefan Luding.
\newblock Towards dense, realistic granular media in 2d.
\newblock {\em Nonlinearity}, 22:101--146, 2009.

\bibitem{granL3}
Y.M.~Jiang and M.~Liu.
\newblock From elasticity to hypoplasticity: Dynamics of granular solids.
\newblock {\em Phys. Rev. Lett.}, 99(10):105501, 2007.

\bibitem{Houlsby}
G.~T. Houlsby and A.~M. Puzrin.
\newblock {\em Principles of Hyperplasticity}.
\newblock Springer (2006).

\bibitem{Houlsby2}
I.~F. Collins and G.~T. Houlsby.
\newblock Application of thermomechanical principles to the modelling of
  geotechnical materials.
\newblock {\em Proc. R. Soc. Lond. A}, 453:1975--2001, 1997.

\bibitem{rubin} 
M.B: Rubin, Physical reasons for abandoning plastic deformation measures in plasticity 
and viscoplasticity theory. Arch. Mech. 53 (4–5), 519–553 (2001).



\bibitem{Bocquet}
L.~Bocquet, W.~Losert, D.~Schalk, T.~C. Lubensky, and J.~P. Gollub.
\newblock Granular shear flow dynamics and forces: Experiment and continuum
  theory.
\newblock {\em Phys. Rev. E}, 65(1):011307, Dec 2001.


\bibitem{ge1}
D.~O. Krimer, M.~Pfitzner, K.~Br\"auer, Y.M.~Jiang, and M.~Liu.
\newblock Granular elasticity: General considerations and the stress dip in
  sand piles.
\newblock {\em Phys. Rev. E)}, 74(6):061310, 2006.

\bibitem{ge2}
K.~Br\"auer, M.~Pfitzner, D.~O. Krimer, M.~Mayer, Y.M.~Jiang, and M.~Liu.
\newblock Granular elasticity: Stress distributions in silos and under point
  loads.
\newblock {\em Phys. Rev. E (Statistical, Nonlinear, and Soft Matter Physics)},
  74(6):061311, 2006.

\bibitem{granR1} Y.M. Jiang, M. Liu. Eur. A brief review of granular elasticity. Phys. J. \textbf{E~22,} 255 (2007).

\bibitem{kuwano2002}
R.~Kuwano and R.~J. Jardine.
\newblock On the applicability of cross-anisotropic elasticity to granular
  materials at very small strains.
\newblock {\em Geotechnique}, 52(10):727--749, Dec 2002.

\bibitem{ge3}
Y.M.~Jiang and M.~Liu.
\newblock Incremental stress-strain relation from granular elasticity:
  Comparison to experiments.
\newblock {\em Phys. Rev. E (Statistical, Nonlinear, and Soft Matter Physics)},
  77(2):021306, 2008.
  
\bibitem{jia2009}
Y.~Khidas and X.~Jia.
\newblock Anisotropic nonlinear elasticity in a spherical-bead pack: Influence
  of the fabric anisotropy.
\newblock {\em Phys. Rev. E}, 81:021303, Feb. 2010.

\bibitem{3inv} Y.M. Jiang, H.P. Zheng, Z. Peng, L.P. Fu, S.X. Song, Q.C. Sun, M. Mayer, and M. Liu, \newblock Expression for the granular elastic energy.
\newblock  Phys. Rev. E {\bf 85}, 051304 (2012)    


\bibitem{hardin}
B.O. Hardin and F.E. Richart.
\newblock Elastic wave velocities in granular soils.
\newblock {\em J. Soil Mech. Found. Div. ASCE}, 89: SM1:33--65, 1963.

\bibitem{denseFlow}
Stefan~Mahle, Yimin~Jiang and Mario~Liu.
\newblock Granular solid hydrodynamics: Dense flow, fluidization and jamming.
\newblock {\em arXiv:1010.5350v1 [cond-mat.soft]}, 2010.


  \bibitem{midi}
GDR MiDi.
\newblock On dense granular flows.
\newblock {\em The European Physical Journal E}, 14(4):341--365 (2004).

\bibitem{roux} J.-N. Roux.  How granular materials deform in quasistatic conditions.  AIP Conf. Proc. 1227, pp. 260-270; doi:http://dx.doi.org/10.1063/1.3435396; Quasistatic behaviour of granular materials: Some things we learned from DEM studies. AIP Conf. Proc. 1542, 46 (2013); http://dx.doi.org/10.1063/1.4811865
\bibitem{h^2} I.~Einav. 
\newblock The unification of hypo-plastic and elasto-plastic theories.
\newblock{\it International Journal of Solid and Structure}, {\bf 49}(2012) 1305-1315


\bibitem{critState}
Stefan Mahle, Yimin Jiang, and Mario Liu.
\newblock The critical state and the steady-state solution in granular solid
  hydrodynamics.
\newblock {\em arXiv:1006.5131v3 [physics.geo-ph]}, 2010.
\bibitem{GSH&Barodesy} Yimin Jiang, and Mario Liu. Proportional Path, Barodesy, and Granular Solid Hydrodynamics.  {\it Granular Matter} {\bf 15}, 237 (2013).
\bibitem{barodesy} Kolymbas D. Barodesy: a new constitutive frame for soils. Geotechnique Letters 2, 17–23,  (2012), http://dx.doi.org/10.1680/geolett.12.00004; 
Barodesy: A new hypoplastic approach. International Journal for Numerical and Analytical Methods in Geomechanics (2011). doi:10.1002/nag.1051;
Sand as an archetypical natural solid. In Mechanics of Natural Solids, Kolymbas D, Viggiani G (eds.). Springer: Berlin, (2009); 1–26; 
\bibitem{wichtmann} T. Wichtmann, Schriftreihe Inst. Grundbau u. Boden\-me\-cha\-nik, Univ. Bochum, Heft~38, (2005), Fig~4.17.
\bibitem{thornton} C. Thorn\-ton, S.J. Antony, {\it Phil.Trans.R.Soc.A: Mathematical, Physical and Engineering Sciences}, 
{\bf 356}, 
No. 1747, Mechanics of Granular Materials in Engineering and Earth Sciences (Nov. 15, 1998), 2763-2782 (1998).
\bibitem{SJ1}D.P. Bi, J. Chang, B. Chakraborty, R.P. Behringer, {\it Nature}, {\bf 480}, 355 (2011) 
\bibitem{SJ2}N. Kumar, Stefan Luding, to be published 
\bibitem{vHecke2011}
J.A. Dijksman, G.H. Wortel, L.T.H. van Dellen,
O. Dauchot, and M. van Hecke. 
\newblock Jamming, yielding, and rheology of weakly vibrated granular media.
\newblock {\em Phys. Rev. Lett.}, {\bf 107},
108303(2011).
\bibitem{StressDip} D. Krimer, S. Mahle and M. Liu, Dip of the Granular Shear Stress, {\it Phys. Rev.} {\bf E86}, 061312 (2012)

\bibitem{P&G2009}  
  Y.M.~Jiang and M.~Liu.
  \newblock GSH, or Granular Solid Hydrodynamics:
on the Analogy between Sand and Polymers.
\newblock {\em AIP Conf. Proc.} 7/1/2009, Vol. 1145 Issue 1, p1096.

\bibitem{aging}
Van~Bau Nguyen, Thierry Darnige, Ary Bruand, and Eric Clement.
\newblock Creep and fluidity of a real granular packing near jamming.
\newblock {\em Phys. Rev. Lett}, 107:138303, 2011.


\bibitem{aranson}
I.~S. Aranson and L.~S. Tsimring.
\newblock {\em Phys. Rev. E}, 65:061303, 2002.
\bibitem{Aranson1}
I.~S. Aranson and L.~S. Tsimring.
\newblock {\em Rev. Mod. Phys.}, 78:641, 2006.




\bibitem{komatsu}
T.S. Komatsu, S.~Inagaki, N.~Nakagawa, and S.~Nasuno.
\newblock Creep motion in a granular pile exhibiting steady surface flow.
\newblock {\em Phys. Rev. Lett.}, 86:1757�1760, 2001.

\bibitem{crassous}
J~Crassous, J-F Metayer, P~Richard, and C.~Laroche.
\newblock Experimental study of a creeping granular flow at very low velocity.
\newblock {\em J. Stat. Mech.}, 2008:P03009, 2008.


\bibitem{kamrin}D.L. Henann and K. Kamrin. {\it Proceedings of  the National  Academy of  Sciences},{\bf 110}, 6730 (2012). http://www.pnas.org/content/110/17/6730.full. 
\bibitem{kamrin2} K. Kamrin and G. Koval. {\it Phys.Rev.Lett.} {\bf108},178301 (2012)

\bibitem{fenistein} D. Fenistein, J.W. van de Meent, M.van Hecke, {\it Nature}, {\bf 425} 695 (2003); {\it Phys.Rev.Lett.} {\bf 96}, 118001 (2004); {\bf 96}, 038001 (2006). 


\bibitem{nichol2010}
Kiri Nichol, Alexey Zanin, Renaud Bastien, Elie Wandersman, and Martin van
  Hecke.
\newblock Flow-induced agitation creates a granular fluid.
\newblock {\em Phys. Rev. Lett.}, 104:078302, 2010.

\bibitem{reddy2011}
K.A. Reddy, Y.~Forterre, and O.~Pouliquen.
\newblock Evidence of mechanical activated processes in slow granular flows.
\newblock {\em Phys. Rev. Lett.}, 106:108301, 2011.

\bibitem{huerta2005} D. A. Huerta, Victor Sosa, M. C. Vargas, and J. C. Ruiz-Suárez.
\newblock Archimedes’ principle in fluidized granular systems.
\newblock {\em Phys. Rev.}, {\bf E72}, 031307(2005).
\bibitem{caballero2009} G.A. Caballero-Robledo and E. Clement.
\newblock Rheology of a sonofluidized granular packing.
\newblock {\em Eur. Phys. J. } {\bf E 30}, 395–401 (2009).


\bibitem{wu2}Wei Wu. {\em On high-order hypoplastic models for granular materials}. Journal of Engineering Mathematics {\bf 56}: 23–34 (2006) 

\bibitem{wu1}Tejchman, J. and Wu, W. {\it FE-investigations of micro-polar boundary conditions along interface between soil and structure}, Granular Matter, {\bf 12}, 399 (2010)



\bibitem{Bagnold}
R.~A. Bagnold.
\newblock Experiments on a gravity-free dispersion of large solid spheres in a
  {N}ewtonian fluid under shear.
\newblock {\em Proceedings of the Royal Society of London. Series A.
  Mathematical and Physical Sciences}, 225(1160):49--63, 1954.

\bibitem{P&G2013}  
  Y.M.~Jiang and M.~Liu.
  \newblock  Stress- and rate-controlled granular rheology
\newblock {\em AIP Conf. Proc.} 1542, 52 (2013); doi: 10.1063/1.4811867


\bibitem{kin1} J. T. Jenkins and S. B. Savage, {\it J. Fluid Mech.} {\bf 130}, 187 (1983).
\bibitem{kin2} 
S. B. Savage, {\it Adv. Appl. Mech.} {\bf 24}, 289 (1984).
\bibitem{kin3} 
C.S. Campbell, {\it Ann. Rev. Fluid Mech.} {\bf 22}, 57 (1990).
\bibitem{kin4} I. Goldhirsch, {\it Chaos} {\bf 9}, 659 (1999) and  {\it Annu. Rev. Fluid Mech.} {\bf 35}, 267 (2003).
  
\bibitem{interpolate2}P.C. Johnson, R. Jackson. Frictional-collisional constitutive relations for granular materials, with application to plane shearing.
{\it J. Fluid Mech.} {\bf 176}, 67–93  (1987).
\bibitem{interpolate3}
M.Y. Louge. Model for dense granular flows down bumpy inclines.
{\it Phys. Rev.} {\bf E 67}, 061303  (2003).
\bibitem{interpolate4} C. Josserand, P.Y. Lagre, D. Lhuillier. Granular pressure and the thickness of a layer jamming on a rough incline. {\it Europhys. Lett.} {\bf 73}, 363–69  (2006).
\bibitem{interpolate5} D. Berzi, C. G. di Prisco, D. Vescovi. {\it Phys. Rev.} {\bf E 84}, 031301 (2011)

    
\bibitem{pouliquen1}
Pierre Jop, Yo\"{e}l Forterre, and Olivier Pouliquen.
\newblock A constitutive law for dense granular flows.
\newblock {\em Nature}, 441:727--730, 2006.

\bibitem{pouliquen2}
Yo\"{e}l Forterre and Olivier Pouliquen.
\newblock Flows of dense granular media.
\newblock {\em Annu. Rev. Fluid Mech.}, 40:1--24, 2008.

\bibitem{pouliquen4} 
F. Boyer, E. Guazzelli, O. Pouliquen, {\it Phys. Rev. Lett.} {\bf 107}, 188301 (2011).

\bibitem{lois}G. Lois, A. Lemaitre, J. Carlson. {\it Phys. Rev.} {\bf E 72}, 051303  (2005).
\bibitem{campbell} C. S. Campbell, {\it J. Fluid Mech.} {\bf 465}, 261 (2002).


\bibitem{Brodsky2}
K.~Lu, E.E. Brodsky, and H.P. Kavehpour.
\newblock {\em J. Fluid. Mech.}, 587:347, 2007.

\bibitem{Brodsky1}
K.~Lu, E.E. Brodsky, and H.P. Kavehpour.
\newblock {\em Nature Letters}, 4:404, 2008.

\bibitem{daCruz} F. da Cruz, S. Emam, M. Prochnow, J. N. Roux, and F. Chevoir,
{\it Phys. Rev.} {\bf E 72}, 021309 (2005).
\bibitem{Opouli} O. Pouliquen, {\it Phys. Fluids} {\bf 11}, 542 (1999)




\bibitem{jia1999}
X.~Jia, C.~Caroli, and B.~Velicky.
\newblock Ultrasound propagation in externally stressed granular media.
\newblock {\em Phys. Rev. Lett.}, 82(9):1863--1866, Mar 1999.

\bibitem{jia2004}
X.~Jia.
\newblock Codalike multiple scattering of elastic waves in dense granular
  media.
\newblock {\em Phys. Rev. Lett.}, 93(15):154303, Oct 2004.

\bibitem{zhang2012}
Q.~Zhang, Y.C. Li, M.Y. Hou, Y.M. Jiang, and M.~Liu.
\newblock Elastic waves in the presence of a granular shear band formed by
  direct shear.
\newblock {\em Phys. Rev. E}, 85:031306, 2012.






\bibitem{mem}C. Josserand, A.V. Tkachenko, D.M.
Mueth, H.M. Jaeger, Phys. Rev. Lett., \textbf{85}, 3632
(2000)


\bibitem{1nico} P. Richard, M. Nicodemi, R. Delannay,
P. Ribiere, D. Bideau, Nature, {\bf 4}, 121 (2005)


\bibitem{Edw} S.F. Edwards, R.B.S. Oakeshott,
Physica A{\bf 157}, 1080 (1989); S.F. Edwards, D.V.
Grinev, Granular Matter,  {\bf 4}, 147 (2003).

\bibitem{compaction} Yimin Jiang, and Mario Liu.
\newblock The critical state and the steady-state solution in granular solid
  hydrodynamics.
\newblock {\em arXiv:0911.2199v2 [cond-mat.soft]}, 2010.


\end{thebibliography}
  \end{document}